\documentclass[aps,prresearch,reprint,groupedaddress]{revtex4-2}
\usepackage[hidelinks]{hyperref}
\usepackage{orcidlink}
\usepackage{amsmath}
\usepackage{amssymb}
\usepackage{siunitx}

\usepackage{tikz}
\usepackage{mathrsfs,pgf,pgfplots,pgfplotstable}
\usetikzlibrary{shapes, positioning, chains, arrows, decorations.text, decorations.markings,
    arrows.meta, calc, shapes.geometric, angles, quotes, intersections,
    datavisualization.formats.functions, datavisualization.polar, external, fadings, tikzmark}
\pgfplotsset{compat=1.18} % Set the compatibility level for pgfplots

\usepgfplotslibrary{polar}

\usepackage{caption}
\usepackage{subcaption}
\usepackage{booktabs}

\definecolor{blue}{RGB}{31,119,180}
\definecolor{orange}{RGB}{255,127,14}
\definecolor{green}{RGB}{44,160,44}
\definecolor{red}{RGB}{214,39,40}
\definecolor{purple}{RGB}{148,103,189}
\definecolor{brown}{RGB}{140,86,75}
\definecolor{pink}{RGB}{227,119,194}
\definecolor{gray}{RGB}{127,127,127}
\definecolor{olive}{RGB}{188,189,34}
\definecolor{cyan}{RGB}{23,190,207}

\newlength\xTemp
\newlength\yTemp
\newlength\aTemp
\newlength\yOffset
\newlength\figW

\tikzset{
    waveArrow/.style={
            -stealth, very thick, shorten >=\s, shorten <=\s, opacity=1,
        },
    tikzPlot/.style = {
            trim axis left,
            trim axis right,
            inner frame sep=0,
        },
}

\pgfplotsset{axisStyle/.style={
            axis lines=none,
            xmin=-1, xmax=1,
            ymin=-1.2, ymax=1.2,
            samples=500,
            scale=0.05,
            anchor=center,}
}

\newcolumntype{M}[1]{>{\centering\arraybackslash}m{#1}}

\pgfmathsetmacro{\xmin}{-1.1}
\pgfmathsetmacro{\xmax}{1.1}
\pgfmathsetmacro{\ymin}{-1.1}
\pgfmathsetmacro{\ymax}{1.1}

\pgfplotsset{
    resultsAxis/.style = {
            width=\xTemp,
            height=\yTemp,
            at={(0,0)},
            scale only axis,
            minor tick num=1,
            axis on top,
            xmin=-0.8,
            xmax=0.8,
            ymin=-0.8,
            ymax=0.8,
            xtick=\empty, % Remove x-axis ticks
            ytick=\empty, % Remove y-axis ticks
            xticklabels={}, % Remove x-axis tick labels
            yticklabels={}, % Remove y-axis tick labels
            xlabel={},
            ylabel={},
            axis lines=none,
            anchor=center,
        },
}

\tikzset{
    textNode/.style={
            anchor=center,
            opacity=0.8,
            font=\bfseries,
            overlay,
        }
}

\usepackage[normalem]{ulem}
\def\IpeakSphereTruth{5e-06}%
\def\IpeakCubeTruth{4.056e-06}%
\def\IpeakCrossTruth{4.83e-06}%
\def\IpeakSphereImpulsive{5.968e-06}%
\def\IpeakCubeImpulsive{4.414e-06}%
\def\IpeakCrossImpulsive{4.452e-06}%
\def\IpeakSphereMdd{1.678e-05}%
\def\IpeakCubeMdd{1.154e-05}%
\def\IpeakCrossMdd{1.222e-05}%

\begin{document}

\title{Acoustic disguising: a unified framework for cloaking and holography}

\author{Jonas Müller\orcidlink{0000-0003-4530-2486}}
\email{Contact author: jmuller.research@gmail.com}
\homepage{https://jmullerresearch.ch/}
\author{Dirk-Jan van Manen\orcidlink{0000-0002-8250-2826}}
% \author{Johan O. A. Robertsson\orcidlink{0000-0002-3292-385X}}
\affiliation{%
    Institute of Geophysics, ETH Zürich, Switzerland
}%T 

\date{\today}

\begin{abstract}
    Cloaking and holography---usually treated as distinct problems---are two limits of a single operation that we call \emph{acoustic disguising}, realized here using immersive boundary conditions on a closed surface. Driving the boundary with homogeneous Green's functions suppresses any incident field inside the enclosed volume and cloaks unknown objects broadband; driving it with scattering Green's functions synthesizes a holographic scatterer indistinguishable from a target for arbitrary illuminations. Combining the two, using heterogeneous Green's functions, replaces the scattering signature of one object with that of another, transforming its acoustic identity. We demonstrate the framework in three-dimensional FDTD simulations driven by impulsive Green's functions, complemented by data-driven Green's-function retrieval, establishing a direct route to real-time 3D acoustic cloaking, holography, cloning, and disguising.
\end{abstract}

% Keywords (APS lets you include these; PRResearch accepts them)

\maketitle
% Manipulating wave scattering through cloaking has been extensively studied in electromagnetics \citep{fleury_invisibility_2015} and acoustics \citep{nelson_active_2000,norris_acoustic_2008, norris_acoustic_2015,song_broadband_2019}. In acoustics, the significantly lower wave speed enables not only passive but also active real-time implementations \cite{norris_acoustic_2015,lasri_active_2023}. These active strategies can adapt to unknown scatterers and dynamically suppress scattering \cite{friotRealtimeActiveSuppression2004}. However, such adaptive methods typically do not achieve complete spatiotemporally broadband manipulation.
Manipulating wave scattering is a central challenge in acoustics, with implications for imaging, sensing, and wave-based control. Cloaking, in particular, aims to suppress the scattering signature of an object, rendering it acoustically invisible~\cite{fleury_invisibility_2015,nelson_active_2000,norris_acoustic_2008,norris_acoustic_2015,song_broadband_2019}. Passive strategies---transformation acoustics and metamaterials---typically impose stringent constraints on material properties and bandwidth~\cite{chen_acoustic_2010,zigoneanu_three-dimensional_2014,cummer_controlling_2016}. Active strategies instead record, manipulate, and re-emit wave fields in real time~\cite{friotRealtimeActiveSuppression2004,guevara_vasquez_exterior_2011,popa_active_2015,ma_shaping_2018,lasri_active_2023}, and in principle allow broadband control without restrictive material constraints. Among active approaches, boundary-control formulations based on the Kirchhoff integral show that a closed distribution of sources can suppress or reconstruct wave fields inside a finite volume~\cite{nelson_active_2000,miller_perfect_2006}. In practice, however, such implementations are limited by incomplete aperture coverage, the spatial separation between sensing and actuation, and the difficulty of maintaining broadband accuracy.

Immersive boundary conditions (IBCs) address several of these limits by holographically reproducing and manipulating scattering from recorded boundary data, demonstrated experimentally in one and two dimensions~\cite{van_manen_exact_2007,vasmel_immersive_2013,van_manen_broadband_2015,beckerImmersiveWavePropagation2018,borsingCloakingHolographyExperiments2019,brogginiImmersiveBoundaryConditions2017,becker_broadband_2021,muller_acoustic_2023}. Extending the approach to three dimensions---where practical applications reside---is challenging: achieving full control of the wave field inside a finite volume while simultaneously shaping the scattered field outside it is difficult in principle, and a practical 3D realization raises numerous further implementation obstacles~\cite{muller_acoustic_2023}.

In this work we introduce a three-dimensional framework for acoustic cloaking, holography, and disguising based on immersive boundary conditions. We show that (i) incident wave fields can be completely suppressed within a closed surface using homogeneous Green's functions, enabling broadband cloaking of arbitrary and unknown objects; (ii) scattering Green's functions generate holographic scatterers that reproduce the response of arbitrary objects under arbitrary illumination; and (iii) combining both, using heterogeneous Green's functions, enables acoustic disguising, in which one object is made to appear as another by replacing its scattering response. We demonstrate the framework in a numerical setting that closely reflects experimental conditions, with realistic source--receiver configurations. Along the way we address concrete obstacles to a 3D realization, among them the discretization of a closed immersive surface and the data-driven retrieval of Green's functions. Together, these establish a viable path to real-time broadband acoustic wave manipulation in three dimensions and to experimental implementation. We first revisit the Kirchhoff integral and its dual role in numerical propagation and physical wave-field synthesis; build immersive boundary conditions from that duality; analyze the roles of homogeneous, scattering, and heterogeneous Green's functions; and finally demonstrate cloaking, holography, and disguising in 3D simulations.

\begin{figure}
    \scalebox{1}[1]{\includegraphics[width=0.95\columnwidth,trim={3cm 2.9cm 1cm 1cm}, clip]{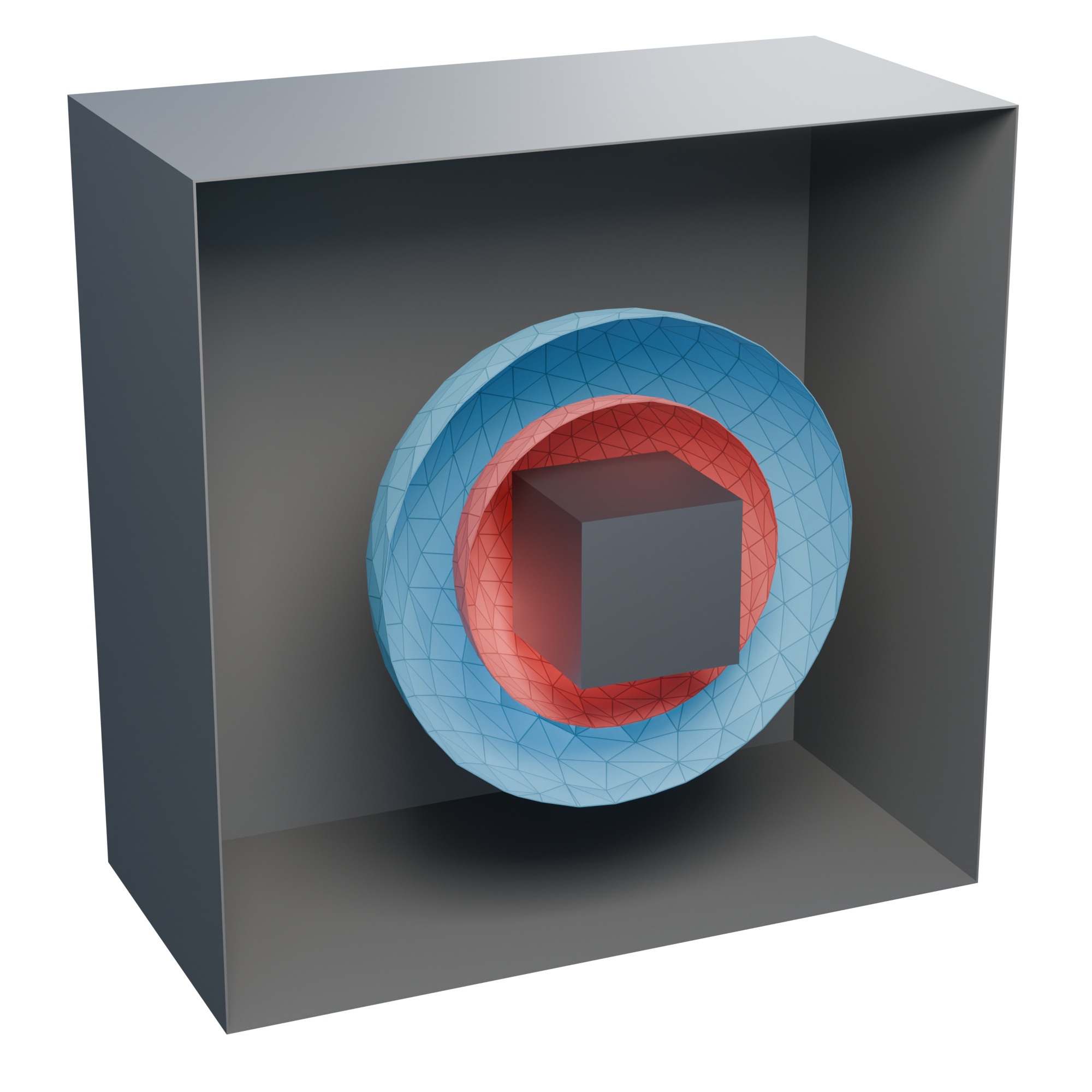} }
    \caption{Acoustic disguising setup. A rigid cubic scatterer (center) is surrounded by a blue emitting surface $\mathcal{S}^\mathrm{I}$ and a red recording surface $\mathcal{S}^\mathrm{O}$. The acoustic space is bounded by a rectangular rigid box; this outer rigid box also provides the reverberant environment used for data-driven Green's-function retrieval via MDD (Supplemental Material).}
    \label{fig:3DdisguisingSetup}
\end{figure}

% We demonstrate the feasibility of the approach in a numerical three dimensional space (Fig.~\ref{fig:3DdisguisingSetup}). A numerical space has the advantage of providing ideal source and receiver transfer functions and being noise-free. The placements of the sources and receivers are exact, and their quantity is effectively unconstrained. Green's functions can be calculated in a straightforward manner using impulsive sources and absorbing boundaries (i.e. perfectly matched layers \cite{berenger_perfectly_1994,aloisi_full_2023}). In contrast to an experimental setup, we also have direct access to measure the normal particle velocity and the pressure field at any point in space.

%outline
% We first show how the Kirchhoff integral can be exploited in two complementary ways: to propagate physical source terms in a numerical domain, and to drive numerically derived source terms in a physical domain. Building on this duality, we construct immersive boundary conditions that bridge physical and numerical wave fields. We then compare the roles of homogeneous, heterogeneous, and scattering Green's functions in shaping the resulting boundary conditions. Finally, we demonstrate how these ingredients enable the transformation of the acoustic response of one object into that of another—realizing acoustic disguising.

\textit{The Kirchhoff Integral.}
Consider a space where acoustic waves can propagate. Take any closed surface $\mathcal{S}^\mathrm{O}$ in this space that separates an interior from the exterior subspace. Denote the wave field on this surface as $\{p,v\}(x,t) \ \forall x \in \mathcal{S}^\mathrm{O}$, or, for short, $\{p,v\}(\mathcal{S}^\mathrm{O})$, where $x$ denotes a three-dimensional position (written without boldface for readability), $p$ the pressure, and $v$ the normal particle velocity ($v = \boldsymbol{v} \cdot \boldsymbol{n}$). The normal vector $\boldsymbol{n}$ on this surface is defined to point outward. The same convention applies to every vector surface quantity introduced below: only its component along $\boldsymbol{n}$ is retained, written as a scalar.

% Fokkema and van den Berg; equation (7.63) gives the representation theorem for extrapolating outside the surface. equation (7.82) gives the representation theorem for extrapolating inside the surface. In both cases the normal vector is defined outwards. For (7.82) we get a negative sign for the extrapolated pressure.
We can find a surface-source representation of the wave field at any point $x$ using a Kirchhoff integral~\cite{barton_elements_1989,fokkema_seismic_1993,hoopHandbookRadiationScattering1995}. The representation is valid in any source-free region bounded by a closed surface $\mathcal{S}$. With $\boldsymbol{n}$ pointing outward, the integral yields $p(x,t)$ when $x$ lies outside $\mathcal{S}$ and the exterior is source-free, and $-p(x,t)$ when $x$ lies inside $\mathcal{S}$ and the interior is source-free; the sign change follows from the fixed outward orientation of $\boldsymbol{n}$ used here. Writing the canonical exterior form,
% \begin{equation}
%     \begin{aligned}
%         \pm p(x,t)= \int_{t_0}^t dt' \int_{S}dS' [                           & G^{\:p\,\mid\, q}(x,t -t',x') v(x',t')
%                                                      +                                                                                                              \\        & G^{\:p\,\mid\, f}(x,t - t',x') p(x',t') ]
%         \label{eq:surface_source_representation}
%     \end{aligned}
% \end{equation}
\begin{equation}
    \begin{split}
        p(x,t) = \int_{t_0}^{t}\!\mathrm{d}t' \int_{\mathcal{S}}\!\mathrm{d}S' \Big[
                                                                                 & G^{\:p\,\mid\, q}(x,t;x',t')\, v(x',t') \\
                                                                              {}+{} & G^{\:p\,\mid\, f}(x,t;x',t')\, p(x',t') \Big]
    \end{split}
    \label{eq:surface_source_representation}
\end{equation}
where $G^{\:p\,\mid\, q}(x,t;x',t')$ denotes the pressure $p$ at $(x,t)$ due to a monopole source $q$ at $(x',t')$, and $G^{\:p\,\mid\, f}$ the pressure due to a dipole source $f$. An analogous expression holds for the normal particle velocity $v(x,t)$ with kernels $G^{\:v\,\mid\, q}$ and $G^{\:v\,\mid\, f}$. Equation~\ref{eq:surface_source_representation} propagates the wave field from $\mathcal{S}$ to any point $x$ in the source-free region on one side of $\mathcal{S}$; the corresponding expression on the other side is obtained by an overall sign flip. In other words, the wave field in a source-free region bounded by $\mathcal{S}$ is fully determined by the wave field passing through $\mathcal{S}$ and the Green's functions $G$ propagating within that region. For the time-invariant media considered here, $G$ depends on the source and observation times only through their difference $t-t'$, and the time integral in Eq.~\ref{eq:surface_source_representation} reduces to a temporal convolution.

The convolution of $G^{\:p\,\mid\, q}$ with the normal particle velocity $v$ and of $G^{\:p\,\mid\, f}$ with the pressure $p$ on $\mathcal{S}$ lets us interpret the boundary data as a surface source distribution on $\mathcal{S}$: the normal velocity $v$ plays the role of a monopole source strength $q$, and the pressure $p$ that of a dipole source strength $f$, with the dipole aligned along the outward normal ($f = \boldsymbol{f}\cdot\boldsymbol{n}$, per the convention introduced above). Given the surface source distribution $\{q,f\}(\mathcal{S})$, Eq.~\ref{eq:surface_source_representation} yields the field at any point $x$ in the source-free subspace on one side of $\mathcal{S}$, due to any volume-source distribution (or scattering) contained in the other subspace.

The Kirchhoff integral, Eq.~\ref{eq:surface_source_representation}, can be used in two ways: computed numerically when the Green's functions $G$ are known, or computed physically when the source field $\{q,f\}(\mathcal{S})$ is known. In the first case, a physical source field $\{q,f\}(\mathcal{S})$ is convolved with the Green's functions $G$ to propagate the wave field in a numerical space. In the second case, a numerically calculated source field $\{q,f\}(\mathcal{S})$ is used to drive physical sources so that the wave field propagates in a physical space. Both use cases of the Kirchhoff integral are needed to construct Immersive Boundary Conditions.

\textit{Immersive Boundary Conditions.}
Take $\mathcal{S}^\mathrm{O}$ to be a transparent recording surface. On this surface, the physical wave field $\{p,v\}(\mathcal{S}^\mathrm{O})$ can be recorded without perturbing the field itself. This wave field continues to propagate through $\mathcal{S}^\mathrm{O}$ unchanged. While the field propagates physically, we can simultaneously predict the wave field at any point $x$ inside the volume enclosed by $\mathcal{S}^\mathrm{O}$ evaluating Eq.~\ref{eq:surface_source_representation} with the corresponding Green's functions $G$ of the medium inside $\mathcal{S}^\mathrm{O}$ numerically.

Take $\mathcal{S}^\mathrm{I}$ to be a closed, transparent surface that is fully contained within $\mathcal{S}^\mathrm{O}$ such that we can evaluate Eq.~\ref{eq:surface_source_representation} on $\mathcal{S}^\mathrm{O}$ for any point $x \in \mathcal{S}^\mathrm{I}$. For the implementation below we take $\mathcal{S}^\mathrm{I}$ and $\mathcal{S}^\mathrm{O}$ to be concentric spherical shells, but the construction does not require this.

% Consider two states, a numerical and a physical state. They differ only inside $\mathcal{S}^\mathrm{I}$ and are identical outside of $\mathcal{S}^\mathrm{I}$. The physical state $P$ corresponds to the medium where the physical wave field $\{p,v\}_\mathrm{P}$ propagates. The numerical state $N$ corresponds to a wave field $\{p,v\}_\mathrm{N}$ that is numerically extrapolated using Eq.~\ref{eq:surface_source_representation}, the physical wave field $\{p,v\}_\mathrm{P}(\mathcal{S}^\mathrm{O})$ and the Green's functions $G_\mathrm{N}$ of the numerical state.

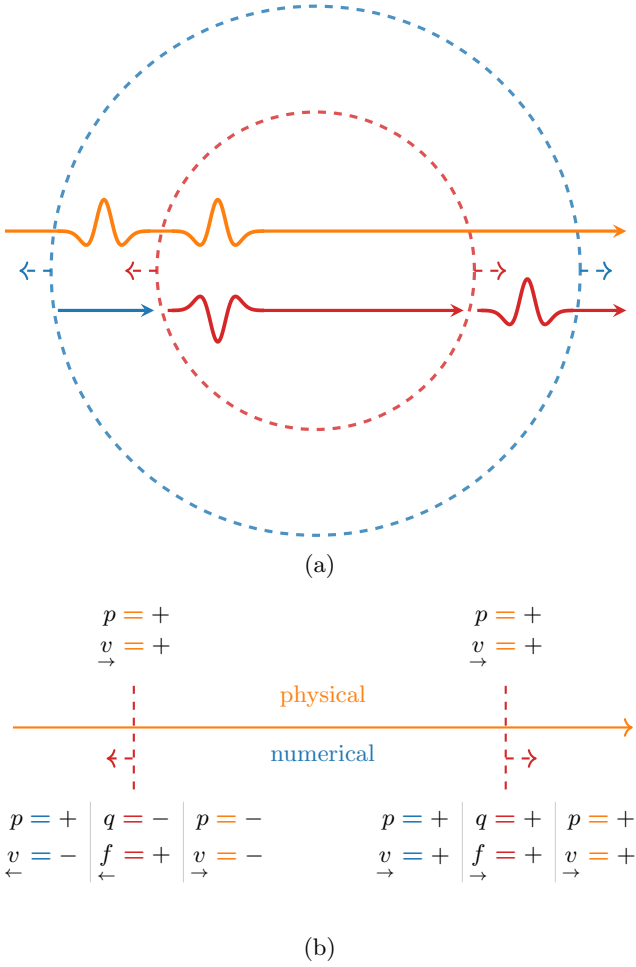
\begin{figure}
    \centering
    \begin{subfigure}[b]{\columnwidth}
        \centering
        % Pin a symmetric bounding box to the true content so pdfLaTeX and LuaLaTeX
% measure this figure identically. (pdfLaTeX otherwise adds ~41pt of phantom
% width from the embedded pgfplots axes, overflowing the column and shifting
% the drawing right.) Keeps Fig 2a centered at its natural size on either engine.
\begin{tikzpicture}[scale=3.5]
    \useasboundingbox (-1.21,-1.005) rectangle (1.21,1.005);

    \def\routerradius{1.0}
    \def\rinnerradius{0.6}
    \draw[blue, dashed, very thick, opacity=0.8] (0,0) circle (\routerradius);
    \draw[red, dashed, very thick, opacity=0.8] (0,0) circle (\rinnerradius);

    \foreach \angle in {-90, 90} {
            \def\normalAngle{90+\angle}
            \draw[blue, ->, dashed, thick] (\normalAngle :\routerradius) -- (\normalAngle :\routerradius+0.12);
            \draw[red, ->, dashed, thick] (\normalAngle :\rinnerradius) -- (\normalAngle :\rinnerradius+0.12);
        }

    \def\yshift{0.15cm}
    \def\s{0.09cm}
    \def\xShiftWavelet{-0.37cm}

    % draw wave arrows
    \draw[orange, yshift=+\yshift, waveArrow] (180:1.2) -- (0:1.2);

    \draw[blue, yshift=-\yshift, waveArrow] (180:\routerradius) -- (180:\rinnerradius-0.015);
    \draw[red, yshift=-\yshift, waveArrow] (180:\rinnerradius-0.015) -- (0:\rinnerradius-0.015);
    \draw[red, yshift=-\yshift, waveArrow] (0:\rinnerradius-0.015) -- (0:1.2);

    \def\equation{(1 - 2*pi^2*x^2)*exp(-pi^2*x^2)}

    % draw masking rectangles
    \fill[white, opacity=1.0,xshift=\xShiftWavelet] (-0.170,+0.2) rectangle (0.170,-0.2);

    \fill[white, opacity=1.0,xshift=-(\routerradius cm /2 + \rinnerradius cm /2),] (-0.170,+0.2) rectangle (0.170,+0.1);

    \fill[white, opacity=1.0,xshift=(\routerradius cm /2 + \rinnerradius cm /2),] (-0.170,-0.2) rectangle (0.170,-0.1);

    % numerical wavelets
    \begin{axis}[
            yshift=-\yshift,
            xshift=\xShiftWavelet,
            axisStyle,
        ]
        \addplot[thin, red] {-\equation};
    \end{axis}

    \begin{axis}[
            yshift=-\yshift,
            xshift=(\routerradius cm /2 + \rinnerradius cm /2),
            axisStyle,
        ]
        \addplot[thin, red] {\equation};
    \end{axis}

    % physical wavelets
    \begin{axis}[
            yshift=+\yshift,
            xshift=\xShiftWavelet,
            axisStyle,
        ]
        \addplot[thin, orange] {\equation};
    \end{axis}

    \begin{axis}[
            yshift=+\yshift,
            xshift=-(\routerradius cm /2 + \rinnerradius cm /2),
            axisStyle,
        ]
        \addplot[thin, orange] {\equation};
    \end{axis}

\end{tikzpicture}
        \caption{}
        \label{fig:homogeneous_ibc}
    \end{subfigure}

    \vspace{0.4em}

    \begin{subfigure}[b]{\columnwidth}
        \centering
        \begin{tikzpicture}[scale=0.82]

    \draw [->,orange,thick] (-5,0) -- (5,0) {};

    % physical
    \node[anchor=center,orange] at (0,0.5) {physical};

    \node[] (p_plus_left) at (-3,1.8) {$p\mathcolor{orange}{\boldsymbol{=}}+$};
    \node[] (p_minus_left) at (-3-0.05,1.2) {$\underset{\rightarrow}{v}\mathcolor{orange}{\boldsymbol{=}}+$};

    \node[] (p_plus_right) at (+3,1.8) {$p\mathcolor{orange}{\boldsymbol{=}}+$};
    \node[] (p_minus_right) at (+3-0.05,1.2) {$\underset{\rightarrow}{v}\mathcolor{orange}{\boldsymbol{=}}+$};

    % numerical
    \node[anchor=center,blue] at (0,-0.4) {numerical};
    % \shade[left color=transparent!0, right color=transparent!100] (-2,-1) rectangle (2,-0.1);

    % left
    \node[] (p_plus_left_num) at (-4.5,-1.5) {$p\mathcolor{blue}{\boldsymbol{=}}+$};
    \node[] (p_minus_left_num) at(-4.5-0.05,-2.2) {$\underset{\leftarrow}{v}\mathcolor{blue}{\boldsymbol{=}}-$};

    \node[] (p_plus_left_num) at (-3,-1.5) {$q\mathcolor{red}{\boldsymbol{=}}-$};
    \node[] (p_minus_left_num) at (-3-0.05,-2.2) {$\underset{\leftarrow}{f}\mathcolor{red}{\boldsymbol{=}}+$};

    \node[] (p_plus_left_num) at (-1.5,-1.5) {$p\mathcolor{orange}{\boldsymbol{=}}-$};
    \node[] (p_minus_left_num) at (-1.5-0.05,-2.2) {$\underset{\rightarrow}{v}\mathcolor{orange}{\boldsymbol{=}}-$};

    \draw [-, opacity=0.2] (-3.75,-2.5) -- (-3.75,-1.2) {};
    \draw [-, opacity=0.2] (-2.25,-2.5) -- (-2.25,-1.2) {};

    % right
    \node[] (p_plus_right_num) at (+1.5,-1.5) {$p\mathcolor{blue}{\boldsymbol{=}}+$};
    \node[] (p_minus_right_num) at (+1.5-0.05,-2.2) {$\underset{\rightarrow}{v}\mathcolor{blue}{\boldsymbol{=}}+$};

    \node[] (p_plus_right_num) at (+3,-1.5) {$q\mathcolor{red}{\boldsymbol{=}}+$};
    \node[] (p_minus_right_num) at (+3-0.05,-2.2) {$\underset{\rightarrow}{f}\mathcolor{red}{\boldsymbol{=}}+$};
    \node[] (p_plus_right_num) at (+4.5,-1.5) {$p\mathcolor{orange}{\boldsymbol{=}}+$};
    \node[] (p_minus_right_num) at (+4.5-0.05,-2.2) {$\underset{\rightarrow}{v}\mathcolor{orange}{\boldsymbol{=}}+$};

    \draw [-, opacity=0.2] (+2.25,-2.5) -- (+2.25,-1.2) {};
    \draw [-, opacity=0.2] (+3.75,-2.5) -- (+3.75,-1.2) {};

    % source boundary
    \draw [-, red, dashed, thick] (-3 - 0.05,-1) -- (-3 - 0.05,0.8) {};
    \draw [-, red, dashed, thick] (+3 - 0.05,-1) -- (+3 - 0.05,0.8) {};

    % boundary normal
    \draw [<-, red, dashed,thick] (-3.49,-0.5) -- (-3 - 0.05,-0.5) {};
    \draw [->, red, dashed,thick] (+3 - 0.05,-0.5) -- (+3.45,-0.5) {};

\end{tikzpicture}
        \caption{}
        \label{fig:homogeneous_sign}
    \end{subfigure}

    \caption{Homogeneous IBC mechanism. (a) Wavelet propagating through a domain enclosed with immersive boundary conditions using homogeneous Green's functions. The wave field inside the inner surface (red) is suppressed through a superposition of the physical wave field (orange) and an equal-magnitude, opposite-sign numerical wave field emitted through the inner surface; outside the inner surface on the opposite side, the wave field is reconstructed. (b) Sign convention: incoming waves on $\mathcal{S}^\mathrm{I}$ result in a negative source field $\{q,f\}(\mathcal{S}^\mathrm{I})$ due to the outward-pointing normal vector (left), while outgoing waves yield a positive source field (right). Colors code the extrapolated field (blue), the source field (red), and the emitted field (orange).}
\end{figure}

We define a state as a specification of source distributions, material properties, and the resulting wave field within a domain~\cite{fokkema_seismic_1993}. Two states are considered: a physical state $P$ and a numerical state $N$. Source distributions are non-zero only in the physical state outside $\mathcal{S}^\mathrm{O}$. The two states are identical in the region between $\mathcal{S}^\mathrm{O}$ and $\mathcal{S}^\mathrm{I}$, and can differ within $\mathcal{S}^\mathrm{I}$. The physical state $P$ describes the medium in which the wave field $\{p,v\}_\mathrm{P}$ propagates. The numerical state $N$ defines a wave field $\{p,v\}_\mathrm{N}$ obtained by numerical extrapolation via Eq.~\ref{eq:surface_source_representation}, using the boundary data $\{p,v\}_\mathrm{P}(\mathcal{S}^\mathrm{O})$ and the Green's functions $G^\mathrm{N}$ of the numerical state.

% Sign convention: f = +p on S^I (scalar, per the line-104 surface
% convention; the n_hat direction is implicit). The surface pressure
% plays the role of the dipole-source-strength density. Sign chosen to
% match the code's cancellation injection (positive C_f * p at the FDTD
% register); this is one flip from the textbook De Hoop reproduction
% form -- see the sign-convention history in lib/sup.tex. C_f, C_q are positive
% discrete injection coefficients, introduced in the Discretization
% section and derived in SI \S\ref{app:source-equivalence}.
Finally, we equip $\mathcal{S}^\mathrm{I}$ with monopole and dipole source distributions $\{q,f\}(\mathcal{S}^\mathrm{I})$ driven by the extrapolated wave field of the numerical state on this surface, $\{p,v\}_\mathrm{N}(\mathcal{S}^\mathrm{I})$: the extrapolated pressure drives the dipole strength and the normal velocity the monopole strength, $p_\mathrm{N}\mapsto f$ and $v_\mathrm{N}\mapsto q$.

This source distribution then emits a wave field into the physical state. The total wave field inside and outside of $\mathcal{S}^\mathrm{I}$ is then a superposition of the physical propagating wave field and the numerical wave field emitted through the source distribution $\{q,f\}(\mathcal{S}^\mathrm{I})$. We have immersed the numerical state $N$ into the physical state $P$ using immersive boundary conditions (IBCs)~\cite{van_manen_exact_2007,van_manen_broadband_2015,brogginiImmersiveBoundaryConditions2017}.

\begin{figure}
    \begin{tikzpicture}[scale=1.4]
    \def\numberOfArrows{9}
    \def\radius{0.89}
    \def\outerRadius{1.3}
    \def\innerRadius{0.7}
    \def\segmentLength{0.29}

    \def\shift{1.5}

    \node[] at (-\shift, 1.7) {\footnotesize physical};
    \node[] at (\shift, 1.7) {\footnotesize numerical};

    % waves

    % direct wave
    \draw[red,thick,decorate,decoration={expanding waves,angle=13,segment length=\segmentLength cm},xshift=-\shift cm] (130:1.55) -- (-50:1.3)  ;

    % illumination
    \draw[orange,thick,decorate,decoration={expanding waves,angle=20,segment length=\segmentLength cm},xshift=-\shift cm] (130:1.55) -- (130:0.5) ;

    % scattering
    \draw[red,thick,decorate,decoration={expanding waves,angle=20,segment length=\segmentLength cm},xshift=-\shift cm] (45:0) -- (45:1.3) ;

    % background
    \fill[white] (-2*\shift,\shift) rectangle (0,\shift);
    \fill[white] (-\shift,0) circle (\radius cm);
    \fill[opacity=0.05] (-\shift,0) circle (\radius cm);

    \fill[opacity=0.05] (0,-\shift) rectangle (2*\shift,\shift);
    \fill[white] (\shift,0) circle (\radius cm);

    % Only this part is clipped
    \begin{scope}
        \clip (\shift,0) circle (\radius cm);

        % direct wave
        \draw[
            blue,
            thick,
            decorate,
            decoration={expanding waves, angle=13, segment length=\segmentLength cm},
            xshift=\shift cm
        ] (130:1.55) -- (-50:1.3);

        % scattering
        \draw[blue,thick,decorate,decoration={expanding waves,angle=20,segment length=\segmentLength cm},xshift=\shift cm] (45:0) -- (45:1.3) ;

    \end{scope}

    % mask 

    \filldraw[fill=white,opacity=1.0, draw=none,xshift=\shift cm]
    (-0.6,0.16) -- (0.0,0.6) -- (0.6,0.1) -- (0.2,-0.4) -- (-0.3,-0.6) -- cycle;

    % \foreach \r in {0.4, 0.45, 0.5, 0.55, 0.6,0.65, 0.7,0.75} {
    %     \fill[white,opacity=0.3,xshift=\shift cm] (0,0) circle (\r cm);
    %   }

    % Draw circle
    \draw[dashed] (-\shift,0) circle (\radius cm);
    \draw[opacity=0.2] (0, -1.5) -- (0, 1.5);
    \draw[dashed] (\shift,0) circle (\radius cm);

    \draw[->, blue, thick] (-0.2,0.9) -- (0.2,0.9);
    \draw[->, red, thick] (0.2,-0.9) -- (-0.2,-0.9);

    % Draw circle at (-\shift,0)
    \fill[blue!30!white,xshift=\shift cm] (0,0) circle (0.4 cm);

    % Draw square at (\shift,0)
    \fill[gray,opacity=0.2,xshift=-\shift cm] (-0.34,-0.34) rectangle (0.34,0.34);

\end{tikzpicture}
    \caption{Acoustic disguising illustration. The incoming wavefield (orange) is suppressed in the physical space (within the sphere $\mathcal{S}^\mathrm{I}$) and therefore not scattered by the cubic physical scatterer. It is scattered by a spherical scatterer in the numerical domain (blue) and the scattering is holographically created in the physical domain (red).}
    \label{fig:disguising_illustration}
\end{figure}
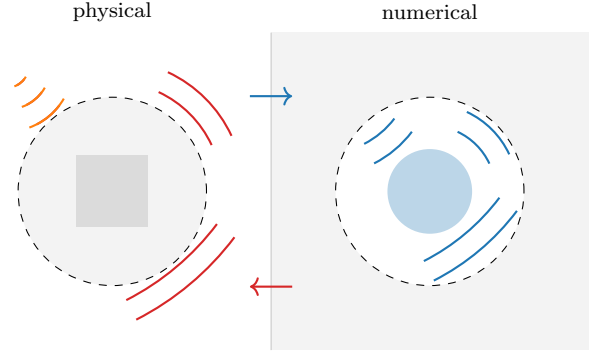

\textit{Green's Functions.}
The choice of Green's functions used in Eq.~\ref{eq:surface_source_representation} determines the effect of the immersive boundary condition on the physical field. For cloaking applications, we wish to remove the physical scattering $G_\mathrm{S_P}$ (i.e., add the negative $-G_\mathrm{S_P}$). For holography applications, we wish to add a target numerical scattering $G_\mathrm{S_N}$. For disguising applications, we wish to simultaneously do both, i.e., replace $G_\mathrm{S_P}$ by $G_\mathrm{S_N}$ (add $G_\mathrm{S_N}-G_\mathrm{S_P}$). See Fig.~\ref{fig:disguising_illustration} for an illustration of this process.

To make this precise, we decompose the heterogeneous Green's functions $G$ into a homogeneous part $G_\mathrm{H}$ and a scattering part $G_\mathrm{S}$, $G = G_\mathrm{H} + G_\mathrm{S}$. The homogeneous part $G_\mathrm{H}$ extrapolates the field $\{p,v\}(\mathcal{S}^\mathrm{O})$ to $\{p,v\}_\mathrm{H}(\mathcal{S}^\mathrm{I})$ in the absence of any scatterer inside $\mathcal{S}^\mathrm{I}$ (direct arrival), while the scattering part $G_\mathrm{S}$ extrapolates only the field scattered inside $\mathcal{S}^\mathrm{I}$, yielding $\{p,v\}_\mathrm{S}(\mathcal{S}^\mathrm{I})$ (indirect arrival). Since the physical and numerical states coincide outside $\mathcal{S}^\mathrm{I}$, the homogeneous part is common to both, and the Green's functions of the two states decompose as $G_\mathrm{P} = G_\mathrm{H} + G_\mathrm{S_P}$ and $G_\mathrm{N} = G_\mathrm{H} + G_\mathrm{S_N}$.

Consider the same homogeneous physical and numerical state $G_\mathrm{N} = G_\mathrm{P} = G_\mathrm{H}$. The physical propagating wave field and the numerically extrapolated wave field are identical in $\mathcal{S}^\mathrm{I}$. A plane wave traveling along an axis through the inner domain is incoming on one side of $\mathcal{S}^\mathrm{I}$ and outgoing on the other (Fig.~\ref{fig:homogeneous_ibc}).

On the incoming side of the plane wave on $\mathcal{S}^\mathrm{I}$, the normal particle velocity and the dipole source direction oppose the propagation direction. This leads to destructive interference between the physical and numerical wave fields, effectively suppressing the wave field inside $\mathcal{S}^\mathrm{I}$ (Fig.~\ref{fig:homogeneous_sign}, left). In other words, inside $\mathcal{S}^\mathrm{I}$ the wave field is a superposition of the physical wave field and a negative numerical copy of the same wave field emitted through the inner surface $\mathcal{S}^\mathrm{I}$ (Fig.~\ref{fig:homogeneous_ibc}).

On the outgoing side of the plane wave on $\mathcal{S}^\mathrm{I}$, the normal particle velocity and the dipole source align with the propagation direction, reconstructing the wave field outside $\mathcal{S}^\mathrm{I}$ (Fig.~\ref{fig:homogeneous_sign}, right). This creates a numerical bridge for physical waves that completely bypasses the physical space inside $\mathcal{S}^\mathrm{I}$.

The suppression--reconstruction pattern above is a direct consequence of the outward-pointing normal convention on $\mathcal{S}^\mathrm{I}$: a field component travelling \emph{into} $\mathcal{S}^\mathrm{I}$ contributes a source distribution $\{q,f\}(\mathcal{S}^\mathrm{I})$ of opposite sign to a field component travelling \emph{out} (Fig.~\ref{fig:homogeneous_sign}). The IBC therefore cancels the incoming part of the driving field inside $\mathcal{S}^\mathrm{I}$ and reconstructs the outgoing part outside $\mathcal{S}^\mathrm{I}$---the same local mechanism in both half-spaces. This single sign-flip rule accounts for both ends of the framework: for the homogeneous Green's function $G_\mathrm{H}$ the driving field has both incoming and outgoing components, producing a \emph{cloak} (interior suppressed, exterior reconstructed); for the scattering Green's function $G_\mathrm{S}$ the driving field is purely outgoing on $\mathcal{S}^\mathrm{I}$, producing a \emph{hologram} (no interior suppression, only exterior emission). Cloaking and holography are thus the two limits of the same boundary operation, distinguished solely by which part of the Green's function is fed into it.

Cloaking can be realized in two distinct ways: (a) cancelling the physical scattering with a negative numerical hologram $G_\mathrm{S_N} = -G_\mathrm{S_P}$~\cite{van_manen_broadband_2015}, or (b) suppressing the interior field entirely via $G_\mathrm{N} = G_\mathrm{H}$ which is what we demonstrate here.

In the first case, the homogeneous field is not extrapolated. Any physical object present inside $\mathcal{S}^\mathrm{I}$ scatters the incident wave field. However, the scattering is then cloaked when propagating out of $\mathcal{S}^\mathrm{I}$ due to the addition of a negative hologram of the scatterer $-G_\mathrm{S_P}$. The scatterer therefore needs to be known in advance.

In the second case, cloaking is achieved by setting the field inside $\mathcal{S}^\mathrm{I}$ to zero by extrapolating the homogeneous field. The field is then suppressed where it is incoming on $\mathcal{S}^\mathrm{I}$, and reconstructed where it is outgoing on $\mathcal{S}^\mathrm{I}$. No scattering can occur on any object inside $\mathcal{S}^\mathrm{I}$, effectively cloaking the space and all unknown objects contained inside $\mathcal{S}^\mathrm{I}$ (removing the source field of the secondary sources).

The Green's functions can be found using different approaches. In a numerical setup, they can conveniently be calculated using impulsive point sources, while in a physical setup a data-driven approach like multidimensional deconvolution (MDD)~\cite{amundsen_elimination_2001,wapenaar_seismic_2011} is more appropriate~\cite{liClosedapertureUnboundedAcoustics2021,muller_acoustic_2023}, since it removes room reverberations numerically in post-processing, obviating the need for an anechoic chamber. This enables \emph{acoustic cloning}: by retrieving the physical Green's functions $G_\mathrm{P}$ via MDD and using them as $G_\mathrm{N}$, an unknown physical scatterer is reproduced holographically in its absence. With the appropriate sources and boundary conditions, both routes are available in either state. As a proof of concept, we demonstrate MDD-based retrieval in a reverberant 3D environment in the Supplemental Material (\S\ref{app:mdd}).

% \textit{Methods.}
\textit{Discretization.}
We test the framework using the finite-difference time-domain (FDTD) method to simulate wave propagation in a homogeneous medium on a staggered grid. Scatterers are represented by a staircase mask on the staggered grid; the resulting stair-step approximation of curved boundaries could be reduced, at the cost of a lower Courant factor, by adopting a locally conformal scheme~\cite{tolan_locally_2003}.

To evaluate Eq.~\ref{eq:surface_source_representation} numerically, we need to discretize all continuous quantities (integration surfaces and time). We discretize the spherical surfaces with $N = 400$ Fibonacci-sphere sample points per surface~\cite{gonzalez_measurement_2010,marques_spherical_2013}, which gives a near-uniform angular distribution with constant per-point solid angle.

The discretized outer surface is then represented by the set of $N$ Fibonacci-distributed points $\mathcal{X}^\mathrm{O} = \{x_1^\mathrm{O}, x_2^\mathrm{O}, \dots, x_N^\mathrm{O}\} \subset \mathcal{S}^\mathrm{O}$, each carrying a local area weight $\Delta S = 4\pi r^2 / N$ with $r$ the radius of the spherical surface. The same discretization is used for the inner surface $\mathcal{S}^\mathrm{I} \approx \mathcal{X}^\mathrm{I}$. Since the discretized surface points do not generally coincide with the Cartesian FDTD grid, we use trilinear interpolation to map to and from the neighboring Cartesian grid points, for recording the field on $\mathcal{X}^\mathrm{O}$ and for injecting sources on $\mathcal{X}^\mathrm{I}$.

We can write Eq.~\ref{eq:surface_source_representation} as a discrete approximation~\cite{van_manen_exact_2007}
\begin{equation}
    \begin{split}
        p[x,t_n] \approx \sum_{x' \in \mathcal{X}^\mathrm{O}}\!\sum_{k=0}^{n}\big[
                                                                              & G^{\:p\,\mid\, q}[x,\,t_k\mid x']\, v[x',t_{n-k}] \\
                                                                           {}+{} & G^{\:p\,\mid\, f}[x,\,t_k\mid x']\, p[x',t_{n-k}] \big]\,\Delta S\Delta t
    \end{split}
    \label{eq:extrap_p_discret}
\end{equation}
Equation~\ref{eq:extrap_p_discret}, with the matching expression for the normal velocity $v$, is the two-way extrapolation scheme used in previous immersive-boundary implementations~\cite{van_manen_exact_2007,vasmel_immersive_2013,van_manen_broadband_2015,brogginiImmersiveBoundaryConditions2017,beckerImmersiveWavePropagation2018,borsingCloakingHolographyExperiments2019,becker_broadband_2021,muller_acoustic_2023}: the recorded field $\{p,v\}$ on $\mathcal{X}^\mathrm{O}$ is propagated to the inner surface $\mathcal{X}^\mathrm{I}$ with four Green's-function kernels. Here we instead use a one-way Kirchhoff integral~\cite{berkhout_one-way_1989,wapenaar_reciprocity_2020} of the incoming constituent $p^\mathrm{in} = (p - z_0 v)/2$, with $z_0 = \rho_0 c_0$ the acoustic impedance: a single constituent propagated with two kernels, halving both the number of convolutions and the required memory. The convolution is evaluated by summing over stored past boundary samples rather than forward-extrapolating each recorded sample into future output times~\cite{van_manen_exact_2007}; the write count per extrapolated sample then drops from $\mathcal{O}(K)$ to $\mathcal{O}(1)$, independent of the temporal kernel length $K$~(Supplemental Material, \S\ref{app:extrapcost}).

On the sampling points of the outer surface $\mathcal{X}^\mathrm{O}$, we record the pressure and normal particle velocity of the physical state during the FDTD simulation $\{p,v\}_\mathrm{P}(\mathcal{X}^\mathrm{O})$. Using the Green's functions of the numerical state $G_\mathrm{N}$ in Eq.~\ref{eq:extrap_p_discret} we extrapolate the recorded field to the inner surface $\mathcal{X}^\mathrm{I}$ to get $\{p,v\}_\mathrm{N}(\mathcal{X}^\mathrm{I})$.

These extrapolated fields are injected back into the FDTD simulation as monopole and dipole source-strength densities on $\mathcal{X}^\mathrm{I}$, $\{q,f\}(\mathcal{X}^\mathrm{I}) = (\Delta S^\mathrm{I}/\Delta x^3)\,\{v,p\}_\mathrm{N}(\mathcal{X}^\mathrm{I})$, and added to the staggered-grid pressure and velocity registers via the standard Yee acoustic update prefactors~\cite{schneider_understanding_nodate, botteldooren_finite-difference_1995, sheaffer_physically-constrained_2012}, with $\Delta S^\mathrm{I}$ the area weight of the inner surface sampling points and $\Delta x^3$ the FDTD cell volume.

\begin{figure*}
    \input{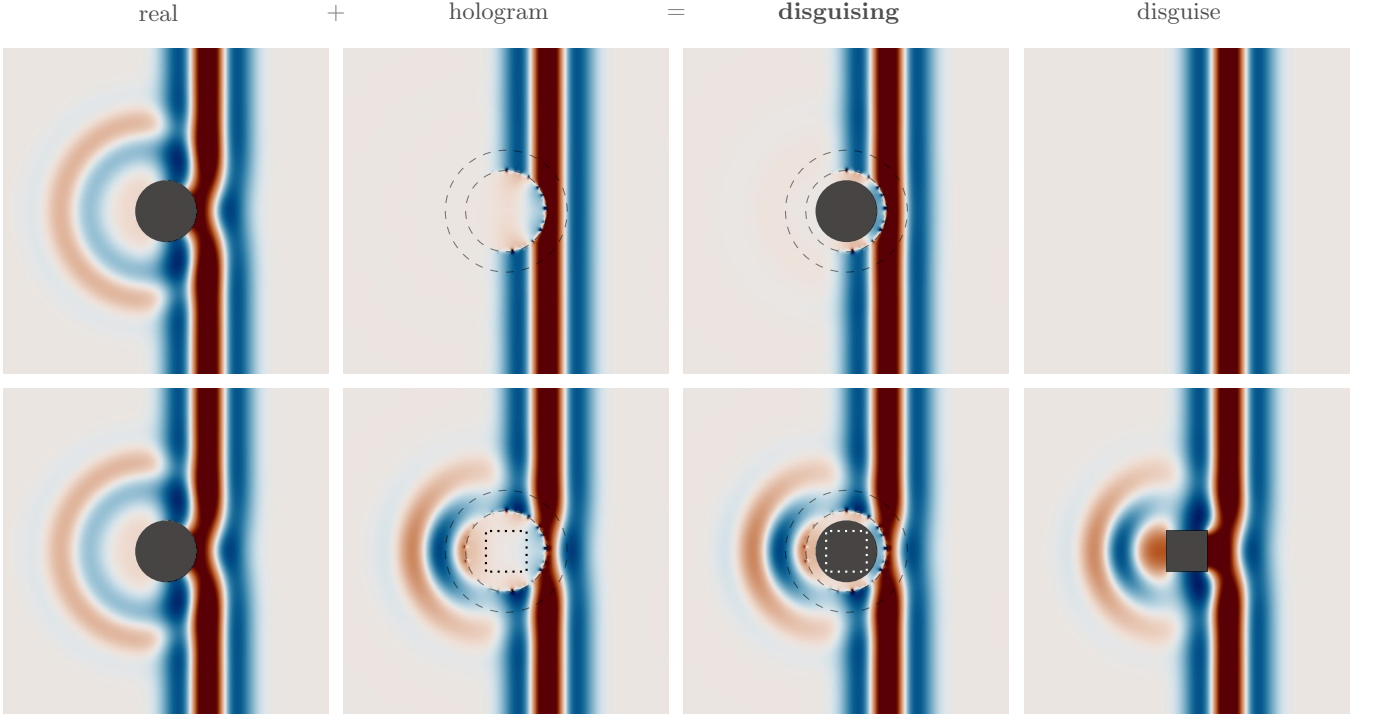}
    \caption{Pressure-field slices of a 3D acoustic plane wave interacting with four scattering configurations, one per column. \textbf{Real} (left): the real spherical scatterer (top and bottom identical). \textbf{Hologram} (middle-left): a holographic scatterer generated by IBCs with an empty interior. \textbf{Disguising} (middle-right): the real spherical scatterer placed inside the hologram. \textbf{Disguise} (right): the real-scatterer response corresponding to the Green's functions used in the hologram (target disguise; spherical in the top row; cubic in the bottom row). Top row uses $G_\mathrm{N}=G_\mathrm{H}$ (cloaking only); bottom row uses $G_\mathrm{N}=G_\mathrm{H}+G_\mathrm{cubic}$ (cloaking + holography).}
    \label{fig:3DdisguisingResults_disguising_t55}
\end{figure*}

\textit{Cloaking, Holography, and Disguising in 3D.}% Results
The scattering behavior of the real and holographic scatterers is here assessed by illuminating them with an incident Ricker plane wave.  The resulting pressure field is recorded on cut planes aligned with the propagation direction of the incident wave and shown in Fig.~\ref{fig:3DdisguisingResults_disguising_t55} during the interaction with the scatterers, and in Fig.~\ref{fig:3DdisguisingResults_disguising_t70} after the interaction. Fig.~\ref{fig:3DdisguisingResults_disguising_t55} shows how a physical scatterer (first column) is disguised as a homogeneous medium (cloaking, top row) and as a cubic scatterer (cloaking + holography, bottom row). The second column presents holographic results obtained with $G_\mathrm{N}=G_\mathrm{H}$ (top) and $G_\mathrm{N}=G_\mathrm{H}+G_\mathrm{cubic}$ (bottom). The pressure field is displayed on a cut plane aligned in the propagation direction of the incident plane wave, passing through the center of the scatterer. The rightmost column shows the real-scatterer response of the target object whose Green's functions drive the hologram.

In the top row of Fig.~\ref{fig:3DdisguisingResults_disguising_t55} we see that using the homogeneous Green's functions $G_\mathrm{N} = G_\mathrm{H}$, the field is fully suppressed within $\mathcal{S}^\mathrm{I}$ (second column, top). Outside $\mathcal{S}^\mathrm{I}$ the field is fully reconstructed at later times (Fig.~\ref{fig:3DdisguisingResults_disguising_t70}), despite never propagating through the (physical) interior. With no incident field inside $\mathcal{S}^\mathrm{I}$ to scatter from, an inserted real scatterer (disguising, third column) is silent and the medium appears homogeneous from outside (cloaking).

In the bottom row of Fig.~\ref{fig:3DdisguisingResults_disguising_t55} we see that using the heterogeneous Green's functions with a cubic scatterer $G_\mathrm{N} = G_\mathrm{H} + G_\mathrm{cubic}$, the field is still fully suppressed within $\mathcal{S}^\mathrm{I}$ and the real spherical scatterer placed inside $\mathcal{S}^\mathrm{I}$ is effectively cloaked. In addition, a holographic scatterer is created that scatters the incident wave field as if a cubic scatterer were present (Fig.~\ref{fig:3DdisguisingResults_disguising_t70}). The spherical scatterer is therefore disguised as a cubic scatterer (cloaking + holography) and appears to be a cube from outside $\mathcal{S}^\mathrm{I}$.

The scattering of the wave field is thus entirely determined by the Green's functions chosen for the hologram, independently of the real scatterer placed inside it. Any probe field applied from outside can therefore only recover the scattering imprint of the chosen disguise---not any information about the real object it hides.

The holograms above were driven by Green's functions computed from impulsive point sources. We next show that they can equally be retrieved in a fully data-driven way, by multidimensional deconvolution (MDD) in a reverberant three-dimensional environment; a detailed description of the 3D MDD procedure is given in the Supplemental Material (\S\ref{app:mdd}).

\begin{figure*}
    % https://tex.stackexchange.com/questions/267938/polar-plot-using-pgfplots-tikz

\pgfplotstableread[col sep=space]{data/polar_sphere_truth_theta0.txt}\tableSphereTruth
\pgfplotstableread[col sep=space]{data/polar_sphere_impulsive_theta0.txt}\tableSphereImpulsive
\pgfplotstableread[col sep=space]{data/polar_sphere_mdd_theta0.txt}\tableSphereMDD

\pgfplotstableread[col sep=space]{data/polar_cube_truth_theta0.txt}\tableCubeTruth
\pgfplotstableread[col sep=space]{data/polar_cube_impulsive_theta0.txt}\tableCubeImpulsive
\pgfplotstableread[col sep=space]{data/polar_cube_mdd_theta0.txt}\tableCubeMDD

\pgfplotstableread[col sep=space]{data/polar_cross_truth_theta0.txt}\tableCrossTruth
\pgfplotstableread[col sep=space]{data/polar_cross_impulsive_theta0.txt}\tableCrossImpulsive
\pgfplotstableread[col sep=space]{data/polar_cross_mdd_theta0.txt}\tableCrossMDD

\setlength{\figW}{\dimexpr(\textwidth)/3/100*97\relax}

\pgfplotsset{
    gfAxis/.style = {
            width=\figW*1.08,
            rotate=-90,
            enlargelimits=false,
            x dir=reverse,
            %xticklabel=$\pgfmathprintnumber{\tick}^\circ$,
            xtick={0,30,...,360},
            ytick={0,0.2,...,1},
            xticklabel style={anchor=-\tick-90,font=\footnotesize,overlay},
            yticklabel style={anchor=west,font=\footnotesize,overlay},
            legend style={at={(+2.6cm,-1.8cm)},anchor=east,font=\small,overlay}, % legend style={at={(-45-180:4cm)},anchor=south west},
            legend cell align={left},
        } % grid=none,ticks=none,axis lines=none
}

\begin{subfigure}[t]{\figW}
    \begin{tikzpicture}[>=latex]
        \begin{polaraxis}[gfAxis]
            \addplot+[name path = Var1,mark=none,very thick,orange] table [x expr = {\thisrow{phi}/pi*180}, y = intensity] {\tableSphereTruth};
            \addplot+[name path = Var1,mark=none,very thick,blue,dashed] table [x expr = {\thisrow{phi}/pi*180}, y = intensity] {\tableSphereImpulsive};
            \addplot+[name path = Var1,mark=none,very thick,green,dotted] table [x expr = {\thisrow{phi}/pi*180}, y = intensity] {\tableSphereMDD};
        \end{polaraxis}
    \end{tikzpicture}
    \hspace{\textheight}
    \caption{Sphere}
    \label{fig:angularSphere}
\end{subfigure}
\hfill
\begin{subfigure}[t]{\figW}
    \begin{tikzpicture}[>=latex]
        \begin{polaraxis}[gfAxis]
            \addplot+[name path = Var1,mark=none,very thick,orange] table [x expr = {\thisrow{phi}/pi*180}, y = intensity] {\tableCubeTruth};
            \addplot+[name path = Var1,mark=none,very thick,blue,dashed] table [x expr = {\thisrow{phi}/pi*180}, y = intensity] {\tableCubeImpulsive};
            \addplot+[name path = Var1,mark=none,very thick,green,dotted] table [x expr = {\thisrow{phi}/pi*180}, y = intensity] {\tableCubeMDD};
            % \legend{Real,Impulsive,MDD}; % redundant with caption (curves named + colour/style coded there)
        \end{polaraxis}
    \end{tikzpicture}
    \hspace{\textheight}
    \caption{Cube}
    \label{fig:angularCube}
\end{subfigure}
\hfill
\begin{subfigure}[t]{\figW}
    \begin{tikzpicture}[>=latex]
        \begin{polaraxis}[gfAxis]
            \addplot+[name path = Var1,mark=none,very thick,orange] table [x expr = {\thisrow{phi}/pi*180}, y = intensity] {\tableCrossTruth};
            \addplot+[name path = Var1,mark=none,very thick,blue,dashed] table [x expr = {\thisrow{phi}/pi*180}, y = intensity] {\tableCrossImpulsive};
            \addplot+[name path = Var1,mark=none,very thick,green,dotted] table [x expr = {\thisrow{phi}/pi*180}, y = intensity] {\tableCrossMDD};
        \end{polaraxis}
    \end{tikzpicture}
    \hspace{\textheight}
    \caption{Cross}
    \label{fig:angularCross}
\end{subfigure}
\hfill
    \caption{
        Angular distribution of the time-integrated scattered intensity $I(\phi)$ [Eq.~\ref{eq:radial_intensity}] along the equatorial plane ($\theta = \pi/2$), recorded on a Fibonacci-sampled far-field sphere of radius $r_\mathrm{ff} = 1.0\,\mathrm{m}$, for the three scatterers studied: (a) sphere, (b) cube, (c) cross. Each panel overlays the real scatterer (\textcolor{orange}{orange}, solid) and the two holographic reconstructions driven by impulsive (\textcolor{blue}{blue}, dashed) and MDD-retrieved (\textcolor{green}{green}, dotted) Green's functions. Each curve is normalized to its own peak; the MDD-retrieved Green's functions are recovered only up to an empirical global scale, so the angular shape, not the absolute magnitude, is what reflects MDD fidelity to the underlying scattering physics (see Supplemental Material, \S\ref{app:resultsAdditional}).
    }
    \label{fig:results_radialIntensity}
\end{figure*}
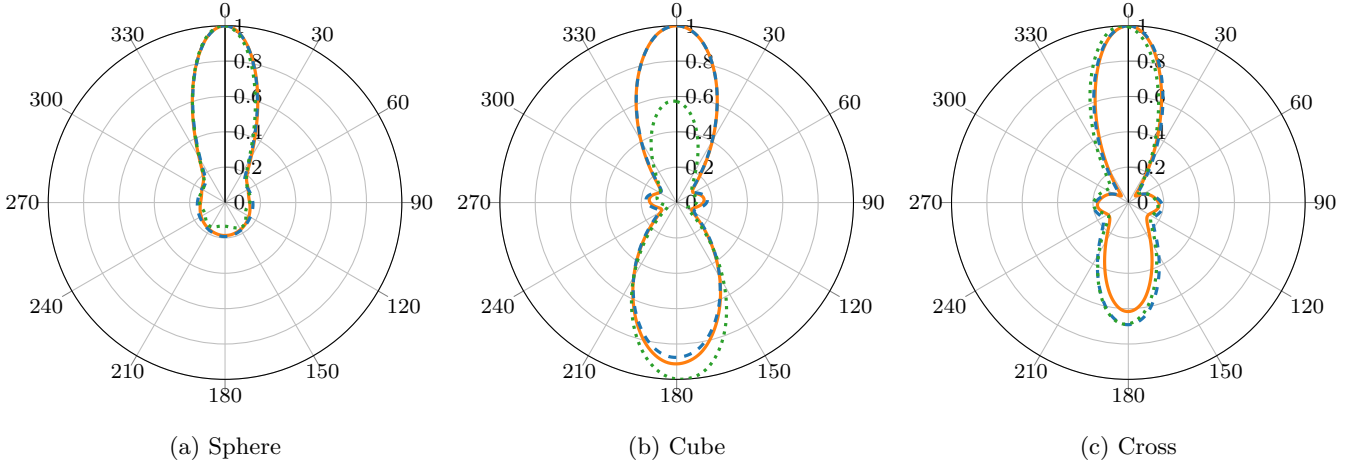

To quantify the angular fidelity of the resulting holograms, and thereby compare the impulsive and data-driven routes, we record the scattered pressure $p_\mathrm{scat} = p_\mathrm{het} - p_\mathrm{hom}$ on a Fibonacci-sampled far-field sphere of radius $r_\mathrm{ff} = 1.0\,\mathrm{m}$ (with $kr_\mathrm{ff} \gtrsim 17$ at the dominant frequencies of the incident wavelet, well into the far field) and evaluate the time-integrated intensity
\begin{equation}
    I(\hat{\boldsymbol{n}}) \;=\; \int \big|p_\mathrm{scat}\big(r_\mathrm{ff}\hat{\boldsymbol{n}},\,t\big)\big|^2\, \mathrm{d}t
    \label{eq:radial_intensity}
\end{equation}
at each receiver direction $\hat{\boldsymbol{n}}$, then extract the equatorial slice $I(\phi) \equiv I(\hat{\boldsymbol{n}}(\theta=\pi/2,\phi))$. At fixed $r_\mathrm{ff}$ in the far field, $p_\mathrm{scat} \sim A(\theta,\phi)/r$, so $I(\phi) \propto |A(\pi/2,\phi)|^2$ and depends only on the angular pattern. Figure~\ref{fig:results_radialIntensity} shows close agreement between the real scatterer and both holographic reconstructions---one driven by Green's functions obtained from an impulsive FDTD source, and one driven by MDD-retrieved Green's functions (acoustic cloning)---confirming the viability of MDD for 3D Green's-function retrieval. 3D renderings of the setup and scattered fields (Figs.~\ref{fig:render_real_het}--\ref{fig:render_ibc_scat}) and additional pressure-field comparisons on multiple cut-planes (Supplemental Material, \S\ref{app:resultsAdditional}) corroborate this conclusion.

\textit{Beyond Laboratory Boundaries.} % Conclusion
Disguising, as introduced and demonstrated here, represents the most general methodology, with cloaking and holography as specific cases. A hologram disguises a homogeneous medium as a heterogeneous medium, effectively introducing a holographic object where none exists. A cloak does the opposite: it disguises a heterogeneous medium as a homogeneous medium, suppressing its scattering signature. A combination of cloaking and holography is best described as disguising; more generally, disguising refers to the transformation of the acoustic identity of one object into that of another.

Beyond cloaking, holography, cloning, and disguising, the framework is extensible. Retrieved Green's functions can be superimposed to combine scattering signatures, spatially weighted to break reciprocity, amplified to realize active gain, or varied in time to morph a scatterer's acoustic identity on the fly.

Within the framework presented here, the assignment of which state is physical and which numerical is formally interchangeable. MDD retrieves Green's functions of either a physical or a numerical object, and a hologram can be constructed in either environment. This permits reversing the setup: rather than immersing a numerical scatterer in a real environment (hologram), a real scatterer can be made to scatter into a numerical one (immersion into a virtual environment). We leave experimental demonstrations of these capabilities for future work.

% A remaining challenge for deploying disguising in practical applications is the physical realization of transparent boundaries, most notably the transparent source surface. Using a transparent source surface allows for a complete separation between the visual and acoustic appearance of an object. A source boundary that imposes Dirichlet or Neumann boundary conditions can still achieve cloaking and holography \cite{becker_broadband_2021}.
%Disguising an object then becomes something like $G^\mathrm{S_N} - G^\mathrm{S_P}$

The 3D implementation presented here closely mirrors a realistic experimental configuration, making physical realization a natural next step. In a laboratory setting, impulsive point sources are ill-suited to Green's-function retrieval because of finite source bandwidth and measurement noise. Additionally, physically damping boundary reflections is both costly and imperfect. Data-driven retrieval via MDD circumvents both limitations: it tolerates arbitrary source waveforms and effectively acts as a virtual absorbing boundary by deconvolving the multiply-scattered illumination data~\cite{liClosedapertureUnboundedAcoustics2021,muller_acoustic_2023}. Our results establish that MDD operates reliably in a fully three-dimensional reverberant setting, removing one of the principal obstacles to a 3D experimental implementation of acoustic disguising.

% \begin{acknowledgments}
\textit{Acknowledgment.}
We thank Johannes Aichele for configuring the compute environment and for proofreading the manuscript, and Johan O. A. Robertsson for guidance and for providing laboratory facilities. We gratefully acknowledge funding from Swiss National Science Foundation (SNSF) grant 197182.
% \end{acknowledgments}

\textit{Author contributions.}
J.M.: Conceptualization; Methodology; Software; Investigation; Formal analysis; Visualization; Writing -- original draft; Writing -- review \& editing.
D.J.v.M.: Conceptualization; Methodology; Supervision; Funding acquisition; Writing -- review \& editing.

\textit{Data availability.}
The code that reproduces all numerical results is archived at \textsc{Zenodo} (DOI: \href{https://doi.org/10.5281/zenodo.20433700}{10.5281/zenodo.20433700}) and developed at \url{https://github.com/Nano560/acoustic-disguising}. The exact snapshot that produced the results in this manuscript is release v1.0.0 (DOI: \href{https://doi.org/10.5281/zenodo.20433701}{10.5281/zenodo.20433701}).

\textit{Competing interests.}
The authors declare no competing interests.

\bibliography{references.bib}

\begin{thebibliography}{40}%
\makeatletter
\providecommand \@ifxundefined [1]{%
 \@ifx{#1\undefined}
}%
\providecommand \@ifnum [1]{%
 \ifnum #1\expandafter \@firstoftwo
 \else \expandafter \@secondoftwo
 \fi
}%
\providecommand \@ifx [1]{%
 \ifx #1\expandafter \@firstoftwo
 \else \expandafter \@secondoftwo
 \fi
}%
\providecommand \natexlab [1]{#1}%
\providecommand \enquote  [1]{``#1''}%
\providecommand \bibnamefont  [1]{#1}%
\providecommand \bibfnamefont [1]{#1}%
\providecommand \citenamefont [1]{#1}%
\providecommand \href@noop [0]{\@secondoftwo}%
\providecommand \href [0]{\begingroup \@sanitize@url \@href}%
\providecommand \@href[1]{\@@startlink{#1}\@@href}%
\providecommand \@@href[1]{\endgroup#1\@@endlink}%
\providecommand \@sanitize@url [0]{\catcode `\\12\catcode `\$12\catcode
  `\&12\catcode `\#12\catcode `\^12\catcode `\_12\catcode `\%12\relax}%
\providecommand \@@startlink[1]{}%
\providecommand \@@endlink[0]{}%
\providecommand \url  [0]{\begingroup\@sanitize@url \@url }%
\providecommand \@url [1]{\endgroup\@href {#1}{\urlprefix }}%
\providecommand \urlprefix  [0]{URL }%
\providecommand \Eprint [0]{\href }%
\providecommand \doibase [0]{https://doi.org/}%
\providecommand \selectlanguage [0]{\@gobble}%
\providecommand \bibinfo  [0]{\@secondoftwo}%
\providecommand \bibfield  [0]{\@secondoftwo}%
\providecommand \translation [1]{[#1]}%
\providecommand \BibitemOpen [0]{}%
\providecommand \bibitemStop [0]{}%
\providecommand \bibitemNoStop [0]{.\EOS\space}%
\providecommand \EOS [0]{\spacefactor3000\relax}%
\providecommand \BibitemShut  [1]{\csname bibitem#1\endcsname}%
\let\auto@bib@innerbib\@empty
%</preamble>
\bibitem [{\citenamefont {Fleury}\ \emph {et~al.}(2015)\citenamefont {Fleury},
  \citenamefont {Monticone},\ and\ \citenamefont
  {Al{\`u}}}]{fleury_invisibility_2015}%
  \BibitemOpen
  \bibfield  {author} {\bibinfo {author} {\bibfnamefont {R.}~\bibnamefont
  {Fleury}}, \bibinfo {author} {\bibfnamefont {F.}~\bibnamefont {Monticone}},\
  and\ \bibinfo {author} {\bibfnamefont {A.}~\bibnamefont {Al{\`u}}},\
  }\bibfield  {title} {\bibinfo {title} {Invisibility and {{Cloaking}}:
  {{Origins}}, {{Present}}, and {{Future Perspectives}}},\ }\href
  {https://doi.org/10.1103/PhysRevApplied.4.037001} {\bibfield  {journal}
  {\bibinfo  {journal} {Physical Review Applied}\ }\textbf {\bibinfo {volume}
  {4}},\ \bibinfo {pages} {037001} (\bibinfo {year} {2015})}\BibitemShut
  {NoStop}%
\bibitem [{\citenamefont {Nelson}\ \emph {et~al.}(2000)\citenamefont {Nelson},
  \citenamefont {Elliott},\ and\ \citenamefont {Nelson}}]{nelson_active_2000}%
  \BibitemOpen
  \bibfield  {author} {\bibinfo {author} {\bibfnamefont {P.~A.}\ \bibnamefont
  {Nelson}}, \bibinfo {author} {\bibfnamefont {S.~J.}\ \bibnamefont
  {Elliott}},\ and\ \bibinfo {author} {\bibfnamefont {P.~A.}\ \bibnamefont
  {Nelson}},\ }\href@noop {} {\emph {\bibinfo {title} {Active Control of
  Sound}}},\ \bibinfo {edition} {paperback ed., 4. print., [nachdr.]}\ ed.\
  (\bibinfo  {publisher} {Academic Press},\ \bibinfo {address} {San Diego},\
  \bibinfo {year} {2000})\BibitemShut {NoStop}%
\bibitem [{\citenamefont {Norris}(2008)}]{norris_acoustic_2008}%
  \BibitemOpen
  \bibfield  {author} {\bibinfo {author} {\bibfnamefont {A.~N.}\ \bibnamefont
  {Norris}},\ }\bibfield  {title} {\bibinfo {title} {Acoustic cloaking
  theory},\ }\href {https://doi.org/10.1098/rspa.2008.0076} {\bibfield
  {journal} {\bibinfo  {journal} {Proceedings of the Royal Society A:
  Mathematical, Physical and Engineering Sciences}\ }\textbf {\bibinfo {volume}
  {464}},\ \bibinfo {pages} {2411} (\bibinfo {year} {2008})},\ \Eprint
  {https://arxiv.org/abs/0805.0080} {arXiv:0805.0080 [cond-mat]} \BibitemShut
  {NoStop}%
\bibitem [{\citenamefont {Norris}(2015)}]{norris_acoustic_2015}%
  \BibitemOpen
  \bibfield  {author} {\bibinfo {author} {\bibfnamefont {A.~N.}\ \bibnamefont
  {Norris}},\ }\bibfield  {title} {\bibinfo {title} {Acoustic cloaking},\
  }\href@noop {} {\bibfield  {journal} {\bibinfo  {journal} {Acoustics Today}\
  } (\bibinfo {year} {2015})}\BibitemShut {NoStop}%
\bibitem [{\citenamefont {Song}\ \emph {et~al.}(2019)\citenamefont {Song},
  \citenamefont {Chen}, \citenamefont {Zhu}, \citenamefont {He},\ and\
  \citenamefont {Liu}}]{song_broadband_2019}%
  \BibitemOpen
  \bibfield  {author} {\bibinfo {author} {\bibfnamefont {X.}~\bibnamefont
  {Song}}, \bibinfo {author} {\bibfnamefont {T.}~\bibnamefont {Chen}}, \bibinfo
  {author} {\bibfnamefont {J.}~\bibnamefont {Zhu}}, \bibinfo {author}
  {\bibfnamefont {Y.}~\bibnamefont {He}},\ and\ \bibinfo {author}
  {\bibfnamefont {Z.}~\bibnamefont {Liu}},\ }\bibfield  {title} {\bibinfo
  {title} {Broadband acoustic cloaking and disguising with full-rangle incident
  angles based on reconfigurable metasurface},\ }\href
  {https://doi.org/10.1142/S0217979219502734} {\bibfield  {journal} {\bibinfo
  {journal} {International Journal of Modern Physics B}\ }\textbf {\bibinfo
  {volume} {33}},\ \bibinfo {pages} {1950273} (\bibinfo {year}
  {2019})}\BibitemShut {NoStop}%
\bibitem [{\citenamefont {Chen}\ and\ \citenamefont
  {Chan}(2010)}]{chen_acoustic_2010}%
  \BibitemOpen
  \bibfield  {author} {\bibinfo {author} {\bibfnamefont {H.}~\bibnamefont
  {Chen}}\ and\ \bibinfo {author} {\bibfnamefont {C.~T.}\ \bibnamefont
  {Chan}},\ }\bibfield  {title} {\bibinfo {title} {Acoustic cloaking and
  transformation acoustics},\ }\href
  {https://doi.org/10.1088/0022-3727/43/11/113001} {\bibfield  {journal}
  {\bibinfo  {journal} {Journal of Physics D: Applied Physics}\ }\textbf
  {\bibinfo {volume} {43}},\ \bibinfo {pages} {113001} (\bibinfo {year}
  {2010})}\BibitemShut {NoStop}%
\bibitem [{\citenamefont {Zigoneanu}\ \emph {et~al.}(2014)\citenamefont
  {Zigoneanu}, \citenamefont {Popa},\ and\ \citenamefont
  {Cummer}}]{zigoneanu_three-dimensional_2014}%
  \BibitemOpen
  \bibfield  {author} {\bibinfo {author} {\bibfnamefont {L.}~\bibnamefont
  {Zigoneanu}}, \bibinfo {author} {\bibfnamefont {B.-I.}\ \bibnamefont
  {Popa}},\ and\ \bibinfo {author} {\bibfnamefont {S.~A.}\ \bibnamefont
  {Cummer}},\ }\bibfield  {title} {\bibinfo {title} {Three-dimensional
  broadband omnidirectional acoustic ground cloak},\ }\href
  {https://doi.org/10.1038/nmat3901} {\bibfield  {journal} {\bibinfo  {journal}
  {Nature Materials}\ }\textbf {\bibinfo {volume} {13}},\ \bibinfo {pages}
  {352} (\bibinfo {year} {2014})}\BibitemShut {NoStop}%
\bibitem [{\citenamefont {Cummer}\ \emph {et~al.}(2016)\citenamefont {Cummer},
  \citenamefont {Christensen},\ and\ \citenamefont
  {Al{\`u}}}]{cummer_controlling_2016}%
  \BibitemOpen
  \bibfield  {author} {\bibinfo {author} {\bibfnamefont {S.~A.}\ \bibnamefont
  {Cummer}}, \bibinfo {author} {\bibfnamefont {J.}~\bibnamefont
  {Christensen}},\ and\ \bibinfo {author} {\bibfnamefont {A.}~\bibnamefont
  {Al{\`u}}},\ }\bibfield  {title} {\bibinfo {title} {Controlling sound with
  acoustic metamaterials},\ }\href {https://doi.org/10.1038/natrevmats.2016.1}
  {\bibfield  {journal} {\bibinfo  {journal} {Nature Reviews Materials}\
  }\textbf {\bibinfo {volume} {1}},\ \bibinfo {pages} {1} (\bibinfo {year}
  {2016})}\BibitemShut {NoStop}%
\bibitem [{\citenamefont {Friot}\ and\ \citenamefont
  {Bordier}(2004)}]{friotRealtimeActiveSuppression2004}%
  \BibitemOpen
  \bibfield  {author} {\bibinfo {author} {\bibfnamefont {E.}~\bibnamefont
  {Friot}}\ and\ \bibinfo {author} {\bibfnamefont {C.}~\bibnamefont
  {Bordier}},\ }\bibfield  {title} {\bibinfo {title} {Real-time active
  suppression of scattered acoustic radiation},\ }\href
  {https://doi.org/10.1016/j.jsv.2003.10.064} {\bibfield  {journal} {\bibinfo
  {journal} {Journal of Sound and Vibration}\ }\textbf {\bibinfo {volume}
  {278}},\ \bibinfo {pages} {563} (\bibinfo {year} {2004})}\BibitemShut
  {NoStop}%
\bibitem [{\citenamefont {Guevara~Vasquez}\ \emph {et~al.}(2011)\citenamefont
  {Guevara~Vasquez}, \citenamefont {Milton},\ and\ \citenamefont
  {Onofrei}}]{guevara_vasquez_exterior_2011}%
  \BibitemOpen
  \bibfield  {author} {\bibinfo {author} {\bibfnamefont {F.}~\bibnamefont
  {Guevara~Vasquez}}, \bibinfo {author} {\bibfnamefont {G.~W.}\ \bibnamefont
  {Milton}},\ and\ \bibinfo {author} {\bibfnamefont {D.}~\bibnamefont
  {Onofrei}},\ }\bibfield  {title} {\bibinfo {title} {Exterior cloaking with
  active sources in two dimensional acoustics},\ }\href
  {https://doi.org/10.1016/j.wavemoti.2011.03.005} {\bibfield  {journal}
  {\bibinfo  {journal} {Wave Motion}\ }\textbf {\bibinfo {volume} {48}},\
  \bibinfo {pages} {515} (\bibinfo {year} {2011})}\BibitemShut {NoStop}%
\bibitem [{\citenamefont {Popa}\ \emph {et~al.}(2015)\citenamefont {Popa},
  \citenamefont {Shinde}, \citenamefont {Konneker},\ and\ \citenamefont
  {Cummer}}]{popa_active_2015}%
  \BibitemOpen
  \bibfield  {author} {\bibinfo {author} {\bibfnamefont {B.-I.}\ \bibnamefont
  {Popa}}, \bibinfo {author} {\bibfnamefont {D.}~\bibnamefont {Shinde}},
  \bibinfo {author} {\bibfnamefont {A.}~\bibnamefont {Konneker}},\ and\
  \bibinfo {author} {\bibfnamefont {S.~A.}\ \bibnamefont {Cummer}},\ }\bibfield
   {title} {\bibinfo {title} {Active acoustic metamaterials reconfigurable in
  real time},\ }\href {https://doi.org/10.1103/PhysRevB.91.220303} {\bibfield
  {journal} {\bibinfo  {journal} {Physical Review B}\ }\textbf {\bibinfo
  {volume} {91}},\ \bibinfo {pages} {220303} (\bibinfo {year}
  {2015})}\BibitemShut {NoStop}%
\bibitem [{\citenamefont {Ma}\ \emph {et~al.}(2018)\citenamefont {Ma},
  \citenamefont {Fan}, \citenamefont {Sheng},\ and\ \citenamefont
  {Fink}}]{ma_shaping_2018}%
  \BibitemOpen
  \bibfield  {author} {\bibinfo {author} {\bibfnamefont {G.}~\bibnamefont
  {Ma}}, \bibinfo {author} {\bibfnamefont {X.}~\bibnamefont {Fan}}, \bibinfo
  {author} {\bibfnamefont {P.}~\bibnamefont {Sheng}},\ and\ \bibinfo {author}
  {\bibfnamefont {M.}~\bibnamefont {Fink}},\ }\bibfield  {title} {\bibinfo
  {title} {Shaping reverberating sound fields with an actively tunable
  metasurface},\ }\href {https://doi.org/10.1073/pnas.1801175115} {\bibfield
  {journal} {\bibinfo  {journal} {Proceedings of the National Academy of
  Sciences}\ }\textbf {\bibinfo {volume} {115}},\ \bibinfo {pages} {6638}
  (\bibinfo {year} {2018})}\BibitemShut {NoStop}%
\bibitem [{\citenamefont {Lasri}\ and\ \citenamefont
  {Sirota}(2023)}]{lasri_active_2023}%
  \BibitemOpen
  \bibfield  {author} {\bibinfo {author} {\bibfnamefont {O.}~\bibnamefont
  {Lasri}}\ and\ \bibinfo {author} {\bibfnamefont {L.}~\bibnamefont {Sirota}},\
  }\bibfield  {title} {\bibinfo {title} {Active control approach to temporal
  acoustic cloaking},\ }\href {https://doi.org/10.1063/5.0152144} {\bibfield
  {journal} {\bibinfo  {journal} {Applied Physics Letters}\ }\textbf {\bibinfo
  {volume} {123}},\ \bibinfo {pages} {032201} (\bibinfo {year}
  {2023})}\BibitemShut {NoStop}%
\bibitem [{\citenamefont {Miller}(2006)}]{miller_perfect_2006}%
  \BibitemOpen
  \bibfield  {author} {\bibinfo {author} {\bibfnamefont {D.~A.~B.}\
  \bibnamefont {Miller}},\ }\bibfield  {title} {\bibinfo {title} {On perfect
  cloaking},\ }\href {https://doi.org/10.1364/OE.14.012457} {\bibfield
  {journal} {\bibinfo  {journal} {Optics Express}\ }\textbf {\bibinfo {volume}
  {14}},\ \bibinfo {pages} {12457} (\bibinfo {year} {2006})}\BibitemShut
  {NoStop}%
\bibitem [{\citenamefont {{van Manen}}\ \emph {et~al.}(2007)\citenamefont {{van
  Manen}}, \citenamefont {Robertsson},\ and\ \citenamefont
  {Curtis}}]{van_manen_exact_2007}%
  \BibitemOpen
  \bibfield  {author} {\bibinfo {author} {\bibfnamefont {D.-J.}\ \bibnamefont
  {{van Manen}}}, \bibinfo {author} {\bibfnamefont {J.~O.~A.}\ \bibnamefont
  {Robertsson}},\ and\ \bibinfo {author} {\bibfnamefont {A.}~\bibnamefont
  {Curtis}},\ }\bibfield  {title} {\bibinfo {title} {Exact wave field
  simulation for finite-volume scattering problems},\ }\href
  {https://doi.org/10.1121/1.2771371} {\bibfield  {journal} {\bibinfo
  {journal} {The Journal of the Acoustical Society of America}\ }\textbf
  {\bibinfo {volume} {122}},\ \bibinfo {pages} {EL115} (\bibinfo {year}
  {2007})}\BibitemShut {NoStop}%
\bibitem [{\citenamefont {Vasmel}\ \emph {et~al.}(2013)\citenamefont {Vasmel},
  \citenamefont {Robertsson}, \citenamefont {{van Manen}},\ and\ \citenamefont
  {Curtis}}]{vasmel_immersive_2013}%
  \BibitemOpen
  \bibfield  {author} {\bibinfo {author} {\bibfnamefont {M.}~\bibnamefont
  {Vasmel}}, \bibinfo {author} {\bibfnamefont {J.~O.~A.}\ \bibnamefont
  {Robertsson}}, \bibinfo {author} {\bibfnamefont {D.-J.}\ \bibnamefont {{van
  Manen}}},\ and\ \bibinfo {author} {\bibfnamefont {A.}~\bibnamefont
  {Curtis}},\ }\bibfield  {title} {\bibinfo {title} {Immersive experimentation
  in a wave propagation laboratory},\ }\href
  {https://doi.org/10.1121/1.4826912} {\bibfield  {journal} {\bibinfo
  {journal} {The Journal of the Acoustical Society of America}\ }\textbf
  {\bibinfo {volume} {134}},\ \bibinfo {pages} {EL492} (\bibinfo {year}
  {2013})}\BibitemShut {NoStop}%
\bibitem [{\citenamefont {{van Manen}}\ \emph {et~al.}(2015)\citenamefont {{van
  Manen}}, \citenamefont {Vasmel}, \citenamefont {Greenhalgh},\ and\
  \citenamefont {Robertsson}}]{van_manen_broadband_2015}%
  \BibitemOpen
  \bibfield  {author} {\bibinfo {author} {\bibfnamefont {D.-J.}\ \bibnamefont
  {{van Manen}}}, \bibinfo {author} {\bibfnamefont {M.}~\bibnamefont {Vasmel}},
  \bibinfo {author} {\bibfnamefont {S.}~\bibnamefont {Greenhalgh}},\ and\
  \bibinfo {author} {\bibfnamefont {J.~O.~A.}\ \bibnamefont {Robertsson}},\
  }\bibfield  {title} {\bibinfo {title} {Broadband cloaking and holography with
  exact boundary conditions},\ }\href {https://doi.org/10.1121/1.4921340}
  {\bibfield  {journal} {\bibinfo  {journal} {The Journal of the Acoustical
  Society of America}\ }\textbf {\bibinfo {volume} {137}},\ \bibinfo {pages}
  {EL415} (\bibinfo {year} {2015})}\BibitemShut {NoStop}%
\bibitem [{\citenamefont {Becker}\ \emph {et~al.}(2018)\citenamefont {Becker},
  \citenamefont {{van Manen}}, \citenamefont {Donahue}, \citenamefont
  {B{\"a}rlocher}, \citenamefont {B{\"o}rsing}, \citenamefont {Broggini},
  \citenamefont {Haag}, \citenamefont {Robertsson}, \citenamefont {Schmidt},
  \citenamefont {Greenhalgh},\ and\ \citenamefont
  {Blum}}]{beckerImmersiveWavePropagation2018}%
  \BibitemOpen
  \bibfield  {author} {\bibinfo {author} {\bibfnamefont {T.~S.}\ \bibnamefont
  {Becker}}, \bibinfo {author} {\bibfnamefont {D.-J.}\ \bibnamefont {{van
  Manen}}}, \bibinfo {author} {\bibfnamefont {C.~M.}\ \bibnamefont {Donahue}},
  \bibinfo {author} {\bibfnamefont {C.}~\bibnamefont {B{\"a}rlocher}}, \bibinfo
  {author} {\bibfnamefont {N.}~\bibnamefont {B{\"o}rsing}}, \bibinfo {author}
  {\bibfnamefont {F.}~\bibnamefont {Broggini}}, \bibinfo {author}
  {\bibfnamefont {T.}~\bibnamefont {Haag}}, \bibinfo {author} {\bibfnamefont
  {J.~O.~A.}\ \bibnamefont {Robertsson}}, \bibinfo {author} {\bibfnamefont
  {D.~R.}\ \bibnamefont {Schmidt}}, \bibinfo {author} {\bibfnamefont {S.~A.}\
  \bibnamefont {Greenhalgh}},\ and\ \bibinfo {author} {\bibfnamefont {T.~E.}\
  \bibnamefont {Blum}},\ }\bibfield  {title} {\bibinfo {title} {Immersive
  {{Wave Propagation Experimentation}}: {{Physical Implementation}} and
  {{One-Dimensional Acoustic Results}}},\ }\href
  {https://doi.org/10.1103/PhysRevX.8.031011} {\bibfield  {journal} {\bibinfo
  {journal} {Physical Review X}\ }\textbf {\bibinfo {volume} {8}},\ \bibinfo
  {pages} {031011} (\bibinfo {year} {2018})}\BibitemShut {NoStop}%
\bibitem [{\citenamefont {B{\"o}rsing}\ \emph {et~al.}(2019)\citenamefont
  {B{\"o}rsing}, \citenamefont {Becker}, \citenamefont {Curtis}, \citenamefont
  {{van Manen}}, \citenamefont {Haag},\ and\ \citenamefont
  {Robertsson}}]{borsingCloakingHolographyExperiments2019}%
  \BibitemOpen
  \bibfield  {author} {\bibinfo {author} {\bibfnamefont {N.}~\bibnamefont
  {B{\"o}rsing}}, \bibinfo {author} {\bibfnamefont {T.~S.}\ \bibnamefont
  {Becker}}, \bibinfo {author} {\bibfnamefont {A.}~\bibnamefont {Curtis}},
  \bibinfo {author} {\bibfnamefont {D.-J.}\ \bibnamefont {{van Manen}}},
  \bibinfo {author} {\bibfnamefont {T.}~\bibnamefont {Haag}},\ and\ \bibinfo
  {author} {\bibfnamefont {J.~O.}\ \bibnamefont {Robertsson}},\ }\bibfield
  {title} {\bibinfo {title} {Cloaking and {{Holography Experiments Using
  Immersive Boundary Conditions}}},\ }\href
  {https://doi.org/10.1103/PhysRevApplied.12.024011} {\bibfield  {journal}
  {\bibinfo  {journal} {Physical Review Applied}\ }\textbf {\bibinfo {volume}
  {12}},\ \bibinfo {pages} {024011} (\bibinfo {year} {2019})}\BibitemShut
  {NoStop}%
\bibitem [{\citenamefont {Broggini}\ \emph {et~al.}(2017)\citenamefont
  {Broggini}, \citenamefont {Vasmel}, \citenamefont {Robertsson},\ and\
  \citenamefont {{van Manen}}}]{brogginiImmersiveBoundaryConditions2017}%
  \BibitemOpen
  \bibfield  {author} {\bibinfo {author} {\bibfnamefont {F.}~\bibnamefont
  {Broggini}}, \bibinfo {author} {\bibfnamefont {M.}~\bibnamefont {Vasmel}},
  \bibinfo {author} {\bibfnamefont {J.~O.~A.}\ \bibnamefont {Robertsson}},\
  and\ \bibinfo {author} {\bibfnamefont {D.-J.}\ \bibnamefont {{van Manen}}},\
  }\bibfield  {title} {\bibinfo {title} {Immersive boundary conditions:
  {{Theory}}, implementation, and examples},\ }\href
  {https://doi.org/10.1190/geo2016-0458.1} {\bibfield  {journal} {\bibinfo
  {journal} {GEOPHYSICS}\ }\textbf {\bibinfo {volume} {82}},\ \bibinfo {pages}
  {T97} (\bibinfo {year} {2017})}\BibitemShut {NoStop}%
\bibitem [{\citenamefont {Becker}\ \emph {et~al.}(2021)\citenamefont {Becker},
  \citenamefont {Van~Manen}, \citenamefont {Haag}, \citenamefont
  {B{\"a}rlocher}, \citenamefont {Li}, \citenamefont {B{\"o}rsing},
  \citenamefont {Curtis}, \citenamefont {{Serra-Garcia}},\ and\ \citenamefont
  {Robertsson}}]{becker_broadband_2021}%
  \BibitemOpen
  \bibfield  {author} {\bibinfo {author} {\bibfnamefont {T.~S.}\ \bibnamefont
  {Becker}}, \bibinfo {author} {\bibfnamefont {D.-J.}\ \bibnamefont
  {Van~Manen}}, \bibinfo {author} {\bibfnamefont {T.}~\bibnamefont {Haag}},
  \bibinfo {author} {\bibfnamefont {C.}~\bibnamefont {B{\"a}rlocher}}, \bibinfo
  {author} {\bibfnamefont {X.}~\bibnamefont {Li}}, \bibinfo {author}
  {\bibfnamefont {N.}~\bibnamefont {B{\"o}rsing}}, \bibinfo {author}
  {\bibfnamefont {A.}~\bibnamefont {Curtis}}, \bibinfo {author} {\bibfnamefont
  {M.}~\bibnamefont {{Serra-Garcia}}},\ and\ \bibinfo {author} {\bibfnamefont
  {J.~O.~A.}\ \bibnamefont {Robertsson}},\ }\bibfield  {title} {\bibinfo
  {title} {Broadband acoustic invisibility and illusions},\ }\href
  {https://doi.org/10.1126/sciadv.abi9627} {\bibfield  {journal} {\bibinfo
  {journal} {Science Advances}\ }\textbf {\bibinfo {volume} {7}},\ \bibinfo
  {pages} {eabi9627} (\bibinfo {year} {2021})}\BibitemShut {NoStop}%
\bibitem [{\citenamefont {M{\"u}ller}\ \emph {et~al.}(2023)\citenamefont
  {M{\"u}ller}, \citenamefont {Becker}, \citenamefont {Li}, \citenamefont
  {Aichele}, \citenamefont {{Serra-Garcia}}, \citenamefont {Robertsson},\ and\
  \citenamefont {Van~Manen}}]{muller_acoustic_2023}%
  \BibitemOpen
  \bibfield  {author} {\bibinfo {author} {\bibfnamefont {J.}~\bibnamefont
  {M{\"u}ller}}, \bibinfo {author} {\bibfnamefont {T.~S.}\ \bibnamefont
  {Becker}}, \bibinfo {author} {\bibfnamefont {X.}~\bibnamefont {Li}}, \bibinfo
  {author} {\bibfnamefont {J.}~\bibnamefont {Aichele}}, \bibinfo {author}
  {\bibfnamefont {M.}~\bibnamefont {{Serra-Garcia}}}, \bibinfo {author}
  {\bibfnamefont {J.~O.}\ \bibnamefont {Robertsson}},\ and\ \bibinfo {author}
  {\bibfnamefont {D.-J.}\ \bibnamefont {Van~Manen}},\ }\bibfield  {title}
  {\bibinfo {title} {Acoustic cloning},\ }\href
  {https://doi.org/10.1103/PhysRevApplied.20.064014} {\bibfield  {journal}
  {\bibinfo  {journal} {Physical Review Applied}\ }\textbf {\bibinfo {volume}
  {20}},\ \bibinfo {pages} {064014} (\bibinfo {year} {2023})}\BibitemShut
  {NoStop}%
\bibitem [{\citenamefont {Barton}(1989)}]{barton_elements_1989}%
  \BibitemOpen
  \bibfield  {author} {\bibinfo {author} {\bibfnamefont {G.}~\bibnamefont
  {Barton}},\ }\href@noop {} {\emph {\bibinfo {title} {Elements of {{Green}}'s
  Functions and Propagation: Potentials, Diffusion, and Waves}}},\ Oxford
  Science Publications\ (\bibinfo  {publisher} {Clarendon Press ; Oxford
  University Press},\ \bibinfo {address} {Oxford : New York},\ \bibinfo {year}
  {1989})\BibitemShut {NoStop}%
\bibitem [{\citenamefont {Fokkema}\ and\ \citenamefont {van~den
  Berg}(1993)}]{fokkema_seismic_1993}%
  \BibitemOpen
  \bibfield  {author} {\bibinfo {author} {\bibfnamefont {J.~T.}\ \bibnamefont
  {Fokkema}}\ and\ \bibinfo {author} {\bibfnamefont {P.~M.}\ \bibnamefont
  {van~den Berg}},\ }\href@noop {} {\emph {\bibinfo {title} {Seismic
  Applications of Acoustic Reciprocity}}}\ (\bibinfo  {publisher} {Elsevier},\
  \bibinfo {address} {Amsterdam ; New York},\ \bibinfo {year}
  {1993})\BibitemShut {NoStop}%
\bibitem [{\citenamefont
  {de~Hoop}(1995)}]{hoopHandbookRadiationScattering1995}%
  \BibitemOpen
  \bibfield  {author} {\bibinfo {author} {\bibfnamefont {A.~T.}\ \bibnamefont
  {de~Hoop}},\ }\href@noop {} {\emph {\bibinfo {title} {Handbook of Radiation
  and Scattering of Waves: Acoustic Waves in Fluid, Elastic Waves in Solids,
  Electromagnetic Waves}}}\ (\bibinfo  {publisher} {Acad. Press},\ \bibinfo
  {address} {London},\ \bibinfo {year} {1995})\BibitemShut {NoStop}%
\bibitem [{\citenamefont {Amundsen}(2001)}]{amundsen_elimination_2001}%
  \BibitemOpen
  \bibfield  {author} {\bibinfo {author} {\bibfnamefont {L.}~\bibnamefont
  {Amundsen}},\ }\bibfield  {title} {\bibinfo {title} {Elimination of
  free-surface related multiples without need of the source wavelet},\ }\href
  {https://doi.org/10.1190/1.1444912} {\bibfield  {journal} {\bibinfo
  {journal} {GEOPHYSICS}\ }\textbf {\bibinfo {volume} {66}},\ \bibinfo {pages}
  {327} (\bibinfo {year} {2001})}\BibitemShut {NoStop}%
\bibitem [{\citenamefont {Wapenaar}\ \emph {et~al.}(2011)\citenamefont
  {Wapenaar}, \citenamefont {{van der Neut}}, \citenamefont {Ruigrok},
  \citenamefont {Draganov}, \citenamefont {Hunziker}, \citenamefont {Slob},
  \citenamefont {Thorbecke},\ and\ \citenamefont
  {Snieder}}]{wapenaar_seismic_2011}%
  \BibitemOpen
  \bibfield  {author} {\bibinfo {author} {\bibfnamefont {K.}~\bibnamefont
  {Wapenaar}}, \bibinfo {author} {\bibfnamefont {J.}~\bibnamefont {{van der
  Neut}}}, \bibinfo {author} {\bibfnamefont {E.}~\bibnamefont {Ruigrok}},
  \bibinfo {author} {\bibfnamefont {D.}~\bibnamefont {Draganov}}, \bibinfo
  {author} {\bibfnamefont {J.}~\bibnamefont {Hunziker}}, \bibinfo {author}
  {\bibfnamefont {E.}~\bibnamefont {Slob}}, \bibinfo {author} {\bibfnamefont
  {J.}~\bibnamefont {Thorbecke}},\ and\ \bibinfo {author} {\bibfnamefont
  {R.}~\bibnamefont {Snieder}},\ }\bibfield  {title} {\bibinfo {title} {Seismic
  interferometry by crosscorrelation and by multidimensional deconvolution: A
  systematic comparison: {{Seismic}} interferometry},\ }\href
  {https://doi.org/10.1111/j.1365-246X.2011.05007.x} {\bibfield  {journal}
  {\bibinfo  {journal} {Geophysical Journal International}\ }\textbf {\bibinfo
  {volume} {185}},\ \bibinfo {pages} {1335} (\bibinfo {year}
  {2011})}\BibitemShut {NoStop}%
\bibitem [{\citenamefont {Li}\ \emph {et~al.}(2021)\citenamefont {Li},
  \citenamefont {Becker}, \citenamefont {Ravasi}, \citenamefont {Robertsson},\
  and\ \citenamefont {{van Manen}}}]{liClosedapertureUnboundedAcoustics2021}%
  \BibitemOpen
  \bibfield  {author} {\bibinfo {author} {\bibfnamefont {X.}~\bibnamefont
  {Li}}, \bibinfo {author} {\bibfnamefont {T.}~\bibnamefont {Becker}}, \bibinfo
  {author} {\bibfnamefont {M.}~\bibnamefont {Ravasi}}, \bibinfo {author}
  {\bibfnamefont {J.}~\bibnamefont {Robertsson}},\ and\ \bibinfo {author}
  {\bibfnamefont {D.-J.}\ \bibnamefont {{van Manen}}},\ }\bibfield  {title}
  {\bibinfo {title} {Closed-aperture unbounded acoustics experimentation using
  multidimensional deconvolution},\ }\href {https://doi.org/10.1121/10.0003706}
  {\bibfield  {journal} {\bibinfo  {journal} {The Journal of the Acoustical
  Society of America}\ }\textbf {\bibinfo {volume} {149}},\ \bibinfo {pages}
  {1813} (\bibinfo {year} {2021})}\BibitemShut {NoStop}%
\bibitem [{\citenamefont {Tolan}\ and\ \citenamefont
  {Schneider}(2003)}]{tolan_locally_2003}%
  \BibitemOpen
  \bibfield  {author} {\bibinfo {author} {\bibfnamefont {J.~G.}\ \bibnamefont
  {Tolan}}\ and\ \bibinfo {author} {\bibfnamefont {J.~B.}\ \bibnamefont
  {Schneider}},\ }\bibfield  {title} {\bibinfo {title} {Locally conformal
  method for acoustic finite-difference time-domain modeling of rigid
  surfaces},\ }\href {https://doi.org/10.1121/1.1616576} {\bibfield  {journal}
  {\bibinfo  {journal} {The Journal of the Acoustical Society of America}\
  }\textbf {\bibinfo {volume} {114}},\ \bibinfo {pages} {2575} (\bibinfo {year}
  {2003})}\BibitemShut {NoStop}%
\bibitem [{\citenamefont {Gonz{\'a}lez}(2010)}]{gonzalez_measurement_2010}%
  \BibitemOpen
  \bibfield  {author} {\bibinfo {author} {\bibfnamefont {{\'A}.}~\bibnamefont
  {Gonz{\'a}lez}},\ }\bibfield  {title} {\bibinfo {title} {Measurement of
  {{Areas}} on a {{Sphere Using Fibonacci}} and {{Latitude}}--{{Longitude
  Lattices}}},\ }\href {https://doi.org/10.1007/s11004-009-9257-x} {\bibfield
  {journal} {\bibinfo  {journal} {Mathematical Geosciences}\ }\textbf {\bibinfo
  {volume} {42}},\ \bibinfo {pages} {49} (\bibinfo {year} {2010})}\BibitemShut
  {NoStop}%
\bibitem [{\citenamefont {Marques}\ \emph {et~al.}(2013)\citenamefont
  {Marques}, \citenamefont {Bouville}, \citenamefont {Ribardi{\`e}re},
  \citenamefont {Santos},\ and\ \citenamefont
  {Bouatouch}}]{marques_spherical_2013}%
  \BibitemOpen
  \bibfield  {author} {\bibinfo {author} {\bibfnamefont {R.}~\bibnamefont
  {Marques}}, \bibinfo {author} {\bibfnamefont {C.}~\bibnamefont {Bouville}},
  \bibinfo {author} {\bibfnamefont {M.}~\bibnamefont {Ribardi{\`e}re}},
  \bibinfo {author} {\bibfnamefont {L.~P.}\ \bibnamefont {Santos}},\ and\
  \bibinfo {author} {\bibfnamefont {K.}~\bibnamefont {Bouatouch}},\ }\bibfield
  {title} {\bibinfo {title} {Spherical {{Fibonacci Point Sets}} for
  {{Illumination Integrals}}},\ }\href {https://doi.org/10.1111/cgf.12190}
  {\bibfield  {journal} {\bibinfo  {journal} {Computer Graphics Forum}\
  }\textbf {\bibinfo {volume} {32}},\ \bibinfo {pages} {134} (\bibinfo {year}
  {2013})}\BibitemShut {NoStop}%
\bibitem [{\citenamefont {Berkhout}\ and\ \citenamefont
  {Wapenaar}(1989)}]{berkhout_one-way_1989}%
  \BibitemOpen
  \bibfield  {author} {\bibinfo {author} {\bibfnamefont {A.~J.}\ \bibnamefont
  {Berkhout}}\ and\ \bibinfo {author} {\bibfnamefont {C.~P.~A.}\ \bibnamefont
  {Wapenaar}},\ }\bibfield  {title} {\bibinfo {title} {One-way versions of the
  {{Kirchhoff}} integral},\ }\href {https://doi.org/10.1190/1.1442672}
  {\bibfield  {journal} {\bibinfo  {journal} {Geophysics}\ }\textbf {\bibinfo
  {volume} {54}},\ \bibinfo {pages} {460} (\bibinfo {year} {1989})}\BibitemShut
  {NoStop}%
\bibitem [{\citenamefont {Wapenaar}(2020)}]{wapenaar_reciprocity_2020}%
  \BibitemOpen
  \bibfield  {author} {\bibinfo {author} {\bibfnamefont {K.}~\bibnamefont
  {Wapenaar}},\ }\bibfield  {title} {\bibinfo {title} {Reciprocity and
  {{Representation Theorems}} for {{Flux-}} and {{Field-Normalised Decomposed
  Wave Fields}}},\ }\href {https://doi.org/10.1155/2020/9540135} {\bibfield
  {journal} {\bibinfo  {journal} {Advances in Mathematical Physics}\ }\textbf
  {\bibinfo {volume} {2020}},\ \bibinfo {pages} {1} (\bibinfo {year}
  {2020})}\BibitemShut {NoStop}%
\bibitem [{\citenamefont {Schneider}()}]{schneider_understanding_nodate}%
  \BibitemOpen
  \bibfield  {author} {\bibinfo {author} {\bibfnamefont {J.~B.}\ \bibnamefont
  {Schneider}},\ }\href@noop {} {\emph {\bibinfo {title} {Understanding the
  {{Finite-Difference Time-Domain Method}}}}}\ (\bibinfo  {publisher}
  {Washington State University})\BibitemShut {NoStop}%
\bibitem [{\citenamefont
  {Botteldooren}(1995)}]{botteldooren_finite-difference_1995}%
  \BibitemOpen
  \bibfield  {author} {\bibinfo {author} {\bibfnamefont {D.}~\bibnamefont
  {Botteldooren}},\ }\bibfield  {title} {\bibinfo {title} {Finite-difference
  time-domain simulation of low-frequency room acoustic problems},\ }\href
  {https://doi.org/10.1121/1.413817} {\bibfield  {journal} {\bibinfo  {journal}
  {The Journal of the Acoustical Society of America}\ }\textbf {\bibinfo
  {volume} {98}},\ \bibinfo {pages} {3302} (\bibinfo {year}
  {1995})}\BibitemShut {NoStop}%
\bibitem [{\citenamefont {Sheaffer}\ \emph {et~al.}(2012)\citenamefont
  {Sheaffer}, \citenamefont {Walstijn},\ and\ \citenamefont
  {Fazenda}}]{sheaffer_physically-constrained_2012}%
  \BibitemOpen
  \bibfield  {author} {\bibinfo {author} {\bibfnamefont {J.}~\bibnamefont
  {Sheaffer}}, \bibinfo {author} {\bibfnamefont {M.~V.}\ \bibnamefont
  {Walstijn}},\ and\ \bibinfo {author} {\bibfnamefont {B.}~\bibnamefont
  {Fazenda}},\ }\bibfield  {title} {\bibinfo {title} {A
  {{Physically-Constrained Source Model}} for {{FDTD Acoustic Simulation}}},\
  }in\ \href@noop {} {\emph {\bibinfo {booktitle} {Proc. of the 15th
  {{International Conference}} on {{Digital Audio Effects}} ({{DAFx-12}})}}}\
  (\bibinfo {address} {York, UK},\ \bibinfo {year} {2012})\BibitemShut
  {NoStop}%
\bibitem [{\citenamefont {Ravasi}\ and\ \citenamefont
  {Vasconcelos}(2020)}]{ravasiPyLopsLinearoperatorPython2020}%
  \BibitemOpen
  \bibfield  {author} {\bibinfo {author} {\bibfnamefont {M.}~\bibnamefont
  {Ravasi}}\ and\ \bibinfo {author} {\bibfnamefont {I.}~\bibnamefont
  {Vasconcelos}},\ }\bibfield  {title} {\bibinfo {title} {{{PyLops}}---{{A}}
  linear-operator {{Python}} library for scalable algebra and optimization},\
  }\href {https://doi.org/10.1016/j.softx.2019.100361} {\bibfield  {journal}
  {\bibinfo  {journal} {SoftwareX}\ }\textbf {\bibinfo {volume} {11}},\
  \bibinfo {pages} {100361} (\bibinfo {year} {2020})}\BibitemShut {NoStop}%
\bibitem [{\citenamefont
  {Roberts}(2018{\natexlab{a}})}]{roberts_unreasonable_2018}%
  \BibitemOpen
  \bibfield  {author} {\bibinfo {author} {\bibfnamefont {M.}~\bibnamefont
  {Roberts}},\ }\href@noop {} {\bibinfo {title} {The {{Unreasonable
  Effectiveness}} of {{Quasirandom Sequences}}}} (\bibinfo {year}
  {2018}{\natexlab{a}})\BibitemShut {NoStop}%
\bibitem [{\citenamefont {Roberts}(2018{\natexlab{b}})}]{roberts_evenly_2018}%
  \BibitemOpen
  \bibfield  {author} {\bibinfo {author} {\bibfnamefont {M.}~\bibnamefont
  {Roberts}},\ }\href@noop {} {\bibinfo {title} {Evenly distributing points on
  a sphere}} (\bibinfo {year} {2018}{\natexlab{b}})\BibitemShut {NoStop}%
\bibitem [{\citenamefont {Crameri}(2023)}]{crameri_scientific_2023}%
  \BibitemOpen
  \bibfield  {author} {\bibinfo {author} {\bibfnamefont {F.}~\bibnamefont
  {Crameri}},\ }\href {https://doi.org/10.5281/zenodo.8409685} {\bibinfo
  {title} {Scientific colour maps}},\ \bibinfo {howpublished} {Zenodo}
  (\bibinfo {year} {2023})\BibitemShut {NoStop}%
\end{thebibliography}%

% Supplementary Material
\appendix
\widetext
% Appendix figure/table numbering: S1, S2, ...
\setcounter{figure}{0}
\renewcommand{\thefigure}{S\arabic{figure}}
\setcounter{table}{0}
\renewcommand{\thetable}{S\arabic{table}}

\section{Supplemental Material}
\subsection{MDD}
\label{app:mdd}

Multidimensional deconvolution (MDD) in three dimensions is similar to the two-dimensional approach \cite{liClosedapertureUnboundedAcoustics2021, muller_acoustic_2023}. In 3D, the closed integration path of the 2D formulation is replaced by a closed surface that fully encloses the scatterer. The multidimensional convolution formalism (MDC) remains applicable, and the volumetric Green's functions are retrieved by solving the associated overdetermined linear system by regularized least-squares inversion. The inversion is carried out with the PyLops MDD operator~\cite{ravasiPyLopsLinearoperatorPython2020}, using damped LSQR with Tikhonov damping $\lambda = 10^{-4}$, a $50$-iteration limit (giving implicit early-stopping regularization), a causality preconditioner, and a frequency-domain bandwidth of $2.5\,f_c = 30\,\mathrm{kHz}$, with $f_c = 12\,\mathrm{kHz}$ the Ricker center frequency and $2.5$ the bandwidth factor.

\begin{figure*}
    \centering
    \includegraphics[width=0.4\textwidth]{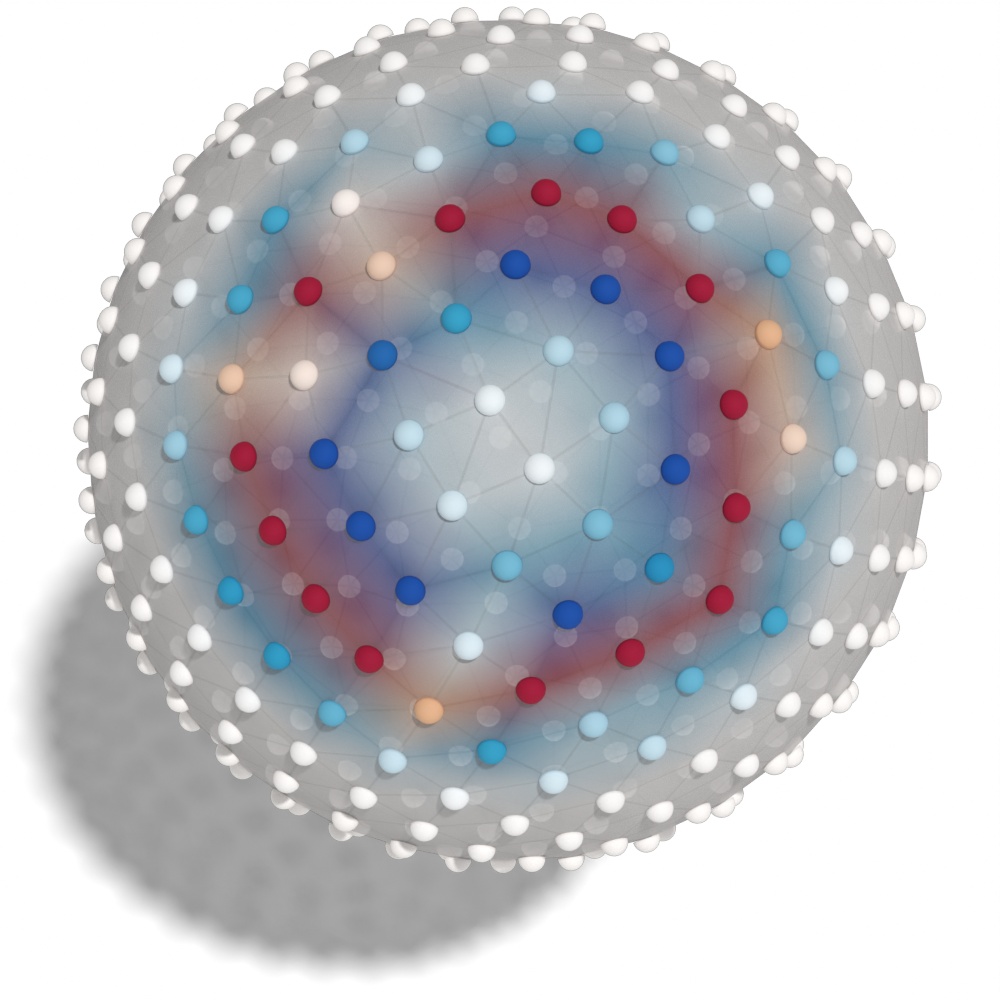}
    \hfil
    \includegraphics[width=0.4\textwidth]{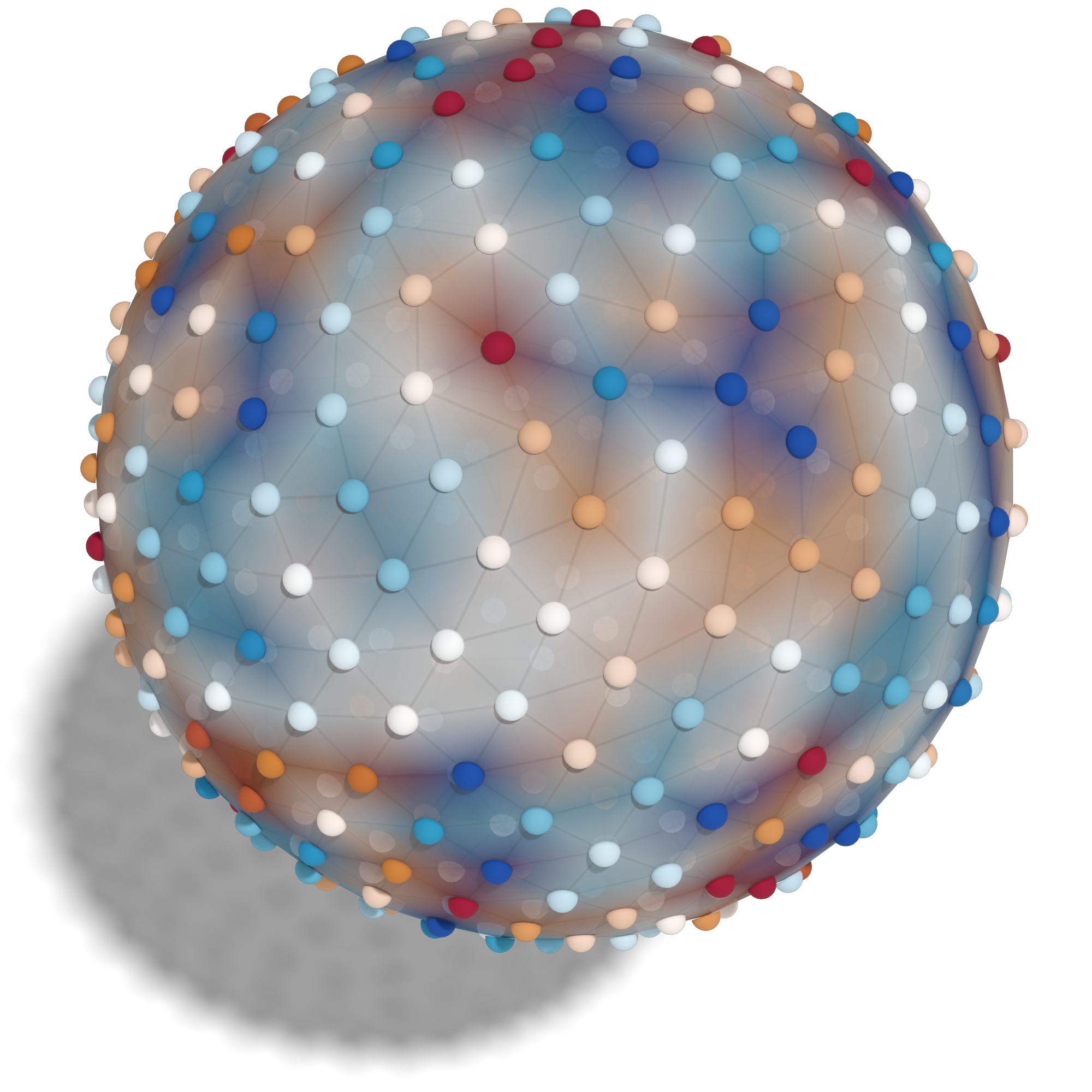}
    \caption{Pressure recorded on the discretised recording surface $\mathcal{S}^\mathrm{O}$ --- a Fibonacci-sampled sphere triangulated via its convex hull --- at two timestamps for a single illumination source (see Fig.~\ref{fig:3Ddisguising_mdd}). \textbf{Left:} just after the wavefront's first arrival. \textbf{Right:} deep in the reverberant tail. The triangulated mesh is Gouraud-shaded by the per-vertex pressure, with a faint wireframe overlay tracing the triangulation; small hemispherical markers at each receiver render in the same tone as the local sample, making the discrete sampling pattern explicit. Each panel is normalised independently.}
    \label{fig:illumination_data}
\end{figure*}

\begin{figure*}
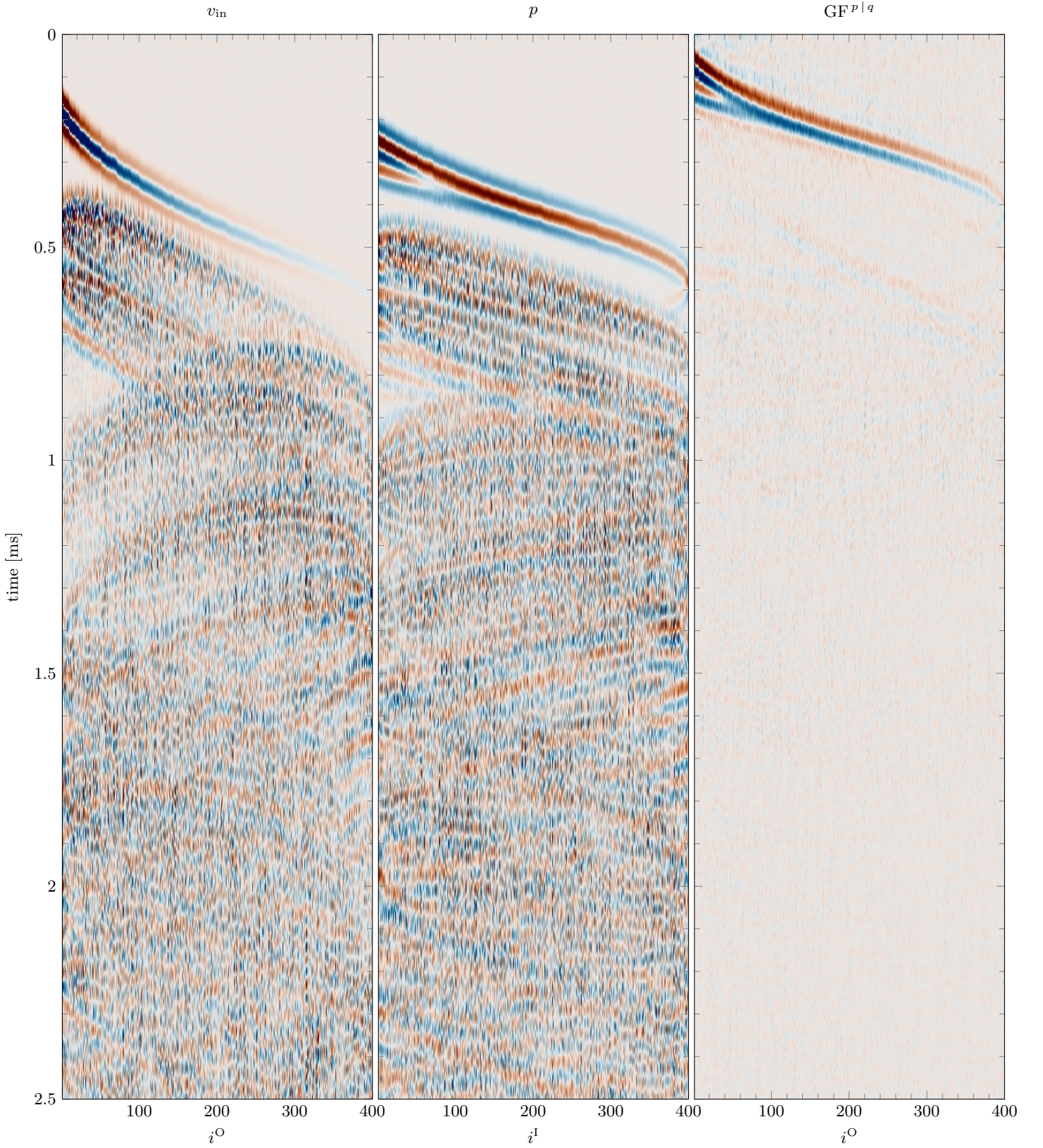

    \centering
    % Each panel reads its axis extents (xmin/xmax/ymin/ymax) from the matching
% <basename>.meta.tex sidecar produced by figures/preprocess/preprocess_mdd_panels.py.
% \begingroup...\endgroup scopes the \def's so panels don't inherit each other.
%
% \tmax (ms) clips the displayed time window for all panels; the image data
% itself extends to \jpgYmax from the sidecars, so the graphics keep that
% extent and the axis just crops to \tmax.
\def\tmax{2.5}%
\begingroup
\input{fig/sim/outer_sphere.meta.tex}%
\begin{tikzpicture}[tikzPlot]
    \begin{axis}[%
            width=0.325\textwidth,
            height=20cm,
            at={(0,0)},
            scale only axis,
            minor tick num=4,
            xtick distance=100,
            axis on top,
            xmin=\jpgXmin,
            xmax=\jpgXmax,
            ymin=\jpgYmin,
            ymax=\tmax,
            ytick distance=0.5,
            y dir=reverse, % Invert the y-axis
            xlabel={$i^\mathrm{O}$},
            ylabel={time [ms]},
            title={$v_\mathrm{in}$},
        ]
        \addplot [forget plot] graphics
            [xmin=\jpgXmin, xmax=\jpgXmax, ymin=\jpgYmin, ymax=\jpgYmax]
            {fig/sim/outer_sphere.jpg};
    \end{axis}
\end{tikzpicture}%
\endgroup
\hfil
\begingroup
\input{fig/sim/inner_sphere.meta.tex}%
\begin{tikzpicture}[tikzPlot]
    \begin{axis}[%
            width=0.325\textwidth,
            height=20cm,
            at={(0,0)},
            scale only axis,
            minor tick num=4,
            xtick distance=100,
            axis on top,
            xmin=\jpgXmin,
            xmax=\jpgXmax,
            ymin=\jpgYmin,
            ymax=\tmax,
            ytick distance=0.5,
            y dir=reverse, % Invert the y-axis
            xlabel={$i^\mathrm{I}$},
            ylabel={},
            yticklabels={},
            title={$p$},
        ]
        \addplot [forget plot] graphics
            [xmin=\jpgXmin, xmax=\jpgXmax, ymin=\jpgYmin, ymax=\jpgYmax]
            {fig/sim/inner_sphere.jpg};
    \end{axis}
\end{tikzpicture}%
\endgroup
\hfil
\begingroup
\input{fig/sim/gf_sphere.meta.tex}%
\begin{tikzpicture}[tikzPlot]
    \begin{axis}[%
            width=0.325\textwidth,
            height=20cm,
            at={(0,0)},
            scale only axis,
            minor tick num=4,
            xtick distance=100,
            axis on top,
            xmin=\jpgXmin,
            xmax=\jpgXmax,
            ymin=\jpgYmin,
            ymax=\tmax,
            ytick distance=0.5,
            y dir=reverse, % Invert the y-axis
            xlabel={$i^\mathrm{O}$},
            ylabel={},
            yticklabels={},
            title={$\mathrm{GF}^{\,p\, |\,  q}$},
        ]
        \addplot [forget plot] graphics
            [xmin=\jpgXmin, xmax=\jpgXmax, ymin=\jpgYmin, ymax=\jpgYmax]
            {fig/sim/gf_sphere.jpg};
    \end{axis}
\end{tikzpicture}%
\endgroup
    \caption{Multidimensional deconvolution (MDD) in three dimensions. The left column shows the ingoing particle velocity $v_\mathrm{in}$ on the recording surface $\mathcal{S}^\mathrm{O}$. The middle column shows the recorded pressure field $p$ on the emitting surface. Both plots are shown for a single illumination source (out of 300). The 30~ms data is truncated to 2.5~ms for improved visualization. The right column shows the deconvolved Green's functions $G_\mathrm{MDD}$ from all positions on the recording surface to a single point on the emitting surface $i^\mathrm{I}$.}
    \label{fig:3Ddisguising_mdd}
\end{figure*}

The multidimensional convolution formalism requires isolating the incoming (incident) component of the wave field on the integration surface. This wave field separation can accurately be performed in the temporal frequency domain \cite{liClosedapertureUnboundedAcoustics2021,muller_acoustic_2023}. Here we adopt a normal-incidence approximation. Under this assumption, the incoming velocity component can be approximated as $v^\mathrm{in} = (p/z_0 - v)/2$ where \(z_0 = \rho\,c\) is the acoustic impedance. Although this ignores oblique arrivals, it provides a useful first-order approximation given our moderately narrow angular aperture.

The validity of this approximation can be evaluated considering the geometry of the recording and emitting surfaces. With an emitting surface radius of \(R^\mathrm{I} = 0.2\,\mathrm{m}\) enclosed within a recording surface of radius \(R^\mathrm{O} = 0.3\,\mathrm{m}\), the maximum incidence angle is $\theta_{\mathrm{max}} = \arctan(R^\mathrm{I} / R^\mathrm{O}) \approx 34^\circ$. This limited angular span ensures that wave incidence angles deviate from normal incidence by at most \(\theta_{\mathrm{max}}\), making the normal-incidence approximation appropriate for this configuration. In the future, a more accurate two-way decomposition could be investigated at the cost of increased computational complexity.

The recorded wave fields shown in Fig. \ref{fig:3Ddisguising_mdd} illustrate only a small portion of the full dataset required for MDD. The complete dataset consists of both \(v_\mathrm{in}\) and \(p\), structured as an array of size \(n_\mathrm{ill} \times n_t \times (n^\mathrm{O} + n^\mathrm{I})\). Consequently, the full data set is bigger than the data visualized in Fig. \ref{fig:3Ddisguising_mdd} by a factor $300 \times (30\,\mathrm{ms}/2.5\,\mathrm{ms}) = 3600$: the figure shows a single illumination of the $n_\mathrm{ill} = 300$, and only the first $2.5\,\mathrm{ms}$ of the $30\,\mathrm{ms}$ record. MDD leverages redundancy across \(n_\mathrm{ill}\) different illuminations to recover Green’s functions of dimension \(n^\mathrm{O} \times n^\mathrm{I} \times n_t\). As evident in the subset of the MDD results shown in Fig. \ref{fig:3Ddisguising_mdd}, deconvolution effectively suppresses boundary reverberations (right column), acting as a cost effective virtual absorbing boundary condition.

\noindent
\begin{figure*}[htbp]
    \centering
    \input{lib/3D_disguising_homogenous_GF.tex}
    \caption{Homogeneous Green's functions $G_\mathrm{H}$ (no scatterer) with $i^\mathrm{O} \in \mathcal{X}^\mathrm{O}$ the index of the point on the recording surface. The impulsive Green's functions $G_\mathrm{imp}$ are calculated using an impulsive source in the FDTD simulation. The MDD Green's functions $G_\mathrm{MDD}$ are calculated using the MDD method on reverberative data recorded in another FDTD simulation.}
    \label{fig:gfs_homogeneous}
\end{figure*}

\noindent
\begin{figure*}[htbp]
    \centering
    \input{lib/3D_disguising_heterogeneous_GF.tex}
    \caption{Heterogeneous Green's functions $G = G_\mathrm{H} + G_\mathrm{S}$ of the spherical scatterer with $i^\mathrm{O} \in \mathcal{X}^\mathrm{O}$ the index of the point on the recording surface. The impulsive Green's functions $G_\mathrm{imp}$ are calculated using an impulsive source in the FDTD simulation. The MDD Green's functions $G_\mathrm{MDD}$ are calculated using the MDD method on reverberative data recorded in another FDTD simulation.}
    \label{fig:gfs_sphere_het}
\end{figure*}

\noindent
\begin{figure*}[htbp]
    \centering
    \input{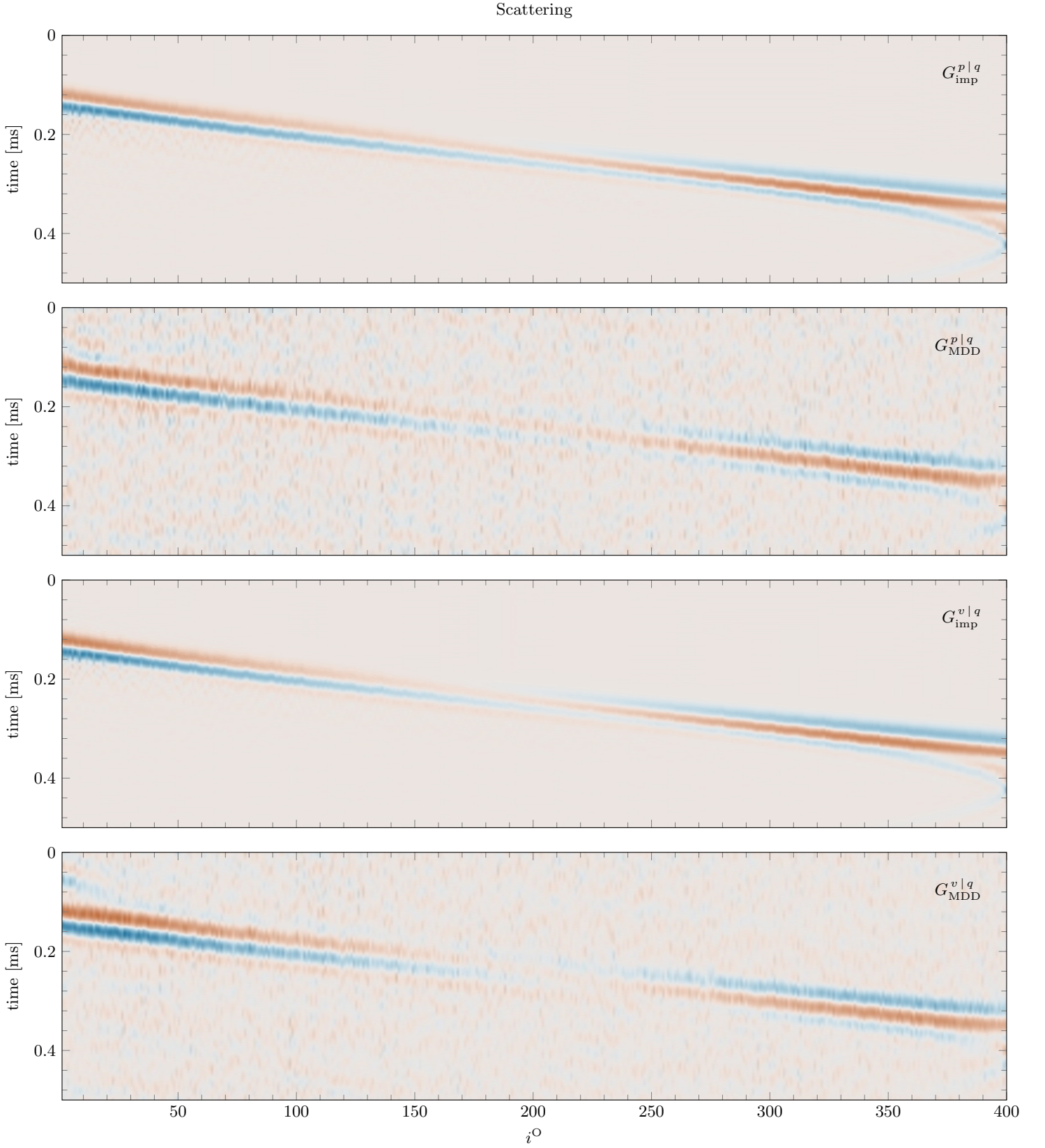}
    \caption{Scattering Green's functions $G_\mathrm{S}$ of the spherical scatterer with $i^\mathrm{O} \in \mathcal{X}^\mathrm{O}$ the index of the point on the recording surface. The scattering Green's functions are calculated by subtracting the homogeneous from the heterogeneous Green's functions $G_\mathrm{S} = G - G_\mathrm{H}$. The impulsive Green's functions $G_\mathrm{imp}$ are calculated using an impulsive source in the FDTD simulation. The MDD Green's functions $G_\mathrm{MDD}$ are calculated using the MDD method on reverberative data recorded in another FDTD simulation.}
    \label{fig:gfs_sphere_scat}
\end{figure*}
\newpage

We observe good agreement between the impulsive Green's functions and those retrieved via MDD (Figs.~\ref{fig:gfs_homogeneous}, \ref{fig:gfs_sphere_het}, \ref{fig:gfs_sphere_scat}). The impulsive Green's functions were computed with a shorter time-step $\mathrm{d}t$ than the illumination data, which is computationally affordable only because the propagation window is kept short enough to keep wall reflections out of the recording. This shorter step preserves higher bandwidth, which explains why the MDD Green's functions appear correspondingly broader in time.
% This behavior is expected given the limited bandwidth of the illumination source.

\begin{figure*}[p]
    \centering
    \vspace{3.2cm}
    \includegraphics[trim=0 0 0 0, clip, width=\textwidth]{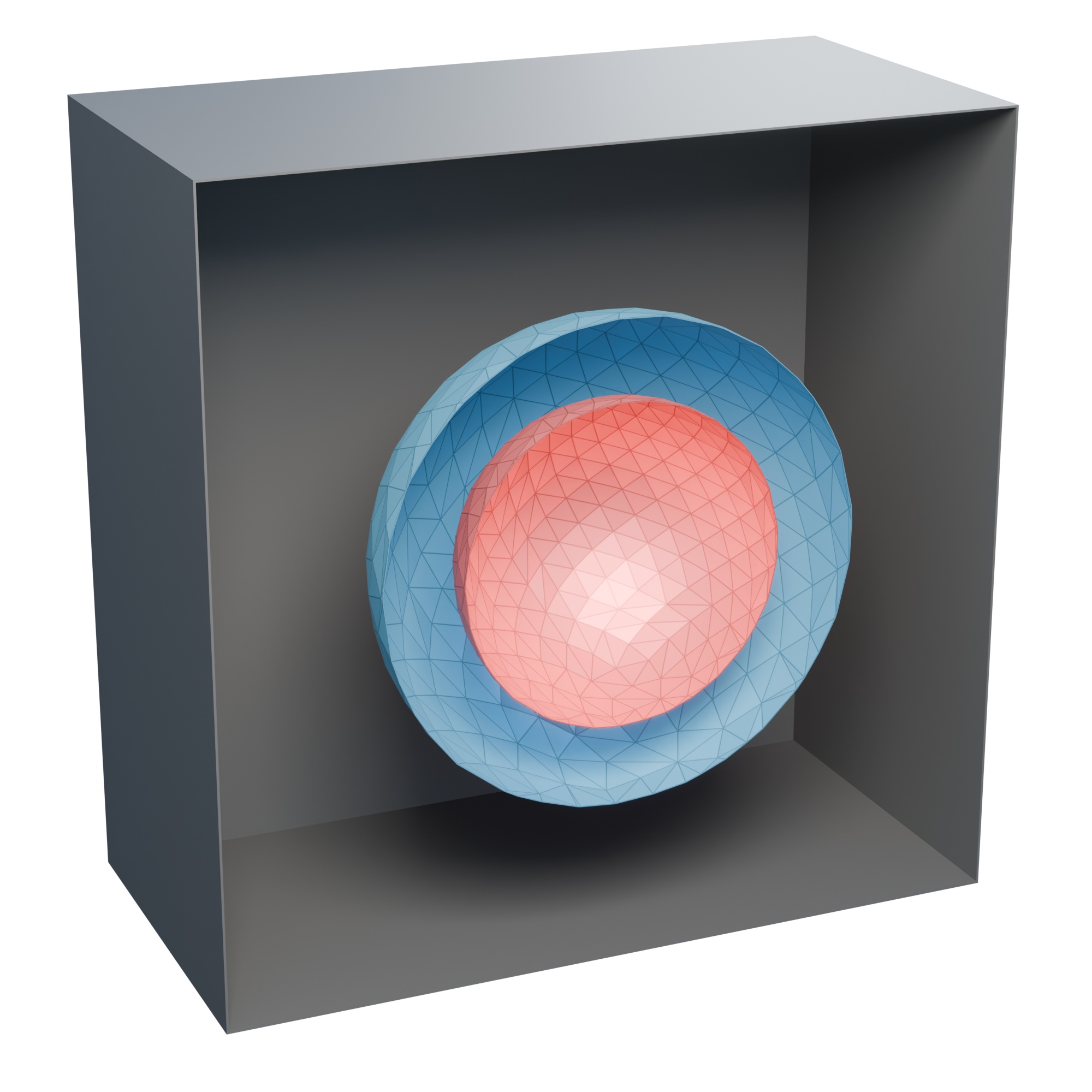}
    \caption{3D setup of the homogeneous configuration, with no interior scatterer. The dark box represents the rigid boundary of the FDTD simulation domain. The two nested Fibonacci-sampled shells correspond to the recording surface $\mathcal{S}^\mathrm{O}$ (outer) and the emitting surface $\mathcal{S}^\mathrm{I}$ (inner); the illumination sphere used for MDD acquisition is omitted from the rendering for clarity. This configuration yields the homogeneous Green's functions $G_\mathrm{H}$ used for cloaking.}
    \label{fig:homogeneous_setup}
\end{figure*}

\begin{figure*}[p]
    \centering
    \vspace{3.2cm}
    \includegraphics[trim=0 0 0 0, clip, width=\textwidth]{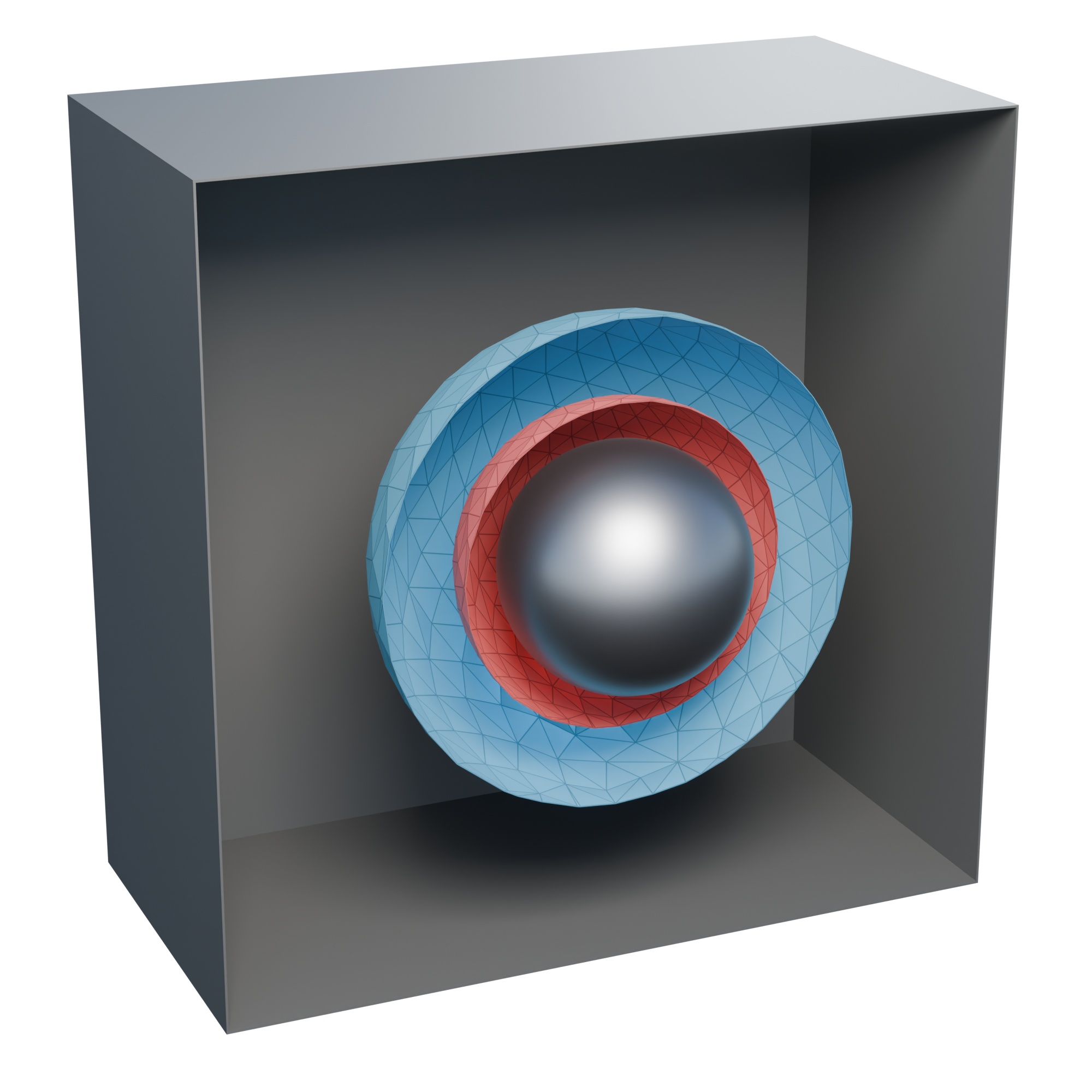}
    \caption{3D setup of the spherical scatterer; surfaces and domain as in Fig.~\ref{fig:homogeneous_setup}.}
    \label{fig:spherical_scatterer_setup}
\end{figure*}

\begin{figure*}[p]
    \centering
    \vspace{3.2cm}
    \includegraphics[trim=0 0 0 0, clip, width=\textwidth]{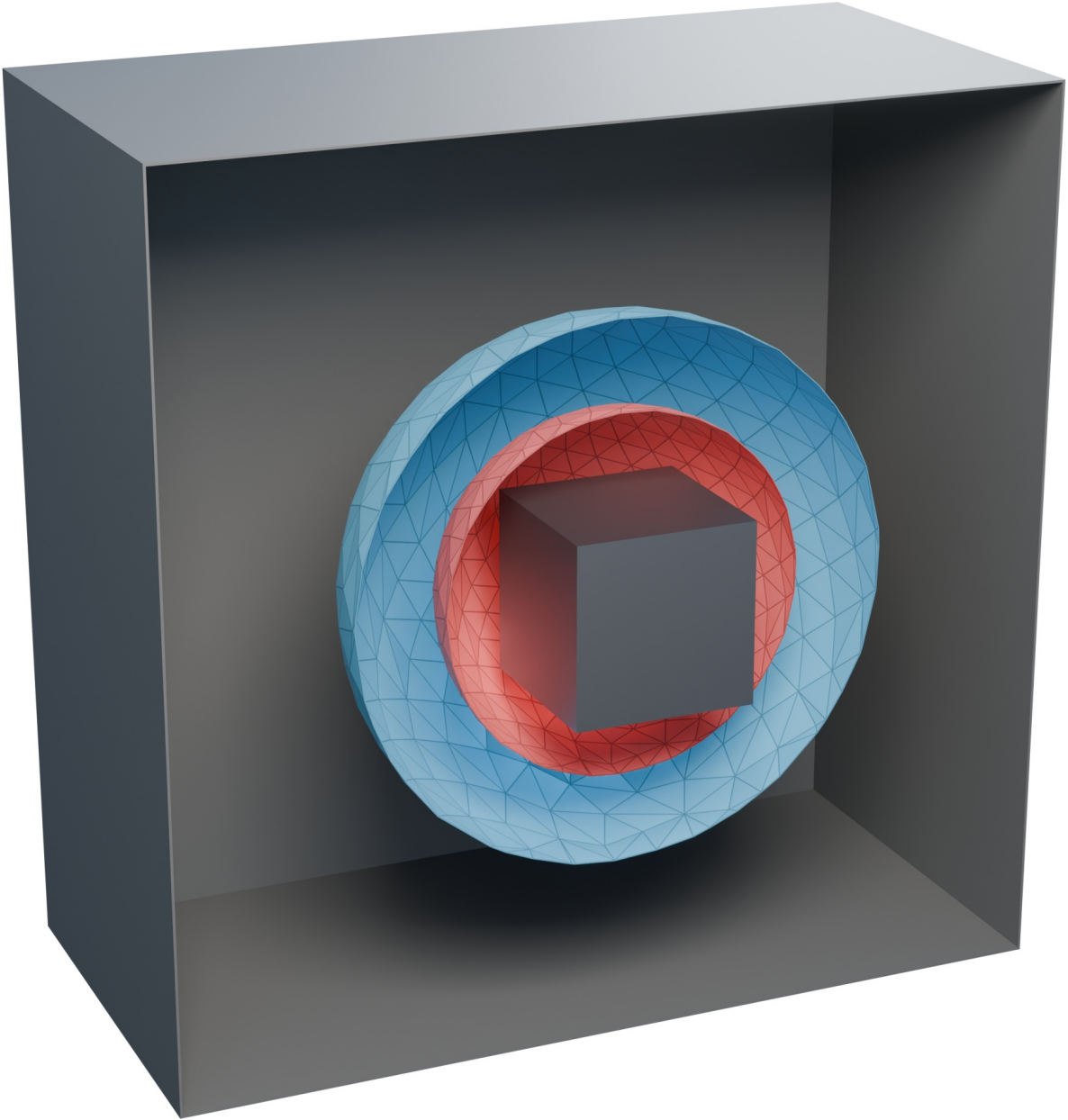}
    \caption{3D setup of the cubic scatterer.}
    \label{fig:cubic_scatterer_setup}
\end{figure*}

\begin{figure*}[p]
    \centering
    \vspace{3.2cm}
    \includegraphics[trim=0 0 0 0, clip, width=\textwidth]{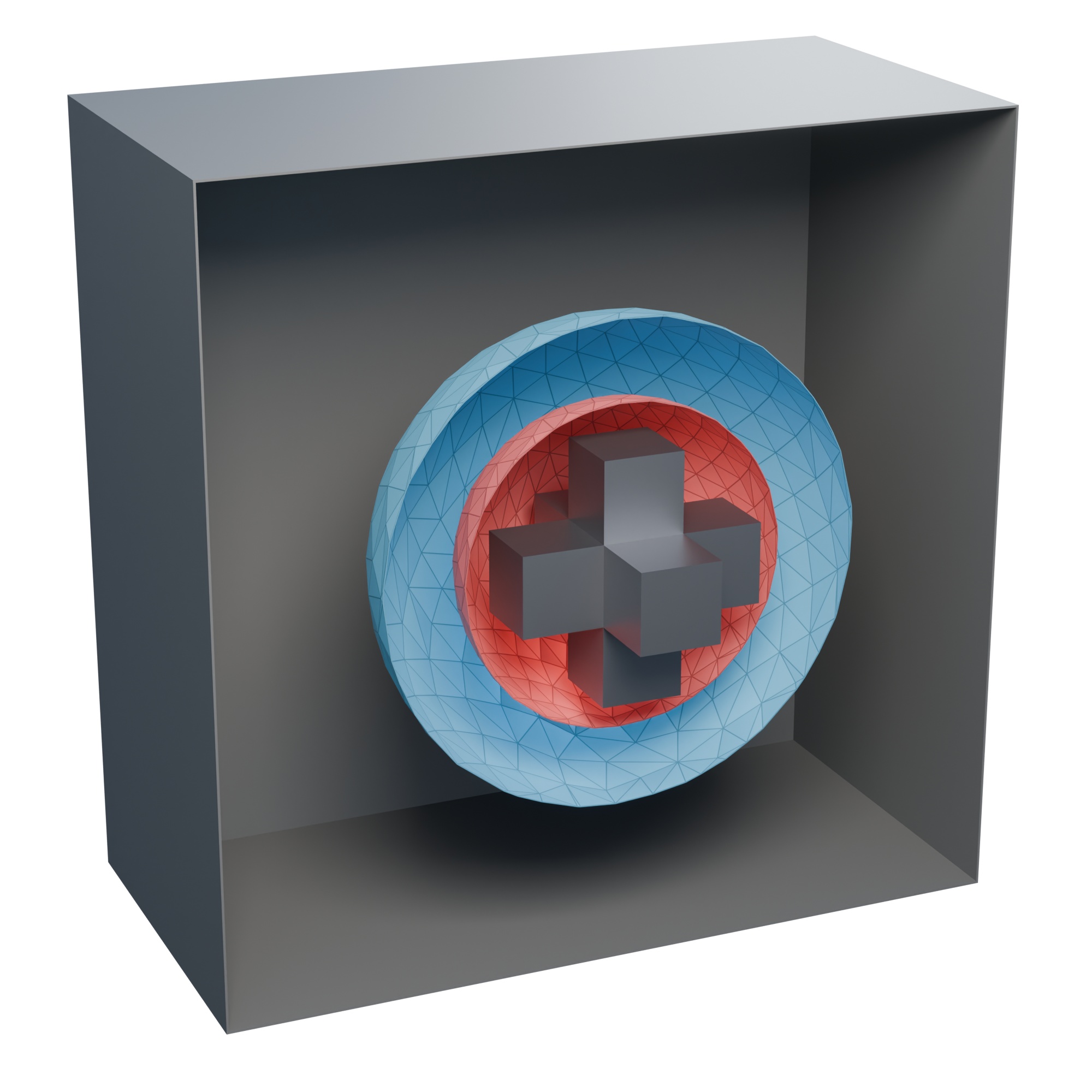}
    \caption{3D setup of the cross scatterer.}
    \label{fig:cross_scatterer_setup}
\end{figure*}

To compute the MDD illumination data, we use a Ricker wavelet with a center frequency of \(f_c = 12\,\mathrm{kHz}\) and record \(t_\mathrm{max} = 30\,\mathrm{ms}\) of data. The reverberant FDTD box is sized to wrap the illumination sphere with a fixed margin: with illumination radius \(r_\mathrm{ill} = 0.45\,\mathrm{m}\) and a four-cell wall clearance, the side length is \(d \approx 0.95\,\mathrm{m}\), as shown in Figs. \ref{fig:homogeneous_setup}, \ref{fig:spherical_scatterer_setup}, \ref{fig:cubic_scatterer_setup}, \ref{fig:cross_scatterer_setup}. Assuming a homogeneous water medium with a sound speed of \(c_0 = 1500\,\mathrm{m/s}\), the geometry gives \(c_0 t_\mathrm{max}/d \approx 47\) round-trip reverberations at the scatterer during the recording window. The reverberant domain is discretized with 19 points per wavelength at \(f_c\) (\(\mathrm{d}x \approx 6.52\,\mathrm{mm}\), \(147^3\) cells), with a Courant factor of \(0.5\). The impulsive Green's functions are computed on a \(301^3\) grid (\(x_\mathrm{max} = 0.85\,\mathrm{m}\), \(\mathrm{d}x \approx 5.67\,\mathrm{mm}\)) with no perfectly matched layer; the propagation window (\(0.70\,\mathrm{ms}\)) ends before the earliest wall reflection reaches the recording surface (\(\approx 0.80\,\mathrm{ms}\) for this geometry), making absorbing boundaries unnecessary. The impulsive-Green's-function source is a Gaussian with $-3\,\mathrm{dB}$ cutoff at \(f_\mathrm{3dB} = 12\,\mathrm{kHz}\), matched to the Ricker bandwidth used for the MDD illumination.

The illumination sources are monopole sources positioned on a sphere of radius \(r_\mathrm{ill} = 0.45\,\mathrm{m}\) with \(n_\mathrm{ill} = 300\) sample points. The source positions are drawn from a Roberts $R_2$ low-discrepancy sequence~\cite{roberts_unreasonable_2018,roberts_evenly_2018} so that increasing $n_\mathrm{ill}$ keeps existing positions fixed and lets the recorded dataset be extended in place. This step is computationally the most expensive due to the large spatiotemporal domain, but it mirrors the constraints of possible future experimental implementations. While an impulsive source is not available experimentally, long-duration recordings are feasible and inexpensive in an experimental setting.

% The relatively high center frequency of the Ricker wavelet increases bandwidth. Although the WaveLab typically operates up to \(f_c = 10\,\mathrm{kHz}\), higher frequencies can still be generated, albeit with potential distortions. Importantly, MDD is robust to the waveform used, as it deconvolves the illumination waveform independently, making it well-suited for experimental application.

Fig. \ref{fig:illumination_data} shows the pressure field recorded on the Fibonacci-sampled recording surface for a single illumination source. Each convex-hull-triangulated face is color-coded by its corresponding pressure value. For \(f_c = 12\,\mathrm{kHz}\), the wavelength is \(\lambda = c_0/f_c = 0.125\,\mathrm{m}\). The $N = 400$ Fibonacci sample points on the recording surface ($R^\mathrm{O} = 0.3\,\mathrm{m}$) have a nearest-neighbour spacing of $50.4\,\mathrm{mm}$ on average and at most $51.7\,\mathrm{mm}$, below $\lambda/2 = 62.5\,\mathrm{mm}$; the discretization is therefore free of spatial aliasing up to $\approx 14.5\,\mathrm{kHz}$, covering the center frequency of the incident wavelet. The emitting surface ($R^\mathrm{I} = 0.2\,\mathrm{m}$) is sampled more finely still.

\begin{figure*}[p]
    \centering
    \vspace*{3.2cm}
    \includegraphics[trim=0 160 0 80, clip, width=\textwidth]{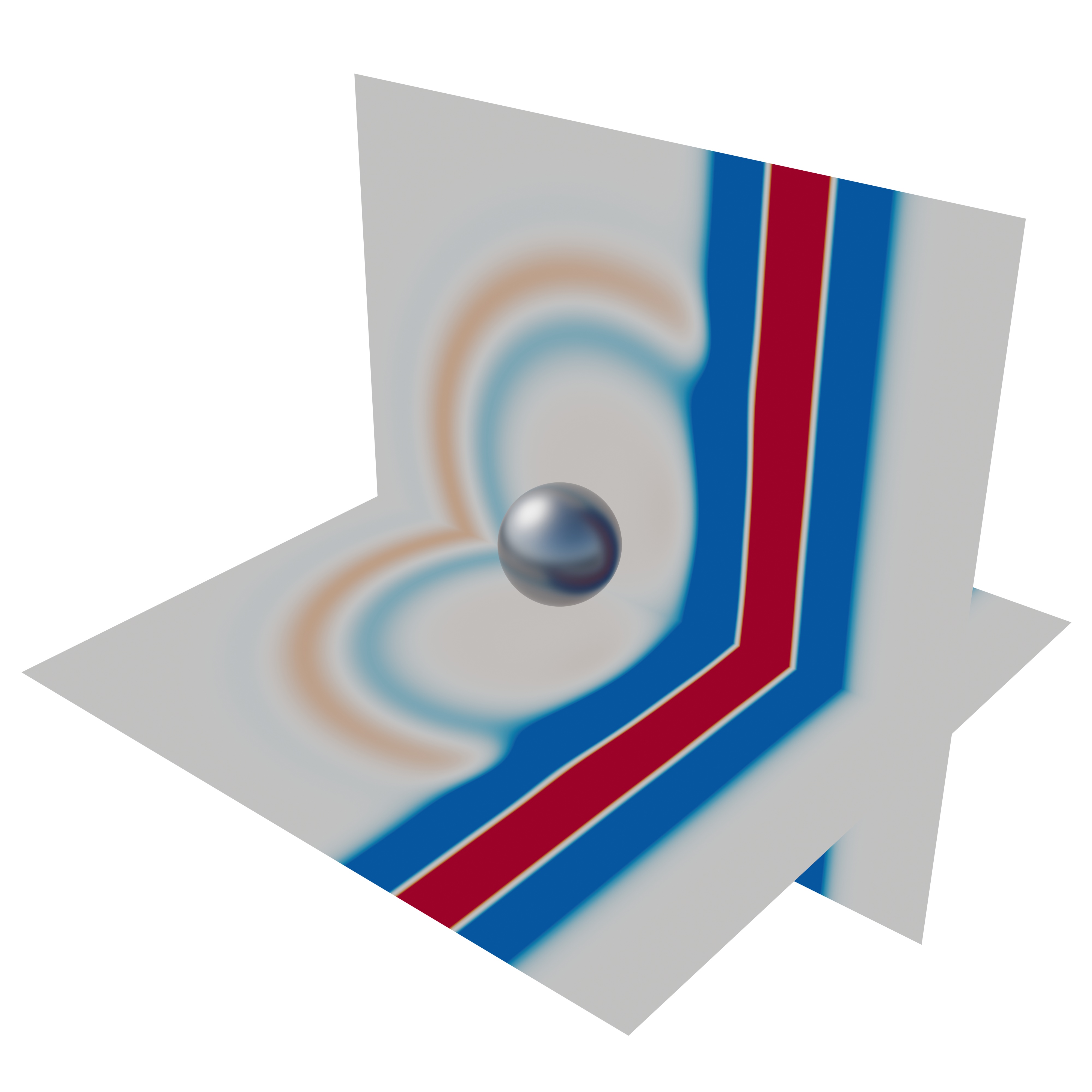}
    \caption{Pressure field of the real spherical scatterer in a homogeneous medium. The spherical object at the center partially backscatters the incident plane Ricker wavelet. The cut planes, positioned 0.3~m from the center, reveal the wave interactions with the scatterer.}
    \label{fig:render_real_het}
\end{figure*}

\begin{figure*}[p]
    \centering
    \vspace*{3.2cm}
    \includegraphics[trim=0 160 0 80, clip, width=1.0\textwidth]{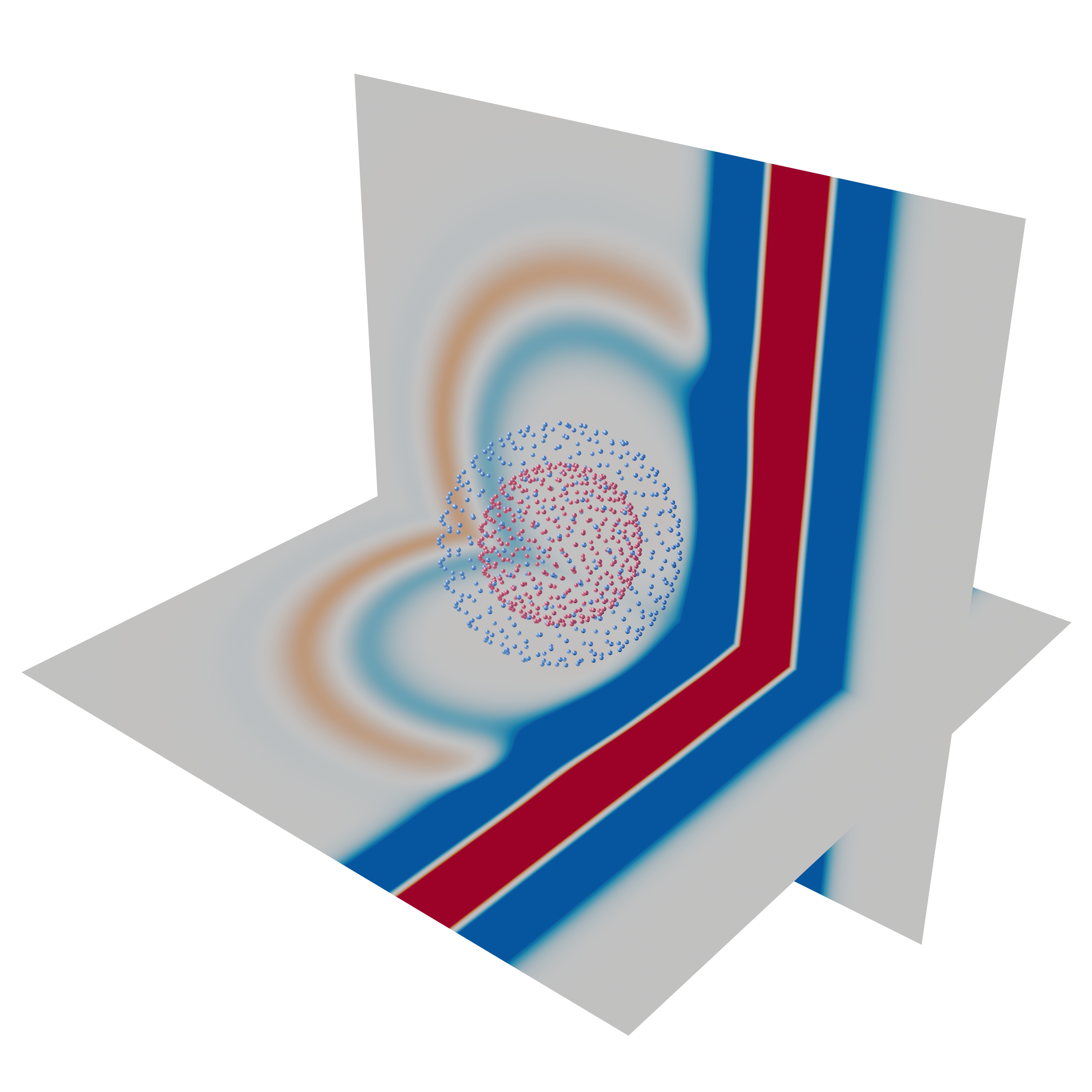}
    \caption{Pressure field of the holographic spherical scatterer in a homogeneous medium. The holography setup is visualized using cones placed on two concentric spherical surfaces. The outer surface represents the recording surface $\mathcal{S}^\mathrm{O}$, enclosing the inner emitting surface $\mathcal{S}^\mathrm{I}$. Each cone points outward, indicating the normal direction of the local surface. The cones are placed at the $N = 400$ Fibonacci sample points of each surface.}
    \label{fig:render_ibc_het}
\end{figure*}

% \begin{figure}[p]
%     \centering
%     \includegraphics[trim=0 160 0 80, clip, width=0.95\textwidth]{fig/ibc_het.jpg}
%     \caption{Pressure field on two cut planes through the 3D plane wave scattered by a spherical acoustic hologram using immersive boundary conditions. The cones indicate the discrete sampling positions and directions of the recording surface $\mathcal{X}^\mathrm{O}$ and emitting surface $\mathcal{X}^\mathrm{I}$.}
%     \label{fig:3D_ibc_het}
% \end{figure}

\begin{figure*}[p]
    \centering
    \vspace*{3.2cm}
    \includegraphics[trim=0 160 0 80, clip, width=\textwidth]{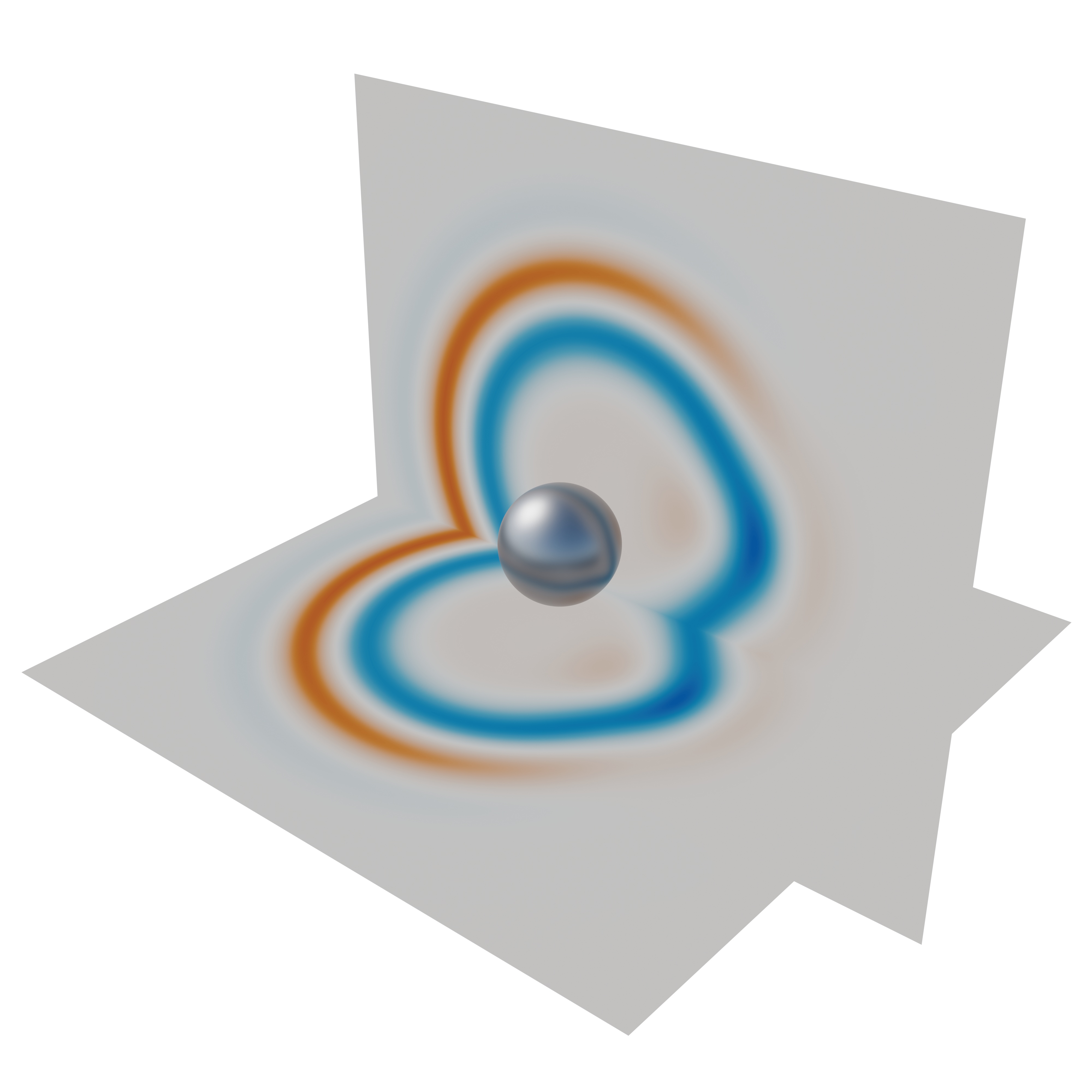}
    \caption{Scattered pressure field of the real spherical scatterer, obtained by subtracting the pressure field in the homogeneous case (without scatterer) from that in the heterogeneous case (with scatterer).}
    \label{fig:render_real_scat}
\end{figure*}

\begin{figure*}[p]
    \centering
    \vspace*{3.2cm}
    \includegraphics[trim=0 160 0 80, clip, width=1.0\textwidth]{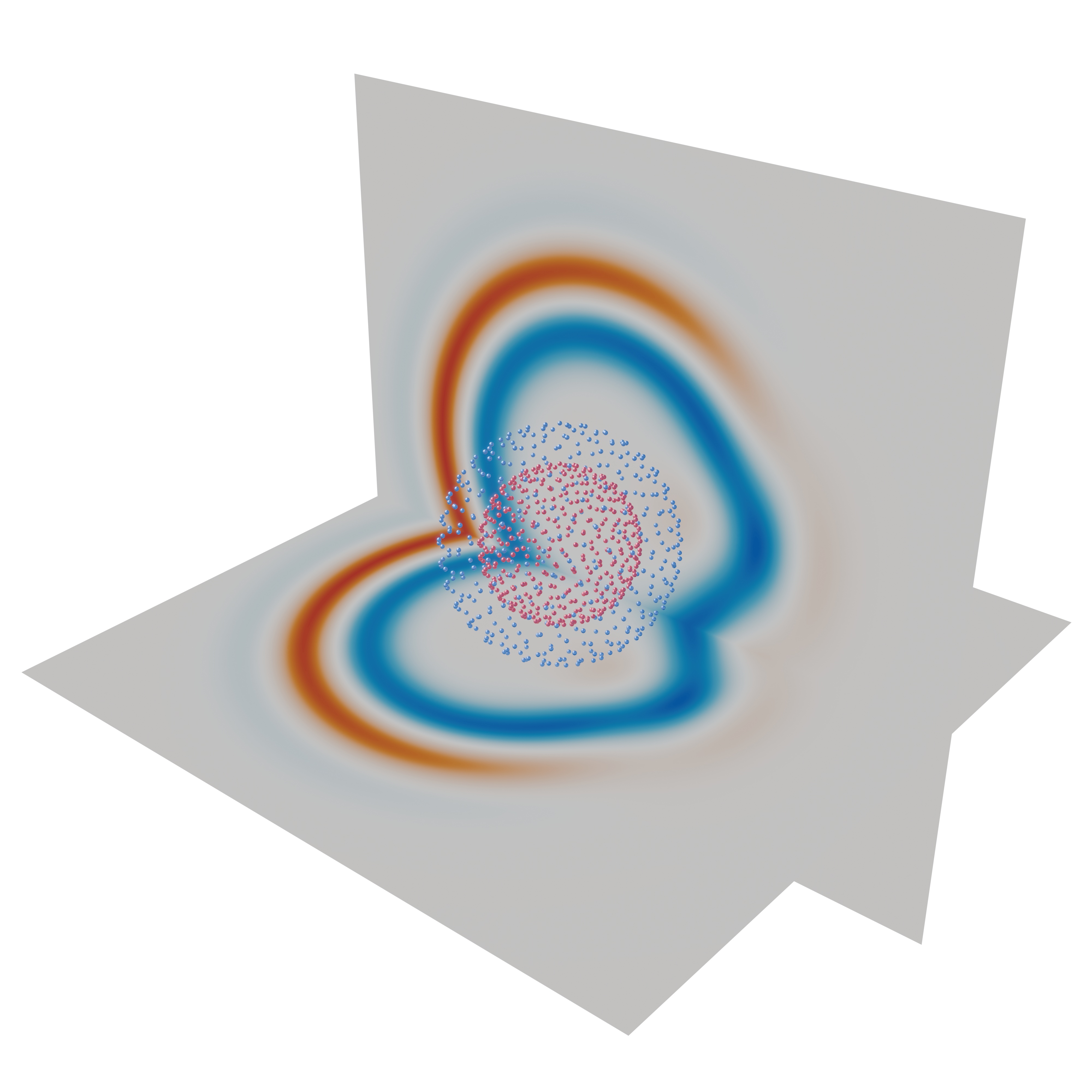}
    \caption{Scattered pressure field of the holographic spherical scatterer, obtained by subtracting the pressure field in the homogeneous case (without scatterer) from that in the heterogeneous case (with scatterer).}
    \label{fig:render_ibc_scat}
\end{figure*}

\subsection{Boundary-extrapolation write count}
\label{app:extrapcost}

The discrete Kirchhoff convolution that extrapolates the recorded outer-surface field to the inner emitting surface, Eq.~\ref{eq:extrap_p_discret}, admits two evaluation schemes. A \emph{scatter} forward-extrapolates each recorded boundary sample into the $K$ future output times reached by its kernel, performing a read--modify--write per kernel tap: $\mathcal{O}(K)$ writes per extrapolated sample. A \emph{gather} instead accumulates, for each output sample, the full sum over source points and kernel taps in a register and commits it with a single store: $\mathcal{O}(1)$ writes per sample. Earlier immersive-boundary implementations used the scatter form~\cite{van_manen_exact_2007}; we use the gather form. A numerical test confirms the two schemes compute the identical convolution to round-off and, by routing each array through an access-counting wrapper, verifies the write counts over sweeps in $K$, $N$, and $n_t$: the gather write count is $n_t N$, flat in $K$, while the scatter write count grows linearly in $K$.

The saving is specifically in writes---memory \emph{loads} are comparable between the schemes (within a factor $\sim 1.1$--$1.5$, both dominated by kernel and boundary-history reads). Its impact is largest under parallel execution: because a scatter has multiple threads writing the same future output slot, a parallel scatter requires atomic accumulation, whereas the gather assigns one output sample per thread and is collision-free. On the multithreaded CPU on which the simulations are run, avoiding this atomic contention is the dominant reason for the gather form.

\subsection{Additional Results}
\label{app:resultsAdditional}

All pressure-field visualizations in this work use the perceptually-uniform Crameri \emph{vik} diverging colormap~\cite{crameri_scientific_2023}, mapped to a fixed symmetric range centered on zero that is shared across all panels of a figure---and across comparable figures---so panels can be compared directly. The scattered-intensity renderings (Figs.~\ref{fig:scattering3Dreal}--\ref{fig:scattering3Dmdd}) instead use the sequential Crameri \emph{batlow} colormap: the time-integrated intensity [Eq.~\ref{eq:radial_intensity}] is non-negative, so a zero-centered diverging map would misleadingly imply a sign change. Each panel is normalized to its own peak intensity, so these renderings convey the angular shape of the scattering pattern rather than absolute magnitude.

We observe that holograms based on impulsive Green’s functions more accurately replicate the target field than those using MDD-derived functions. This discrepancy is likely due to two effects: (i) the slightly narrower bandwidth of the MDD-derived Green’s functions, as also noted in Fig.~\ref{fig:gfs_homogeneous}, \ref{fig:gfs_sphere_het}, \ref{fig:gfs_sphere_scat}, and (ii) the normal-incidence approximation used in the wave-field separation step of MDD, which mis-projects the oblique-incidence components within the angular aperture $\theta_\mathrm{max} \approx 34^\circ$ of our geometry.

This behavior is quantified by the angular scattering patterns in Fig.~\ref{fig:results_radialIntensity}, with the corresponding full three-dimensional scattering directivities rendered in Figs.~\ref{fig:scattering3Dreal}--\ref{fig:scattering3Dmdd}. The raw time-integrated peak intensities $I_\mathrm{peak}$ that normalize each curve are listed in Table~\ref{tab:scattering_peaks}. Before normalization, the impulsive-Green's-function holograms reproduce the absolute peak scattered intensity of the real scatterer to within $20\%$ (peak-intensity ratios of $1.2$, $1.1$, and $0.9$ for the sphere, cube, and cross, respectively). The MDD-retrieved Green's functions, by contrast, are recovered only up to an empirical global scale---visible as the roughly threefold-larger MDD column in Table~\ref{tab:scattering_peaks}---so their absolute intensity carries no physical meaning; each curve in Fig.~\ref{fig:results_radialIntensity} is therefore normalized to its own peak. The angular shape, not the absolute magnitude, is what reflects MDD fidelity to the underlying scattering physics.

The MDD cube shows the largest angular distortion of the three scatterers, consistent with its faceted geometry producing the most strongly directional scattering---narrow specular lobes from flat faces---which the simplified normal-incidence wave-field separation reproduces less faithfully than the smoother angular patterns of the sphere and cross. Concretely, the real cube's forward and back lobes are nearly equal ($I(180^\circ)/I(0^\circ) \approx 0.9$), and under the MDD reconstruction this ratio inverts ($\approx 1.7$), making the back lobe dominant. The sphere and cross have much larger forward-to-back contrasts ($\approx 0.2$ and $\approx 0.6$, respectively), so their dominant lobes are robust to comparable reconstruction errors.

\begin{table}[htbp]
    \centering
    \caption{Raw time-integrated peak scattered intensity $I_\mathrm{peak}$, in units of $10^{-6}\,\mathrm{Pa}^2\,\mathrm{s}$, used as the per-panel normalization constant for each panel of Fig.~\ref{fig:results_radialIntensity}. The \emph{Real} and \emph{Impulsive} columns agree to within $20\%$, whereas the \emph{MDD} column is larger by a roughly constant factor of $\sim\!3$ (see text).}
    \label{tab:scattering_peaks}
    \sisetup{table-format=2.2, round-mode=places, round-precision=2,
        exponent-mode=fixed, fixed-exponent=-6, drop-exponent=true}
    \begin{tabular}{l S S S}
        \toprule
               & {Real}            & {Impulsive}           & {MDD}           \\
        \midrule
        Sphere & \IpeakSphereTruth & \IpeakSphereImpulsive & \IpeakSphereMdd \\
        Cube   & \IpeakCubeTruth   & \IpeakCubeImpulsive   & \IpeakCubeMdd   \\
        Cross  & \IpeakCrossTruth  & \IpeakCrossImpulsive  & \IpeakCrossMdd  \\
        \bottomrule
    \end{tabular}
\end{table}

In Fig.~\ref{fig:render_real_scat} and Fig.~\ref{fig:render_ibc_scat}, we show the scattered pressure field of the real and holographic scatterers, respectively. The scattered field is obtained by subtracting the homogeneous case (no scatterer) from the heterogeneous case (with scatterer). Good agreement is observed between the real and holographic scatterer, indicating that the holographic scatterer is indistinguishable from the real scatterer.

The radial-intensity patterns above are time-integrated and so cannot reveal agreement or disagreement at specific instants. For a more detailed comparison, we examine pressure-field slices at two time instances and on two orthogonal planes. Fig.~\ref{fig:3DdisguisingResults_z_t55_het} shows the heterogeneous field on the xy-slice perpendicular to the incident plane wave, at a time when the incident wave is interacting with the scatterer. The same slice $150\,\mu s$ later (Fig.~\ref{fig:3DdisguisingResults_z_t70_het}), after the wave has passed, shows that the wave field originally suppressed inside $\mathcal{S}^\mathrm{I}$ has been fully reconstructed outside it. Subtracting the homogeneous field from this heterogeneous field isolates the scattered component (Fig.~\ref{fig:3DdisguisingResults_z_t70_scat}), which highlights the differences between the real and holographic scatterers in terms of scattering fidelity. Finally, Fig.~\ref{fig:3DdisguisingResults_x_t70_het} shows the orthogonal yz-slice of the heterogeneous field at the same time instance, providing a complementary view of the reconstruction. Each figure includes results for the homogeneous medium, a real scatterer, and two holographic reconstructions: one driven by Green's functions from impulsive FDTD simulations and one driven by Green's functions retrieved via MDD.

Holographic scatterers consistently suppress the wave field within the emitting surface $\mathcal{S}^\mathrm{I}$, which is expected because the employed heterogeneous Green’s functions include both direct homogeneous $G_\mathrm{H}$ and scattered components $G_\mathrm{S_N}$. Using only the scattering Green’s functions $G_\mathrm{S_N}$ would leave the field inside $\mathcal{S}^\mathrm{I}$ unchanged while reconstructing only the scattered wave field outside of $\mathcal{S}^\mathrm{I}$.

\begin{figure*}
    \input{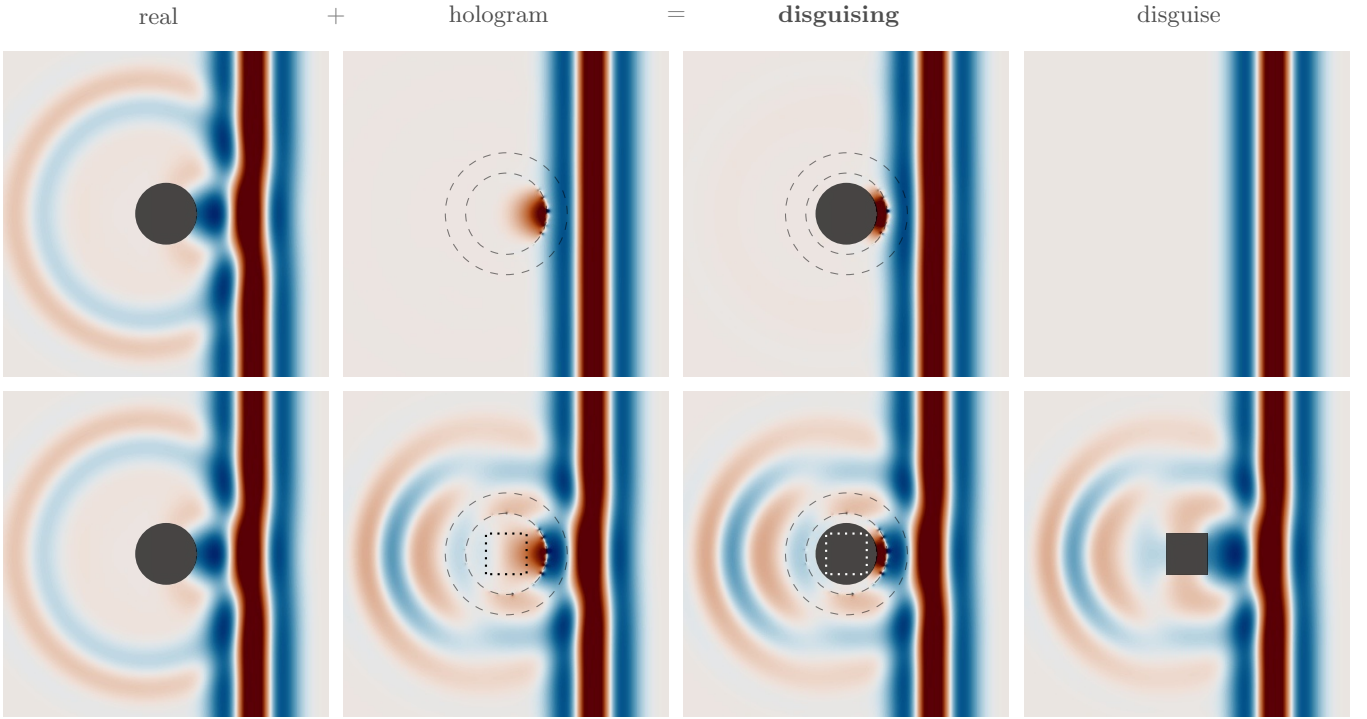}
    \caption{Time delay of 150~$\mu s$ with respect to Fig.~\ref{fig:3DdisguisingResults_disguising_t55} in the main text. The wave field that was suppressed within the emitting surface $\mathcal{S}^\mathrm{I}$ is fully reconstructed outside the emitting surface $\mathcal{S}^\mathrm{I}$, confirming that the IBC-based cloaking and disguising act only inside the enclosed volume.}
    \label{fig:3DdisguisingResults_disguising_t70}
\end{figure*}

\def\fpCondition{het}
\def\fpFileSlug{t5.5_zSlice}
\def\fpCaption{Recorded pressure field for a Ricker plane wave with $f_c=12$~kHz. The columns show the real field and the holographic field. The latter is shown for impulsive Green's functions as well as the Green's functions derived from scattering data using multi-dimensional deconvolution (MDD). The rows show the different scatterers: homogeneous, sphere, cube, and cross. The dashed circles indicate the recording surface and emitting surface with radii of 0.2m and 0.3m, respectively. The plots show an xy-slice at $z=0$ with a domain size of 1.6m x 1.6m.}
\def\fpLabel{fig:3DdisguisingResults_z_t55_het}
% Parametrized 4x3 field-plots grid template.
% Used by lib/disguisingFieldPlots[XZ]_t*_*.tex.
%
% Caller defines these via \def before \input{}-ing this file:
%   \fpCondition  -- "het" or "scat" (subdirectory under each row)
%   \fpFileSlug   -- filename slug after "_slice0.0_", e.g. "t7.0_zSlice"
%   \fpCaption    -- caption text
%   \fpLabel      -- label string (full, e.g. "fig:3DdisguisingResults_z_t70_het")
%
% Rows: hom (label "homogeneous"), sphere, square (label "cube"), cross.
% Cols: real, ref ("impulsive"), mdd ("MDD").
\begin{figure*}[p]
    \centering
    \vspace{2.0cm}
    \begin{tikzpicture}[]

        \coordinate (o) at (0,0);

        % column labels
        \node[textNode] at ([xshift=-1\aTemp, yshift=2.1\aTemp+\yOffset] o) {real};
        \node[textNode] at ([xshift=0\aTemp, yshift=2.1\aTemp+\yOffset] o) {impulsive};
        \node[textNode] at ([xshift=1\aTemp, yshift=2.1\aTemp+\yOffset] o) {MDD};

        % row labels
        \node[textNode,rotate=90] at ([xshift=-1.6\aTemp, yshift=1.5\aTemp+\yOffset] o) {homogeneous};
        \node[textNode,rotate=90] at ([xshift=-1.6\aTemp, yshift=0.5\aTemp+\yOffset] o) {sphere};
        \node[textNode,rotate=90] at ([xshift=-1.6\aTemp, yshift=-0.5\aTemp+\yOffset] o) {cube};
        \node[textNode,rotate=90] at ([xshift=-1.6\aTemp, yshift=-1.5\aTemp+\yOffset] o) {cross};

        % --- ROW 1: homogeneous ---
        \coordinate (a) at ($(o)+ (-1.0\aTemp+0cm, 1.5\aTemp+\yOffset)$);
        \coordinate (b) at ($(a)+ (0*\aTemp,0)$);
        \begin{axis}[resultsAxis, at=(b)]
            \addplot [forget plot] graphics [xmin=\xmin, xmax=\xmax, ymin=\ymin, ymax=\ymax]
                {"fig/sim/hom/\fpCondition/real_slice0.0_\fpFileSlug.jpg"};
            \draw[dashed,opacity=0.5] (0,0) circle [radius=0.2];
            \draw[dashed,opacity=0.5] (0,0) circle [radius=0.3];
        \end{axis}
        \coordinate (b) at ($(a)+ (1*\aTemp,0)$);
        \begin{axis}[resultsAxis, at=(b)]
            \addplot [forget plot] graphics [xmin=\xmin, xmax=\xmax, ymin=\ymin, ymax=\ymax]
                {"fig/sim/hom/\fpCondition/ref_slice0.0_\fpFileSlug.jpg"};
            \draw[dashed,opacity=0.5] (0,0) circle [radius=0.2];
            \draw[dashed,opacity=0.5] (0,0) circle [radius=0.3];
        \end{axis}
        \coordinate (b) at ($(a)+ (2*\aTemp,0)$);
        \begin{axis}[resultsAxis, at=(b)]
            \addplot [forget plot] graphics [xmin=\xmin, xmax=\xmax, ymin=\ymin, ymax=\ymax]
                {"fig/sim/hom/\fpCondition/mdd_slice0.0_\fpFileSlug.jpg"};
            \draw[dashed,opacity=0.5] (0,0) circle [radius=0.2];
            \draw[dashed,opacity=0.5] (0,0) circle [radius=0.3];
        \end{axis}

        % --- ROW 2: sphere ---
        \coordinate (a) at ($(o)+ (-1.0\aTemp+0.0cm, 0.5\aTemp+\yOffset)$);
        \coordinate (b) at ($(a)+ (0*\aTemp,0)$);
        \begin{axis}[resultsAxis, at=(b)]
            \addplot [forget plot] graphics [xmin=\xmin, xmax=\xmax, ymin=\ymin, ymax=\ymax]
                {"fig/sim/sphere/\fpCondition/real_slice0.0_\fpFileSlug.jpg"};
            \fill[opacity=0.7] (0,0) circle [radius=0.1522];
            \draw[dashed,opacity=0.5] (0,0) circle [radius=0.2];
            \draw[dashed,opacity=0.5] (0,0) circle [radius=0.3];
        \end{axis}
        \coordinate (b) at ($(a)+ (1*\aTemp,0)$);
        \begin{axis}[resultsAxis, at=(b)]
            \addplot [forget plot] graphics [xmin=\xmin, xmax=\xmax, ymin=\ymin, ymax=\ymax]
                {"fig/sim/sphere/\fpCondition/ref_slice0.0_\fpFileSlug.jpg"};
            \draw[opacity=1.0,dotted, thick] (0,0) circle [radius=0.15];
            \draw[dashed,opacity=0.5] (0,0) circle [radius=0.2];
            \draw[dashed,opacity=0.5] (0,0) circle [radius=0.3];
        \end{axis}
        \coordinate (b) at ($(a)+ (2*\aTemp,0)$);
        \begin{axis}[resultsAxis, at=(b)]
            \addplot [forget plot] graphics [xmin=\xmin, xmax=\xmax, ymin=\ymin, ymax=\ymax]
                {"fig/sim/sphere/\fpCondition/mdd_slice0.0_\fpFileSlug.jpg"};
            \draw[opacity=1.0,dotted, thick] (0,0) circle [radius=0.15];
            \draw[dashed,opacity=0.5] (0,0) circle [radius=0.2];
            \draw[dashed,opacity=0.5] (0,0) circle [radius=0.3];
        \end{axis}

        % --- ROW 3: square (label "cube") ---
        \coordinate (a) at ($(o)+ (-1.0\aTemp+0.0cm, -0.5\aTemp+\yOffset)$);
        \coordinate (b) at ($(a)+ (0*\aTemp,0)$);
        \begin{axis}[resultsAxis, at=(b)]
            \addplot [forget plot] graphics [xmin=\xmin, xmax=\xmax, ymin=\ymin, ymax=\ymax]
                {"fig/sim/square/\fpCondition/real_slice0.0_\fpFileSlug.jpg"};
            \fill[opacity=0.7,] (-0.1022,-0.1022) rectangle (0.1022,0.1022);
            \draw[dashed,opacity=0.5] (0,0) circle [radius=0.2];
            \draw[dashed,opacity=0.5] (0,0) circle [radius=0.3];
        \end{axis}
        \coordinate (b) at ($(a)+ (1*\aTemp,0)$);
        \begin{axis}[resultsAxis, at=(b)]
            \addplot [forget plot] graphics [xmin=\xmin, xmax=\xmax, ymin=\ymin, ymax=\ymax]
                {"fig/sim/square/\fpCondition/ref_slice0.0_\fpFileSlug.jpg"};
            \draw[opacity=1.0,dotted, thick] (-0.1,-0.1) rectangle (0.1,0.1);
            \draw[dashed,opacity=0.5] (0,0) circle [radius=0.2];
            \draw[dashed,opacity=0.5] (0,0) circle [radius=0.3];
        \end{axis}
        \coordinate (b) at ($(a)+ (2*\aTemp,0)$);
        \begin{axis}[resultsAxis, at=(b)]
            \addplot [forget plot] graphics [xmin=\xmin, xmax=\xmax, ymin=\ymin, ymax=\ymax]
                {"fig/sim/square/\fpCondition/mdd_slice0.0_\fpFileSlug.jpg"};
            \draw[opacity=1.0,dotted, thick] (-0.1,-0.1) rectangle (0.1,0.1);
            \draw[dashed,opacity=0.5] (0,0) circle [radius=0.2];
            \draw[dashed,opacity=0.5] (0,0) circle [radius=0.3];
        \end{axis}

        % --- ROW 4: cross ---
        \coordinate (a) at ($(o)+ (-1.0\aTemp+0.0cm, -1.5\aTemp+\yOffset)$);
        \coordinate (b) at ($(a)+ (0*\aTemp,0)$);
        \begin{axis}[resultsAxis, at=(b)]
            \addplot [forget plot] graphics [xmin=\xmin, xmax=\xmax, ymin=\ymin, ymax=\ymax]
                {"fig/sim/cross/\fpCondition/real_slice0.0_\fpFileSlug.jpg"};
            \fill[opacity=0.7]
            (0.1522,0.0522) -- (0.1522,-0.0522) -- (0.0522,-0.0522) -- (0.0522,-0.1522) -- (-0.0522,-0.1522) -- (-0.0522,-0.0522) -- (-0.1522,-0.0522) -- (-0.1522,0.0522) -- (-0.0522,0.0522) -- (-0.0522,0.1522) -- (0.0522,0.1522) -- (0.0522,0.0522) -- cycle;
            \draw[dashed,opacity=0.5] (0,0) circle [radius=0.2];
            \draw[dashed,opacity=0.5] (0,0) circle [radius=0.3];
        \end{axis}
        \coordinate (b) at ($(a)+ (1*\aTemp,0)$);
        \begin{axis}[resultsAxis, at=(b)]
            \addplot [forget plot] graphics [xmin=\xmin, xmax=\xmax, ymin=\ymin, ymax=\ymax]
                {"fig/sim/cross/\fpCondition/ref_slice0.0_\fpFileSlug.jpg"};
            \draw[opacity=1.0, dotted, thick]
            (0.15,0.05) -- (0.15,-0.05) -- (0.05,-0.05) -- (0.05,-0.15) -- (-0.05,-0.15) -- (-0.05,-0.05) -- (-0.15,-0.05) -- (-0.15,0.05) -- (-0.05,0.05) -- (-0.05,0.15) -- (0.05,0.15) -- (0.05,0.05) -- cycle;
            \draw[dashed,opacity=0.5] (0,0) circle [radius=0.2];
            \draw[dashed,opacity=0.5] (0,0) circle [radius=0.3];
        \end{axis}
        \coordinate (b) at ($(a)+ (2*\aTemp,0)$);
        \begin{axis}[resultsAxis, at=(b)]
            \addplot [forget plot] graphics [xmin=\xmin, xmax=\xmax, ymin=\ymin, ymax=\ymax]
                {"fig/sim/cross/\fpCondition/mdd_slice0.0_\fpFileSlug.jpg"};
            \draw[opacity=1.0, dotted, thick]
            (0.15,0.05) -- (0.15,-0.05) -- (0.05,-0.05) -- (0.05,-0.15) -- (-0.05,-0.15) -- (-0.05,-0.05) -- (-0.15,-0.05) -- (-0.15,0.05) -- (-0.05,0.05) -- (-0.05,0.15) -- (0.05,0.15) -- (0.05,0.05) -- cycle;
            \draw[dashed,opacity=0.5] (0,0) circle [radius=0.2];
            \draw[dashed,opacity=0.5] (0,0) circle [radius=0.3];
        \end{axis}

    \end{tikzpicture}

    \caption{\fpCaption}
    \label{\fpLabel}

\end{figure*}

\def\fpCondition{het}
\def\fpFileSlug{t7.0_zSlice}
\def\fpCaption{The plots show the heterogeneous field as an xy-slice at $z=0$ with a domain size of 1.6m x 1.6m. The time instance is 150~$\mu s$ after the above plots.}
\def\fpLabel{fig:3DdisguisingResults_z_t70_het}
% Parametrized 4x3 field-plots grid template.
% Used by lib/disguisingFieldPlots[XZ]_t*_*.tex.
%
% Caller defines these via \def before \input{}-ing this file:
%   \fpCondition  -- "het" or "scat" (subdirectory under each row)
%   \fpFileSlug   -- filename slug after "_slice0.0_", e.g. "t7.0_zSlice"
%   \fpCaption    -- caption text
%   \fpLabel      -- label string (full, e.g. "fig:3DdisguisingResults_z_t70_het")
%
% Rows: hom (label "homogeneous"), sphere, square (label "cube"), cross.
% Cols: real, ref ("impulsive"), mdd ("MDD").
\begin{figure*}[p]
    \centering
    \vspace{2.0cm}
    \begin{tikzpicture}[]

        \coordinate (o) at (0,0);

        % column labels
        \node[textNode] at ([xshift=-1\aTemp, yshift=2.1\aTemp+\yOffset] o) {real};
        \node[textNode] at ([xshift=0\aTemp, yshift=2.1\aTemp+\yOffset] o) {impulsive};
        \node[textNode] at ([xshift=1\aTemp, yshift=2.1\aTemp+\yOffset] o) {MDD};

        % row labels
        \node[textNode,rotate=90] at ([xshift=-1.6\aTemp, yshift=1.5\aTemp+\yOffset] o) {homogeneous};
        \node[textNode,rotate=90] at ([xshift=-1.6\aTemp, yshift=0.5\aTemp+\yOffset] o) {sphere};
        \node[textNode,rotate=90] at ([xshift=-1.6\aTemp, yshift=-0.5\aTemp+\yOffset] o) {cube};
        \node[textNode,rotate=90] at ([xshift=-1.6\aTemp, yshift=-1.5\aTemp+\yOffset] o) {cross};

        % --- ROW 1: homogeneous ---
        \coordinate (a) at ($(o)+ (-1.0\aTemp+0cm, 1.5\aTemp+\yOffset)$);
        \coordinate (b) at ($(a)+ (0*\aTemp,0)$);
        \begin{axis}[resultsAxis, at=(b)]
            \addplot [forget plot] graphics [xmin=\xmin, xmax=\xmax, ymin=\ymin, ymax=\ymax]
                {"fig/sim/hom/\fpCondition/real_slice0.0_\fpFileSlug.jpg"};
            \draw[dashed,opacity=0.5] (0,0) circle [radius=0.2];
            \draw[dashed,opacity=0.5] (0,0) circle [radius=0.3];
        \end{axis}
        \coordinate (b) at ($(a)+ (1*\aTemp,0)$);
        \begin{axis}[resultsAxis, at=(b)]
            \addplot [forget plot] graphics [xmin=\xmin, xmax=\xmax, ymin=\ymin, ymax=\ymax]
                {"fig/sim/hom/\fpCondition/ref_slice0.0_\fpFileSlug.jpg"};
            \draw[dashed,opacity=0.5] (0,0) circle [radius=0.2];
            \draw[dashed,opacity=0.5] (0,0) circle [radius=0.3];
        \end{axis}
        \coordinate (b) at ($(a)+ (2*\aTemp,0)$);
        \begin{axis}[resultsAxis, at=(b)]
            \addplot [forget plot] graphics [xmin=\xmin, xmax=\xmax, ymin=\ymin, ymax=\ymax]
                {"fig/sim/hom/\fpCondition/mdd_slice0.0_\fpFileSlug.jpg"};
            \draw[dashed,opacity=0.5] (0,0) circle [radius=0.2];
            \draw[dashed,opacity=0.5] (0,0) circle [radius=0.3];
        \end{axis}

        % --- ROW 2: sphere ---
        \coordinate (a) at ($(o)+ (-1.0\aTemp+0.0cm, 0.5\aTemp+\yOffset)$);
        \coordinate (b) at ($(a)+ (0*\aTemp,0)$);
        \begin{axis}[resultsAxis, at=(b)]
            \addplot [forget plot] graphics [xmin=\xmin, xmax=\xmax, ymin=\ymin, ymax=\ymax]
                {"fig/sim/sphere/\fpCondition/real_slice0.0_\fpFileSlug.jpg"};
            \fill[opacity=0.7] (0,0) circle [radius=0.1522];
            \draw[dashed,opacity=0.5] (0,0) circle [radius=0.2];
            \draw[dashed,opacity=0.5] (0,0) circle [radius=0.3];
        \end{axis}
        \coordinate (b) at ($(a)+ (1*\aTemp,0)$);
        \begin{axis}[resultsAxis, at=(b)]
            \addplot [forget plot] graphics [xmin=\xmin, xmax=\xmax, ymin=\ymin, ymax=\ymax]
                {"fig/sim/sphere/\fpCondition/ref_slice0.0_\fpFileSlug.jpg"};
            \draw[opacity=1.0,dotted, thick] (0,0) circle [radius=0.15];
            \draw[dashed,opacity=0.5] (0,0) circle [radius=0.2];
            \draw[dashed,opacity=0.5] (0,0) circle [radius=0.3];
        \end{axis}
        \coordinate (b) at ($(a)+ (2*\aTemp,0)$);
        \begin{axis}[resultsAxis, at=(b)]
            \addplot [forget plot] graphics [xmin=\xmin, xmax=\xmax, ymin=\ymin, ymax=\ymax]
                {"fig/sim/sphere/\fpCondition/mdd_slice0.0_\fpFileSlug.jpg"};
            \draw[opacity=1.0,dotted, thick] (0,0) circle [radius=0.15];
            \draw[dashed,opacity=0.5] (0,0) circle [radius=0.2];
            \draw[dashed,opacity=0.5] (0,0) circle [radius=0.3];
        \end{axis}

        % --- ROW 3: square (label "cube") ---
        \coordinate (a) at ($(o)+ (-1.0\aTemp+0.0cm, -0.5\aTemp+\yOffset)$);
        \coordinate (b) at ($(a)+ (0*\aTemp,0)$);
        \begin{axis}[resultsAxis, at=(b)]
            \addplot [forget plot] graphics [xmin=\xmin, xmax=\xmax, ymin=\ymin, ymax=\ymax]
                {"fig/sim/square/\fpCondition/real_slice0.0_\fpFileSlug.jpg"};
            \fill[opacity=0.7,] (-0.1022,-0.1022) rectangle (0.1022,0.1022);
            \draw[dashed,opacity=0.5] (0,0) circle [radius=0.2];
            \draw[dashed,opacity=0.5] (0,0) circle [radius=0.3];
        \end{axis}
        \coordinate (b) at ($(a)+ (1*\aTemp,0)$);
        \begin{axis}[resultsAxis, at=(b)]
            \addplot [forget plot] graphics [xmin=\xmin, xmax=\xmax, ymin=\ymin, ymax=\ymax]
                {"fig/sim/square/\fpCondition/ref_slice0.0_\fpFileSlug.jpg"};
            \draw[opacity=1.0,dotted, thick] (-0.1,-0.1) rectangle (0.1,0.1);
            \draw[dashed,opacity=0.5] (0,0) circle [radius=0.2];
            \draw[dashed,opacity=0.5] (0,0) circle [radius=0.3];
        \end{axis}
        \coordinate (b) at ($(a)+ (2*\aTemp,0)$);
        \begin{axis}[resultsAxis, at=(b)]
            \addplot [forget plot] graphics [xmin=\xmin, xmax=\xmax, ymin=\ymin, ymax=\ymax]
                {"fig/sim/square/\fpCondition/mdd_slice0.0_\fpFileSlug.jpg"};
            \draw[opacity=1.0,dotted, thick] (-0.1,-0.1) rectangle (0.1,0.1);
            \draw[dashed,opacity=0.5] (0,0) circle [radius=0.2];
            \draw[dashed,opacity=0.5] (0,0) circle [radius=0.3];
        \end{axis}

        % --- ROW 4: cross ---
        \coordinate (a) at ($(o)+ (-1.0\aTemp+0.0cm, -1.5\aTemp+\yOffset)$);
        \coordinate (b) at ($(a)+ (0*\aTemp,0)$);
        \begin{axis}[resultsAxis, at=(b)]
            \addplot [forget plot] graphics [xmin=\xmin, xmax=\xmax, ymin=\ymin, ymax=\ymax]
                {"fig/sim/cross/\fpCondition/real_slice0.0_\fpFileSlug.jpg"};
            \fill[opacity=0.7]
            (0.1522,0.0522) -- (0.1522,-0.0522) -- (0.0522,-0.0522) -- (0.0522,-0.1522) -- (-0.0522,-0.1522) -- (-0.0522,-0.0522) -- (-0.1522,-0.0522) -- (-0.1522,0.0522) -- (-0.0522,0.0522) -- (-0.0522,0.1522) -- (0.0522,0.1522) -- (0.0522,0.0522) -- cycle;
            \draw[dashed,opacity=0.5] (0,0) circle [radius=0.2];
            \draw[dashed,opacity=0.5] (0,0) circle [radius=0.3];
        \end{axis}
        \coordinate (b) at ($(a)+ (1*\aTemp,0)$);
        \begin{axis}[resultsAxis, at=(b)]
            \addplot [forget plot] graphics [xmin=\xmin, xmax=\xmax, ymin=\ymin, ymax=\ymax]
                {"fig/sim/cross/\fpCondition/ref_slice0.0_\fpFileSlug.jpg"};
            \draw[opacity=1.0, dotted, thick]
            (0.15,0.05) -- (0.15,-0.05) -- (0.05,-0.05) -- (0.05,-0.15) -- (-0.05,-0.15) -- (-0.05,-0.05) -- (-0.15,-0.05) -- (-0.15,0.05) -- (-0.05,0.05) -- (-0.05,0.15) -- (0.05,0.15) -- (0.05,0.05) -- cycle;
            \draw[dashed,opacity=0.5] (0,0) circle [radius=0.2];
            \draw[dashed,opacity=0.5] (0,0) circle [radius=0.3];
        \end{axis}
        \coordinate (b) at ($(a)+ (2*\aTemp,0)$);
        \begin{axis}[resultsAxis, at=(b)]
            \addplot [forget plot] graphics [xmin=\xmin, xmax=\xmax, ymin=\ymin, ymax=\ymax]
                {"fig/sim/cross/\fpCondition/mdd_slice0.0_\fpFileSlug.jpg"};
            \draw[opacity=1.0, dotted, thick]
            (0.15,0.05) -- (0.15,-0.05) -- (0.05,-0.05) -- (0.05,-0.15) -- (-0.05,-0.15) -- (-0.05,-0.05) -- (-0.15,-0.05) -- (-0.15,0.05) -- (-0.05,0.05) -- (-0.05,0.15) -- (0.05,0.15) -- (0.05,0.05) -- cycle;
            \draw[dashed,opacity=0.5] (0,0) circle [radius=0.2];
            \draw[dashed,opacity=0.5] (0,0) circle [radius=0.3];
        \end{axis}

    \end{tikzpicture}

    \caption{\fpCaption}
    \label{\fpLabel}

\end{figure*}

\def\fpCondition{scat}
\def\fpFileSlug{t7.0_zSlice}
\def\fpCaption{Scattered pressure field on the xy-slice at $z=0$ at the same time instance as Fig.~\ref{fig:3DdisguisingResults_z_t70_het}, obtained by subtracting the homogeneous field from the heterogeneous case. Domain size 1.6\,m $\times$ 1.6\,m.}
\def\fpLabel{fig:3DdisguisingResults_z_t70_scat}
% Parametrized 4x3 field-plots grid template.
% Used by lib/disguisingFieldPlots[XZ]_t*_*.tex.
%
% Caller defines these via \def before \input{}-ing this file:
%   \fpCondition  -- "het" or "scat" (subdirectory under each row)
%   \fpFileSlug   -- filename slug after "_slice0.0_", e.g. "t7.0_zSlice"
%   \fpCaption    -- caption text
%   \fpLabel      -- label string (full, e.g. "fig:3DdisguisingResults_z_t70_het")
%
% Rows: hom (label "homogeneous"), sphere, square (label "cube"), cross.
% Cols: real, ref ("impulsive"), mdd ("MDD").
\begin{figure*}[p]
    \centering
    \vspace{2.0cm}
    \begin{tikzpicture}[]

        \coordinate (o) at (0,0);

        % column labels
        \node[textNode] at ([xshift=-1\aTemp, yshift=2.1\aTemp+\yOffset] o) {real};
        \node[textNode] at ([xshift=0\aTemp, yshift=2.1\aTemp+\yOffset] o) {impulsive};
        \node[textNode] at ([xshift=1\aTemp, yshift=2.1\aTemp+\yOffset] o) {MDD};

        % row labels
        \node[textNode,rotate=90] at ([xshift=-1.6\aTemp, yshift=1.5\aTemp+\yOffset] o) {homogeneous};
        \node[textNode,rotate=90] at ([xshift=-1.6\aTemp, yshift=0.5\aTemp+\yOffset] o) {sphere};
        \node[textNode,rotate=90] at ([xshift=-1.6\aTemp, yshift=-0.5\aTemp+\yOffset] o) {cube};
        \node[textNode,rotate=90] at ([xshift=-1.6\aTemp, yshift=-1.5\aTemp+\yOffset] o) {cross};

        % --- ROW 1: homogeneous ---
        \coordinate (a) at ($(o)+ (-1.0\aTemp+0cm, 1.5\aTemp+\yOffset)$);
        \coordinate (b) at ($(a)+ (0*\aTemp,0)$);
        \begin{axis}[resultsAxis, at=(b)]
            \addplot [forget plot] graphics [xmin=\xmin, xmax=\xmax, ymin=\ymin, ymax=\ymax]
                {"fig/sim/hom/\fpCondition/real_slice0.0_\fpFileSlug.jpg"};
            \draw[dashed,opacity=0.5] (0,0) circle [radius=0.2];
            \draw[dashed,opacity=0.5] (0,0) circle [radius=0.3];
        \end{axis}
        \coordinate (b) at ($(a)+ (1*\aTemp,0)$);
        \begin{axis}[resultsAxis, at=(b)]
            \addplot [forget plot] graphics [xmin=\xmin, xmax=\xmax, ymin=\ymin, ymax=\ymax]
                {"fig/sim/hom/\fpCondition/ref_slice0.0_\fpFileSlug.jpg"};
            \draw[dashed,opacity=0.5] (0,0) circle [radius=0.2];
            \draw[dashed,opacity=0.5] (0,0) circle [radius=0.3];
        \end{axis}
        \coordinate (b) at ($(a)+ (2*\aTemp,0)$);
        \begin{axis}[resultsAxis, at=(b)]
            \addplot [forget plot] graphics [xmin=\xmin, xmax=\xmax, ymin=\ymin, ymax=\ymax]
                {"fig/sim/hom/\fpCondition/mdd_slice0.0_\fpFileSlug.jpg"};
            \draw[dashed,opacity=0.5] (0,0) circle [radius=0.2];
            \draw[dashed,opacity=0.5] (0,0) circle [radius=0.3];
        \end{axis}

        % --- ROW 2: sphere ---
        \coordinate (a) at ($(o)+ (-1.0\aTemp+0.0cm, 0.5\aTemp+\yOffset)$);
        \coordinate (b) at ($(a)+ (0*\aTemp,0)$);
        \begin{axis}[resultsAxis, at=(b)]
            \addplot [forget plot] graphics [xmin=\xmin, xmax=\xmax, ymin=\ymin, ymax=\ymax]
                {"fig/sim/sphere/\fpCondition/real_slice0.0_\fpFileSlug.jpg"};
            \fill[opacity=0.7] (0,0) circle [radius=0.1522];
            \draw[dashed,opacity=0.5] (0,0) circle [radius=0.2];
            \draw[dashed,opacity=0.5] (0,0) circle [radius=0.3];
        \end{axis}
        \coordinate (b) at ($(a)+ (1*\aTemp,0)$);
        \begin{axis}[resultsAxis, at=(b)]
            \addplot [forget plot] graphics [xmin=\xmin, xmax=\xmax, ymin=\ymin, ymax=\ymax]
                {"fig/sim/sphere/\fpCondition/ref_slice0.0_\fpFileSlug.jpg"};
            \draw[opacity=1.0,dotted, thick] (0,0) circle [radius=0.15];
            \draw[dashed,opacity=0.5] (0,0) circle [radius=0.2];
            \draw[dashed,opacity=0.5] (0,0) circle [radius=0.3];
        \end{axis}
        \coordinate (b) at ($(a)+ (2*\aTemp,0)$);
        \begin{axis}[resultsAxis, at=(b)]
            \addplot [forget plot] graphics [xmin=\xmin, xmax=\xmax, ymin=\ymin, ymax=\ymax]
                {"fig/sim/sphere/\fpCondition/mdd_slice0.0_\fpFileSlug.jpg"};
            \draw[opacity=1.0,dotted, thick] (0,0) circle [radius=0.15];
            \draw[dashed,opacity=0.5] (0,0) circle [radius=0.2];
            \draw[dashed,opacity=0.5] (0,0) circle [radius=0.3];
        \end{axis}

        % --- ROW 3: square (label "cube") ---
        \coordinate (a) at ($(o)+ (-1.0\aTemp+0.0cm, -0.5\aTemp+\yOffset)$);
        \coordinate (b) at ($(a)+ (0*\aTemp,0)$);
        \begin{axis}[resultsAxis, at=(b)]
            \addplot [forget plot] graphics [xmin=\xmin, xmax=\xmax, ymin=\ymin, ymax=\ymax]
                {"fig/sim/square/\fpCondition/real_slice0.0_\fpFileSlug.jpg"};
            \fill[opacity=0.7,] (-0.1022,-0.1022) rectangle (0.1022,0.1022);
            \draw[dashed,opacity=0.5] (0,0) circle [radius=0.2];
            \draw[dashed,opacity=0.5] (0,0) circle [radius=0.3];
        \end{axis}
        \coordinate (b) at ($(a)+ (1*\aTemp,0)$);
        \begin{axis}[resultsAxis, at=(b)]
            \addplot [forget plot] graphics [xmin=\xmin, xmax=\xmax, ymin=\ymin, ymax=\ymax]
                {"fig/sim/square/\fpCondition/ref_slice0.0_\fpFileSlug.jpg"};
            \draw[opacity=1.0,dotted, thick] (-0.1,-0.1) rectangle (0.1,0.1);
            \draw[dashed,opacity=0.5] (0,0) circle [radius=0.2];
            \draw[dashed,opacity=0.5] (0,0) circle [radius=0.3];
        \end{axis}
        \coordinate (b) at ($(a)+ (2*\aTemp,0)$);
        \begin{axis}[resultsAxis, at=(b)]
            \addplot [forget plot] graphics [xmin=\xmin, xmax=\xmax, ymin=\ymin, ymax=\ymax]
                {"fig/sim/square/\fpCondition/mdd_slice0.0_\fpFileSlug.jpg"};
            \draw[opacity=1.0,dotted, thick] (-0.1,-0.1) rectangle (0.1,0.1);
            \draw[dashed,opacity=0.5] (0,0) circle [radius=0.2];
            \draw[dashed,opacity=0.5] (0,0) circle [radius=0.3];
        \end{axis}

        % --- ROW 4: cross ---
        \coordinate (a) at ($(o)+ (-1.0\aTemp+0.0cm, -1.5\aTemp+\yOffset)$);
        \coordinate (b) at ($(a)+ (0*\aTemp,0)$);
        \begin{axis}[resultsAxis, at=(b)]
            \addplot [forget plot] graphics [xmin=\xmin, xmax=\xmax, ymin=\ymin, ymax=\ymax]
                {"fig/sim/cross/\fpCondition/real_slice0.0_\fpFileSlug.jpg"};
            \fill[opacity=0.7]
            (0.1522,0.0522) -- (0.1522,-0.0522) -- (0.0522,-0.0522) -- (0.0522,-0.1522) -- (-0.0522,-0.1522) -- (-0.0522,-0.0522) -- (-0.1522,-0.0522) -- (-0.1522,0.0522) -- (-0.0522,0.0522) -- (-0.0522,0.1522) -- (0.0522,0.1522) -- (0.0522,0.0522) -- cycle;
            \draw[dashed,opacity=0.5] (0,0) circle [radius=0.2];
            \draw[dashed,opacity=0.5] (0,0) circle [radius=0.3];
        \end{axis}
        \coordinate (b) at ($(a)+ (1*\aTemp,0)$);
        \begin{axis}[resultsAxis, at=(b)]
            \addplot [forget plot] graphics [xmin=\xmin, xmax=\xmax, ymin=\ymin, ymax=\ymax]
                {"fig/sim/cross/\fpCondition/ref_slice0.0_\fpFileSlug.jpg"};
            \draw[opacity=1.0, dotted, thick]
            (0.15,0.05) -- (0.15,-0.05) -- (0.05,-0.05) -- (0.05,-0.15) -- (-0.05,-0.15) -- (-0.05,-0.05) -- (-0.15,-0.05) -- (-0.15,0.05) -- (-0.05,0.05) -- (-0.05,0.15) -- (0.05,0.15) -- (0.05,0.05) -- cycle;
            \draw[dashed,opacity=0.5] (0,0) circle [radius=0.2];
            \draw[dashed,opacity=0.5] (0,0) circle [radius=0.3];
        \end{axis}
        \coordinate (b) at ($(a)+ (2*\aTemp,0)$);
        \begin{axis}[resultsAxis, at=(b)]
            \addplot [forget plot] graphics [xmin=\xmin, xmax=\xmax, ymin=\ymin, ymax=\ymax]
                {"fig/sim/cross/\fpCondition/mdd_slice0.0_\fpFileSlug.jpg"};
            \draw[opacity=1.0, dotted, thick]
            (0.15,0.05) -- (0.15,-0.05) -- (0.05,-0.05) -- (0.05,-0.15) -- (-0.05,-0.15) -- (-0.05,-0.05) -- (-0.15,-0.05) -- (-0.15,0.05) -- (-0.05,0.05) -- (-0.05,0.15) -- (0.05,0.15) -- (0.05,0.05) -- cycle;
            \draw[dashed,opacity=0.5] (0,0) circle [radius=0.2];
            \draw[dashed,opacity=0.5] (0,0) circle [radius=0.3];
        \end{axis}

    \end{tikzpicture}

    \caption{\fpCaption}
    \label{\fpLabel}

\end{figure*}

\def\fpCondition{het}
\def\fpFileSlug{t7.0_xSlice}
\def\fpCaption{Heterogeneous pressure field on the yz-slice at $x=0$ at the same time instance as Fig.~\ref{fig:3DdisguisingResults_z_t70_het}. Domain size 1.6\,m $\times$ 1.6\,m.}
\def\fpLabel{fig:3DdisguisingResults_x_t70_het}
% Parametrized 4x3 field-plots grid template.
% Used by lib/disguisingFieldPlots[XZ]_t*_*.tex.
%
% Caller defines these via \def before \input{}-ing this file:
%   \fpCondition  -- "het" or "scat" (subdirectory under each row)
%   \fpFileSlug   -- filename slug after "_slice0.0_", e.g. "t7.0_zSlice"
%   \fpCaption    -- caption text
%   \fpLabel      -- label string (full, e.g. "fig:3DdisguisingResults_z_t70_het")
%
% Rows: hom (label "homogeneous"), sphere, square (label "cube"), cross.
% Cols: real, ref ("impulsive"), mdd ("MDD").
\begin{figure*}[p]
    \centering
    \vspace{2.0cm}
    \begin{tikzpicture}[]

        \coordinate (o) at (0,0);

        % column labels
        \node[textNode] at ([xshift=-1\aTemp, yshift=2.1\aTemp+\yOffset] o) {real};
        \node[textNode] at ([xshift=0\aTemp, yshift=2.1\aTemp+\yOffset] o) {impulsive};
        \node[textNode] at ([xshift=1\aTemp, yshift=2.1\aTemp+\yOffset] o) {MDD};

        % row labels
        \node[textNode,rotate=90] at ([xshift=-1.6\aTemp, yshift=1.5\aTemp+\yOffset] o) {homogeneous};
        \node[textNode,rotate=90] at ([xshift=-1.6\aTemp, yshift=0.5\aTemp+\yOffset] o) {sphere};
        \node[textNode,rotate=90] at ([xshift=-1.6\aTemp, yshift=-0.5\aTemp+\yOffset] o) {cube};
        \node[textNode,rotate=90] at ([xshift=-1.6\aTemp, yshift=-1.5\aTemp+\yOffset] o) {cross};

        % --- ROW 1: homogeneous ---
        \coordinate (a) at ($(o)+ (-1.0\aTemp+0cm, 1.5\aTemp+\yOffset)$);
        \coordinate (b) at ($(a)+ (0*\aTemp,0)$);
        \begin{axis}[resultsAxis, at=(b)]
            \addplot [forget plot] graphics [xmin=\xmin, xmax=\xmax, ymin=\ymin, ymax=\ymax]
                {"fig/sim/hom/\fpCondition/real_slice0.0_\fpFileSlug.jpg"};
            \draw[dashed,opacity=0.5] (0,0) circle [radius=0.2];
            \draw[dashed,opacity=0.5] (0,0) circle [radius=0.3];
        \end{axis}
        \coordinate (b) at ($(a)+ (1*\aTemp,0)$);
        \begin{axis}[resultsAxis, at=(b)]
            \addplot [forget plot] graphics [xmin=\xmin, xmax=\xmax, ymin=\ymin, ymax=\ymax]
                {"fig/sim/hom/\fpCondition/ref_slice0.0_\fpFileSlug.jpg"};
            \draw[dashed,opacity=0.5] (0,0) circle [radius=0.2];
            \draw[dashed,opacity=0.5] (0,0) circle [radius=0.3];
        \end{axis}
        \coordinate (b) at ($(a)+ (2*\aTemp,0)$);
        \begin{axis}[resultsAxis, at=(b)]
            \addplot [forget plot] graphics [xmin=\xmin, xmax=\xmax, ymin=\ymin, ymax=\ymax]
                {"fig/sim/hom/\fpCondition/mdd_slice0.0_\fpFileSlug.jpg"};
            \draw[dashed,opacity=0.5] (0,0) circle [radius=0.2];
            \draw[dashed,opacity=0.5] (0,0) circle [radius=0.3];
        \end{axis}

        % --- ROW 2: sphere ---
        \coordinate (a) at ($(o)+ (-1.0\aTemp+0.0cm, 0.5\aTemp+\yOffset)$);
        \coordinate (b) at ($(a)+ (0*\aTemp,0)$);
        \begin{axis}[resultsAxis, at=(b)]
            \addplot [forget plot] graphics [xmin=\xmin, xmax=\xmax, ymin=\ymin, ymax=\ymax]
                {"fig/sim/sphere/\fpCondition/real_slice0.0_\fpFileSlug.jpg"};
            \fill[opacity=0.7] (0,0) circle [radius=0.1522];
            \draw[dashed,opacity=0.5] (0,0) circle [radius=0.2];
            \draw[dashed,opacity=0.5] (0,0) circle [radius=0.3];
        \end{axis}
        \coordinate (b) at ($(a)+ (1*\aTemp,0)$);
        \begin{axis}[resultsAxis, at=(b)]
            \addplot [forget plot] graphics [xmin=\xmin, xmax=\xmax, ymin=\ymin, ymax=\ymax]
                {"fig/sim/sphere/\fpCondition/ref_slice0.0_\fpFileSlug.jpg"};
            \draw[opacity=1.0,dotted, thick] (0,0) circle [radius=0.15];
            \draw[dashed,opacity=0.5] (0,0) circle [radius=0.2];
            \draw[dashed,opacity=0.5] (0,0) circle [radius=0.3];
        \end{axis}
        \coordinate (b) at ($(a)+ (2*\aTemp,0)$);
        \begin{axis}[resultsAxis, at=(b)]
            \addplot [forget plot] graphics [xmin=\xmin, xmax=\xmax, ymin=\ymin, ymax=\ymax]
                {"fig/sim/sphere/\fpCondition/mdd_slice0.0_\fpFileSlug.jpg"};
            \draw[opacity=1.0,dotted, thick] (0,0) circle [radius=0.15];
            \draw[dashed,opacity=0.5] (0,0) circle [radius=0.2];
            \draw[dashed,opacity=0.5] (0,0) circle [radius=0.3];
        \end{axis}

        % --- ROW 3: square (label "cube") ---
        \coordinate (a) at ($(o)+ (-1.0\aTemp+0.0cm, -0.5\aTemp+\yOffset)$);
        \coordinate (b) at ($(a)+ (0*\aTemp,0)$);
        \begin{axis}[resultsAxis, at=(b)]
            \addplot [forget plot] graphics [xmin=\xmin, xmax=\xmax, ymin=\ymin, ymax=\ymax]
                {"fig/sim/square/\fpCondition/real_slice0.0_\fpFileSlug.jpg"};
            \fill[opacity=0.7,] (-0.1022,-0.1022) rectangle (0.1022,0.1022);
            \draw[dashed,opacity=0.5] (0,0) circle [radius=0.2];
            \draw[dashed,opacity=0.5] (0,0) circle [radius=0.3];
        \end{axis}
        \coordinate (b) at ($(a)+ (1*\aTemp,0)$);
        \begin{axis}[resultsAxis, at=(b)]
            \addplot [forget plot] graphics [xmin=\xmin, xmax=\xmax, ymin=\ymin, ymax=\ymax]
                {"fig/sim/square/\fpCondition/ref_slice0.0_\fpFileSlug.jpg"};
            \draw[opacity=1.0,dotted, thick] (-0.1,-0.1) rectangle (0.1,0.1);
            \draw[dashed,opacity=0.5] (0,0) circle [radius=0.2];
            \draw[dashed,opacity=0.5] (0,0) circle [radius=0.3];
        \end{axis}
        \coordinate (b) at ($(a)+ (2*\aTemp,0)$);
        \begin{axis}[resultsAxis, at=(b)]
            \addplot [forget plot] graphics [xmin=\xmin, xmax=\xmax, ymin=\ymin, ymax=\ymax]
                {"fig/sim/square/\fpCondition/mdd_slice0.0_\fpFileSlug.jpg"};
            \draw[opacity=1.0,dotted, thick] (-0.1,-0.1) rectangle (0.1,0.1);
            \draw[dashed,opacity=0.5] (0,0) circle [radius=0.2];
            \draw[dashed,opacity=0.5] (0,0) circle [radius=0.3];
        \end{axis}

        % --- ROW 4: cross ---
        \coordinate (a) at ($(o)+ (-1.0\aTemp+0.0cm, -1.5\aTemp+\yOffset)$);
        \coordinate (b) at ($(a)+ (0*\aTemp,0)$);
        \begin{axis}[resultsAxis, at=(b)]
            \addplot [forget plot] graphics [xmin=\xmin, xmax=\xmax, ymin=\ymin, ymax=\ymax]
                {"fig/sim/cross/\fpCondition/real_slice0.0_\fpFileSlug.jpg"};
            \fill[opacity=0.7]
            (0.1522,0.0522) -- (0.1522,-0.0522) -- (0.0522,-0.0522) -- (0.0522,-0.1522) -- (-0.0522,-0.1522) -- (-0.0522,-0.0522) -- (-0.1522,-0.0522) -- (-0.1522,0.0522) -- (-0.0522,0.0522) -- (-0.0522,0.1522) -- (0.0522,0.1522) -- (0.0522,0.0522) -- cycle;
            \draw[dashed,opacity=0.5] (0,0) circle [radius=0.2];
            \draw[dashed,opacity=0.5] (0,0) circle [radius=0.3];
        \end{axis}
        \coordinate (b) at ($(a)+ (1*\aTemp,0)$);
        \begin{axis}[resultsAxis, at=(b)]
            \addplot [forget plot] graphics [xmin=\xmin, xmax=\xmax, ymin=\ymin, ymax=\ymax]
                {"fig/sim/cross/\fpCondition/ref_slice0.0_\fpFileSlug.jpg"};
            \draw[opacity=1.0, dotted, thick]
            (0.15,0.05) -- (0.15,-0.05) -- (0.05,-0.05) -- (0.05,-0.15) -- (-0.05,-0.15) -- (-0.05,-0.05) -- (-0.15,-0.05) -- (-0.15,0.05) -- (-0.05,0.05) -- (-0.05,0.15) -- (0.05,0.15) -- (0.05,0.05) -- cycle;
            \draw[dashed,opacity=0.5] (0,0) circle [radius=0.2];
            \draw[dashed,opacity=0.5] (0,0) circle [radius=0.3];
        \end{axis}
        \coordinate (b) at ($(a)+ (2*\aTemp,0)$);
        \begin{axis}[resultsAxis, at=(b)]
            \addplot [forget plot] graphics [xmin=\xmin, xmax=\xmax, ymin=\ymin, ymax=\ymax]
                {"fig/sim/cross/\fpCondition/mdd_slice0.0_\fpFileSlug.jpg"};
            \draw[opacity=1.0, dotted, thick]
            (0.15,0.05) -- (0.15,-0.05) -- (0.05,-0.05) -- (0.05,-0.15) -- (-0.05,-0.15) -- (-0.05,-0.05) -- (-0.15,-0.05) -- (-0.15,0.05) -- (-0.05,0.05) -- (-0.05,0.15) -- (0.05,0.15) -- (0.05,0.05) -- cycle;
            \draw[dashed,opacity=0.5] (0,0) circle [radius=0.2];
            \draw[dashed,opacity=0.5] (0,0) circle [radius=0.3];
        \end{axis}

    \end{tikzpicture}

    \caption{\fpCaption}
    \label{\fpLabel}

\end{figure*}

% =======================================================================
% Incident-wave tag for the 3D scattering belt figures
% (lib/scatteringReal.tex, scatteringRef.tex, scatteringMDD.tex).
%
% Draws a left-to-right arrow (the incident plane wave) striking a
% scatterer marker near an image's west edge, just right of the row
% label. Real scatterers use \incidentTag; the holographic cases
% (impulsive, mdd) use \incidentTagHolo, which rings the marker with two
% concentric dashed circles (IBC surfaces) and draws the marker at
% 30% opacity. Edit the geometry/colours here once to update all three
% figures. Offsets are measured leftward from the image edge (negative
% = into the image) — retune once the JPGs are re-cropped to the lobes.
% =======================================================================
\newcommand{\imageShift}{1cm}        % whole-figure shift (image + its tag)
\newcommand{\typeLabelX}{2cm}        % bold real/impulsive/mdd label, left of edge
\newcommand{\tagMarker}{0.48cm}      % scatterer marker half-extent
\newcommand{\tagArm}{0.18cm}         % cross-arm half-thickness
\newcommand{\tagHoloR}{0.74cm}       % inner dashed hologram-ring radius
\newcommand{\tagHoloRouter}{1.04cm}  % outer dashed hologram-ring radius
\newcommand{\tagMarkerX}{-1.40cm}    % marker centre, left of the image edge
\newcommand{\tagArrowBack}{0.64cm}   % arrow tail,    left of the image edge
\newcommand{\tagArrowTip}{-0.06cm}   % arrow head,    left of the image edge
\newcommand{\tagLabelX}{0.82cm}      % row label,     left of the image edge
\colorlet{tagfill}{black!52}         % scatterer marker shade
\colorlet{tagface}{black!32}         % cross-centre square (3D highlight)

% Scatterer markers, drawn around the coordinate (mk).
\newcommand{\tagmarkersphere}{%
    \fill[tagfill] (mk) circle[radius=\tagMarker];}
\newcommand{\tagmarkercube}{%
    \fill[tagfill] ($(mk)-(\tagMarker,\tagMarker)$) rectangle ($(mk)+(\tagMarker,\tagMarker)$);}
\newcommand{\tagmarkercross}{%
    \fill[tagfill] ($(mk)-(\tagMarker,\tagArm)$) rectangle ($(mk)+(\tagMarker,\tagArm)$);%
    \fill[tagfill] ($(mk)-(\tagArm,\tagMarker)$) rectangle ($(mk)+(\tagArm,\tagMarker)$);%
    \fill[tagface] ($(mk)-(\tagArm,\tagArm)$) rectangle ($(mk)+(\tagArm,\tagArm)$);}

% Shared body: #1 = image node, #2 = sphere|cube|cross,
%               #3 = 1 adds the dashed ring, #4 = marker opacity.
\newcommand{\incidentTagBody}[4]{%
    \begin{scope}[overlay]
        \coordinate (mk) at ($(#1.west)-(\tagMarkerX,0)$);
        \ifnum#3=1
            \draw[dashed,line width=0.8pt,black!55] (mk) circle[radius=\tagHoloR];
            \draw[dashed,line width=0.8pt,black!55] (mk) circle[radius=\tagHoloRouter];
        \fi
        \draw[-{Stealth[length=2.4mm,width=2.8mm]},line width=1.2pt,black!62]
            ($(#1.west)-(\tagArrowBack,0)$) -- ($(#1.west)-(\tagArrowTip,0)$);
        \begin{scope}[opacity=#4, transparency group]
            \csname tagmarker#2\endcsname
        \end{scope}
    \end{scope}%
}

% Real scatterer: opaque marker, no ring.
\newcommand{\incidentTag}[2]{\incidentTagBody{#1}{#2}{0}{1}}

% Holographic scatterer (impulsive, mdd): the real marker colour drawn
% at 30% opacity (70% transparent), inside a dashed hologram ring.
\newcommand{\incidentTagHolo}[2]{\incidentTagBody{#1}{#2}{1}{0.3}}

\def\sPrefix{real}%
\def\sTagMacro{\incidentTag}%
\def\sTypeLabel{real}%
\def\sCaption{The scattered intensity of the real scatterers as a function of the incident angle.}%
\def\sLabel{fig:scattering3Dreal}%
% Parametrized 3-row scattering panel template (sphere/cube/cross).
% Used by lib/scattering{Real,Ref,MDD}.tex.
%
% Caller defines via \def before \input{}-ing this file:
%   \sPrefix    -- jpg filename prefix, e.g. "real", "ref", "mdd"
%   \sTagMacro  -- incident-wave tag macro, e.g. \incidentTag or \incidentTagHolo
%   \sTypeLabel -- text label shown on the middle row, e.g. "real", "impulsive", "mdd"
%   \sCaption   -- caption text
%   \sLabel     -- label string
%
% Native 5:1 belt aspect. \imageSize is the WIDTH; the height computes to
% \imageSize/5 automatically. \imageSeparation is the row-centre distance.
% \providecommand keeps the first \input setting the values and the rest
% finding them already defined (matching the original \newcommand-then-
% \renewcommand pattern across scatteringReal -> scatteringRef -> scatteringMDD).
\providecommand{\imageSize}{\textwidth}
\providecommand{\imageSeparation}{8cm}

\begin{figure*}[p]
    \centering
    \vspace{1cm}
    \hspace*{\imageShift}\begin{tikzpicture}[image/.style={inner sep=0pt}]

        \coordinate (ref) at ($(0,0)+(0,1.7cm)$);
        \coordinate (top) at ($(ref)+(0,+\imageSeparation)$);
        \coordinate (middle) at ($(ref)+(0,0)$);
        \coordinate (bottom) at ($(ref)+(0,-\imageSeparation)$);

        % Row 1 — sphere
        \node[image] (sphereImage) at (top) {\includegraphics[width=\imageSize]{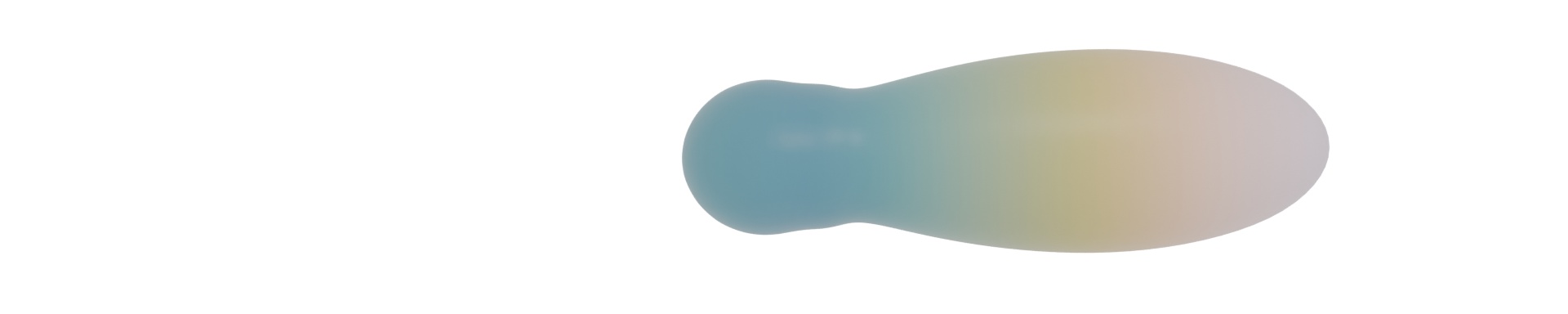}};
        \node[anchor=east,overlay] at ($(sphereImage.west)-(\tagLabelX,0)$) {sphere};
        \sTagMacro{sphereImage}{sphere}

        % Row 2 — cube
        \node[image] (squareImage) at (middle) {\includegraphics[width=\imageSize]{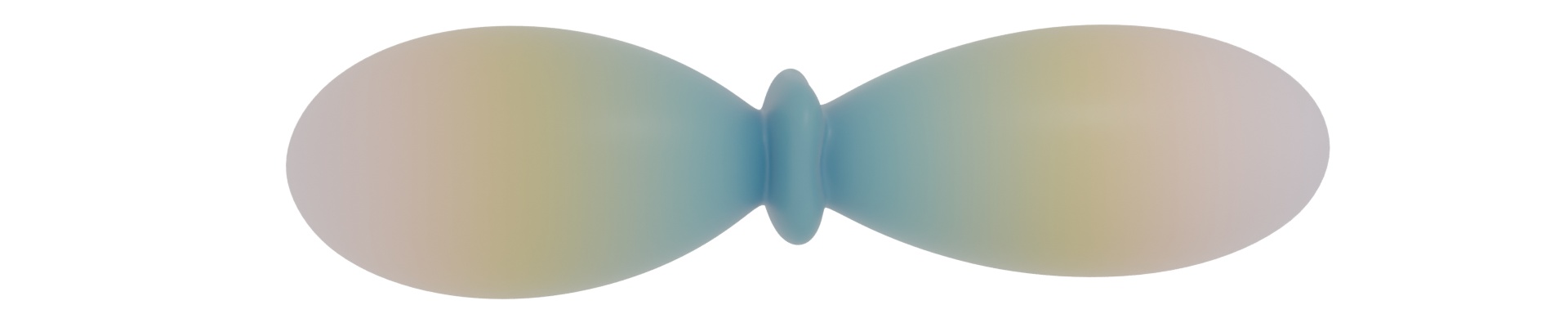}};
        \node[anchor=east,overlay] at ($(squareImage.west)-(\tagLabelX,0)$) {cube};
        \sTagMacro{squareImage}{cube}
        \node[anchor=west,font=\bfseries,overlay] at ($(squareImage.east)-(\typeLabelX,0)$) {\sTypeLabel};

        % Row 3 — cross
        \node[image] (crossImage) at (bottom) {\includegraphics[width=\imageSize]{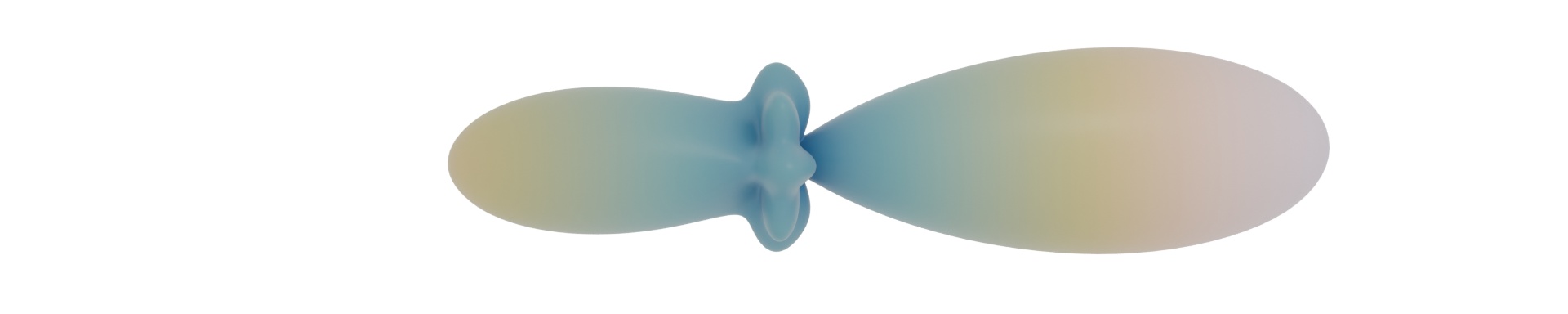}};
        \node[anchor=east,overlay] at ($(crossImage.west)-(\tagLabelX,0)$) {cross};
        \sTagMacro{crossImage}{cross}

    \end{tikzpicture}\hspace*{-\imageShift}

    \vspace{1.5cm}
    \caption{\sCaption}
    \label{\sLabel}

\end{figure*}

\def\sPrefix{ref}%
\def\sTagMacro{\incidentTagHolo}%
\def\sTypeLabel{impulsive}%
\def\sCaption{The scattered intensity of the holographic scatterers driven by impulsive Green's functions, as a function of the incident angle.}%
\def\sLabel{fig:scattering3Dref}%

\def\sPrefix{mdd}%
\def\sTagMacro{\incidentTagHolo}%
\def\sTypeLabel{mdd}%
\def\sCaption{The scattered intensity of the holographic scatterers driven by MDD-retrieved Green's functions, as a function of the incident angle.}%
\def\sLabel{fig:scattering3Dmdd}%

\end{document}